\documentclass[apj]{emulateapj}
\usepackage{apjfonts}
\usepackage{graphicx}

\slugcomment{Accepted for the publication in ApJ}
\shorttitle{Massive YSOs in the Galactic Center}
\shortauthors{An et~al.}

\begin{document}
\title{Massive Young Stellar Objects in the Galactic Center. I.\\
Spectroscopic Identification from {\it Spitzer}/IRS Observations}

\author{Deokkeun An\altaffilmark{1},
Solange V.\ Ram\'irez\altaffilmark{2},
Kris Sellgren\altaffilmark{3},
Richard G. Arendt\altaffilmark{4},
A.\ C.\ Adwin Boogert\altaffilmark{2},\\
Thomas P.\ Robitaille\altaffilmark{5,6},
Mathias Schultheis\altaffilmark{7},
Angela S.\ Cotera\altaffilmark{8},\\
Howard A.\ Smith\altaffilmark{5},
Susan R.\ Stolovy\altaffilmark{9}
}

\altaffiltext{1}{Department of Science Education,
Ewha Womans University, Seoul 120-750, Korea; deokkeun@ewha.ac.kr.}
\altaffiltext{2}{NASA Exoplanet Science Institute,
California Institute of Technology, Mail Stop 100-22, Pasadena, CA 91125.}
\altaffiltext{3}{Department of Astronomy, The Ohio State University,
140 West 18th Avenue, Columbus, OH 43210.}
\altaffiltext{4}{CRESST/UMBC/GSFC, Code 665, NASA/Goddard Space Flight Center,
8800 Greenbelt Road, Greenbelt, MD 20771.}
\altaffiltext{5}{Harvard-Smithsonian Center for Astrophysics, 60 Garden Street,
Cambridge, MA 02138.}
\altaffiltext{6}{Spitzer Postdoctoral Fellow.}
\altaffiltext{7}{Observatoire de Besan\c{c}on, 41bis, avenue de l'Observatoire,
25000 Besan\c{c}on, France.}
\altaffiltext{8}{SETI Institute, 515 North Whisman Road, Mountain View, CA 94043.}
\altaffiltext{9}{Spitzer Science Center, California Institute of Technology,
Mail Code 220-6, 1200 East California Boulevard, Pasadena, CA 91125.}

\begin{abstract}

We present results from our spectroscopic study, using the Infrared Spectrograph 
(IRS) onboard the {\it Spitzer Space Telescope}, designed to identify massive young 
stellar objects (YSOs) in the Galactic Center (GC). Our sample of 107 YSO candidates
was selected based on IRAC colors from the high spatial resolution, high sensitivity
{\it Spitzer}/IRAC images in the Central Molecular Zone (CMZ), which spans the
central $\sim300$~pc region of the Milky Way Galaxy. We obtained IRS spectra over
$5\ \mu$m to $35\ \mu$m using both high- and low-resolution IRS modules. We
spectroscopically identify massive YSOs by the presence of a 15.4 $\mu$m shoulder
on the absorption profile of $15\ \mu$m CO$_2$ ice, suggestive of CO$_2$ ice mixed
with CH$_3$OH ice on grains. This 15.4 $\mu$m shoulder is clearly observed in 16
sources and possibly observed in an additional 19 sources. We show that 9 massive
YSOs also reveal molecular gas-phase absorption from CO$_2$, C$_2$H$_2$, and/or HCN,
which traces warm and dense gas in YSOs. Our results provide the first spectroscopic
census of the massive YSO population in the GC. We fit YSO models to the observed
spectral energy distributions and find YSO masses of $8 - 23\ M_\odot$, which
generally agree with the masses derived from observed radio continuum emission.
We find that about 50\% of photometrically identified YSOs are confirmed with our
spectroscopic study. This implies a preliminary star formation rate of
$\sim0.07\ M_\odot\ yr^{-1}$ at the GC.
\end{abstract}


\section{Introduction}

Our Galactic center (GC), at a distance of $7.9\pm0.8$~kpc \citep{reid:09},
is the closest galactic nucleus, observable at spatial resolutions 
unapproachable in other galaxies (1 pc $\approx$ 26$''$). The extent of the
GC region is defined by a region of relatively high density molecular gas
\citep[$n_{H_2} \sim 10^{4}$ cm$^{-3}$;][]{bally:87}, covering the inner
$200~{\rm pc} \times 50~{\rm pc}$ ($170\arcmin\times40\arcmin$), called
the Central Molecular Zone (CMZ). The CMZ produces 5\%--10\% of the Galaxy's
infrared and Lyman continuum luminosity and contains 10\% of its molecular gas
\citep{smith:78,nishimura:80,bally:87,bally:88,morris:96}. The CMZ exhibits
extreme conditions with high gas temperature, pressure, turbulence, strong
magnetic field strengths and strong tidal shear \citep{serabyn:96,fatuzzo:09}.
As a result, star formation in the CMZ may be altered or suppressed.

The CMZ, nevertheless, shows several signposts of recent massive star formation,  
such as (compact) H II regions and supernova remnants. In addition, there are
massive young stars \citep[ages of $\sim$2-7 Myrs;][]{krabbe:91,figer:99}
in three known discrete star clusters -- the Central, Quintuplet, and Arches
clusters -- which make the CMZ distinctly different from the Galactic bulge with
its predominantly old stellar population \citep{frogel:87}. Nevertheless, it has
been unclear how star formation proceeds in this hostile environment. There have
been several studies in the literature that identified young stellar object (YSO)
candidates in the GC based on infrared  photometry
\citep[e.g.,][]{felli:02,schuller:06,yusefzadeh:09}.
The high and patchy extinction towards the GC ($A_V \approx 30$) and its mix
of young and old stellar populations, however, mean that spectroscopic
observations are required to confirm YSO identifications. This is because red
giants and asymptotic giant branch (AGB) stars (also part of the GC stellar
population) can look like YSOs from broad-band photometry, if they are heavily
dust attenuated \citep[e.g.,][]{schultheis:03}.

The GC provides a unique opportunity to investigate circumnuclear star formation 
with an unprecedented spatial resolution. We announced the first spectroscopic
identification of massive YSOs in the CMZ \citep[][hereafter A09]{an:09}, using
the Infrared Spectrograph \citep[IRS;][]{houck:04} onboard the
{\it Spitzer Space Telescope} \citep{werner:04}. In this paper, we follow up our
initial exploration of the IRS data set in A09 and refine our methods to identify
YSOs in the CMZ, aiming at providing a list of spectroscopically confirmed YSOs
as tracers of the early stages of star formation in the GC. As described and
employed in A09, our selection criteria for YSOs are based on gas- and solid-phase
absorption from mid-IR spectroscopy. This includes  solid-phase absorption from
the CO$_2$ bending mode \citep[e.g.,][]{gerakines:99} and gas-phase absorption
from C$_2$H$_2$, HCN, and CO$_2$ \citep[e.g.,][]{lahuis:00,boonman:03,knez:09}.
We look for signatures of CO$_2$ ice mixed with a large amount of CH$_3$OH ice.
This combination has been observed towards high-mass YSOs and low-mass YSOs
\citep{gerakines:99,pontoppidan:08,zasowski:09,seale:11}, but not toward field
stars behind molecular clouds \citep{gerakines:99,bergin:05,knez:05,whittet:07,
whittet:09}.

In \S~\ref{sec:method} we summarize the IRS target selection criteria and data
reduction. In \S~\ref{sec:results} we describe our spectroscopic identification
of YSOs, showing that 15\%--30\% of our 107 targets are massive YSOs. We measure
the extinction for YSOs and possible YSOs, along with column densities of
solid-phase and gas-phase molecular absorbers. In \S~\ref{sec:analysis} we
examine properties of these YSOs and possible YSOs, and derive a preliminary
estimate of the star formation rate in the GC.

\section{Methods}\label{sec:method}

In this section we describe procedures for the sample selection, spectroscopic
follow-up observations, and IRS data reduction. Parallel information on these
subjects can be found in A09, but here we repeat this for the reader's convenience
with additional details where there has been improvements in the data reduction
steps.

\subsection{{\it Spitzer}/IRS Sample}\label{sec:sample}

Our 107 spectroscopic targets (Table~\ref{tab:tab1}) were selected from the GC
point source catalog \citep{ramirez:08} extracted from the IRAC images of the CMZ
\citep{stolovy:06} made using the Infrared Array Camera \citep[IRAC;][]{fazio:04}
onboard the {\it Spitzer Space Telescope}. These images cover $280$~pc $\times$ $200$~pc
in the four IRAC channels ($3.6\ \mu$m, $4.5\ \mu$m, $5.8\ \mu$m, and $8.0\ \mu$m)
with uniform high sensitivity. Compared to earlier imaging surveys of this region,
such as that from the {\it Midcourse Space Experiment} \citep[{\it MSX};][]{price:01}
or ISOGAL \citep{omont:03}, the IRAC images have a higher spatial resolution
($\approx2\arcsec$ vs.\ $> 6\arcsec$ of earlier surveys), which has led not
only to a better estimate of source fluxes, but also to more accurate source
positions for  follow-up spectroscopic observations.

The spectroscopic sample was selected using IRAC color criteria based on 
the \citet{whitney:04} study of the giant \ion{H}{2} region RCW~49.
\citeauthor{whitney:04} determined the locations of YSOs with $2.5\ M_\odot$,
$3.8\ M_\odot$, and $5.9\ M_\odot$ on the IRAC color-magnitude diagrams, using
radiative transfer models described in \citet{whitney:03}. From this we chose
an initial color criterion \citep[${\rm [3.6] - [8.0]} \geq 2.0$;][]{whitney:04}.
We added a latitudinal constraint ($|b|<15$\arcmin) to increase the probability
that the objects are located at the distance of the GC \citep[$8$~kpc;][]{reid:09} 
rather than in one of the several intervening spiral arms along the line of sight.
We note that the range of this latitude selection is about 5 times larger
than the scale height of photometric YSO candidates ($\sim7$~pc) in \citet{yusefzadeh:09}.
These color and position constraints provided an initial sample of 1207 objects
from the GC point source catalog.

We combined the IRAC photometry with Two Micron All Sky Survey (2MASS)
photometry \citep[$JHK_s$;][]{skrutskie:06} and ISOGAL $7\ \mu$m and $15\ \mu$m
point-source catalogs. Note that $24\ \mu$m Multiband Imaging Photometer for
{\it Spitzer} (MIPS) observations at the GC were not available at the time
when our IRS sample was chosen. Among the initially selected 1207 objects,
336 had photometry in at least 5 bandpasses, which allowed reliable spectral
energy distribution (SED) fitting. We used the SED fitting tool developed by
\citet{robitaille:07}, which makes use of a grid of 200,000 YSO models
\citep{robitaille:06} to estimate YSO parameters, such 
as the mass of a central object (see \S~\ref{sec:model}).
For those objects with photometry in fewer than five bandpasses,
we instead applied color constraints based on the
work of \citet{whitney:04} of
${\rm [3.6]-[4.5]} \geq 0.5$, ${\rm [4.5]-[5.8]} \geq 0.5$, and
${\rm [5.8]-[8.0]} \geq 1.0$.

The SED fitting and additional IRAC color constraints narrowed our sample down to
about 200 objects, which were then further inspected using IRAC three-color
images. 
The sources were evaluated by their distinctiveness (i.e., whether sources are 
easily distinguishable from the background) 
and their local background emission. 
Among 200 objects examined, 112 were found to exhibit the necessary distinctiveness 
within the IRS slit widths.

\begin{figure*}
\epsscale{1.1}
\plotone{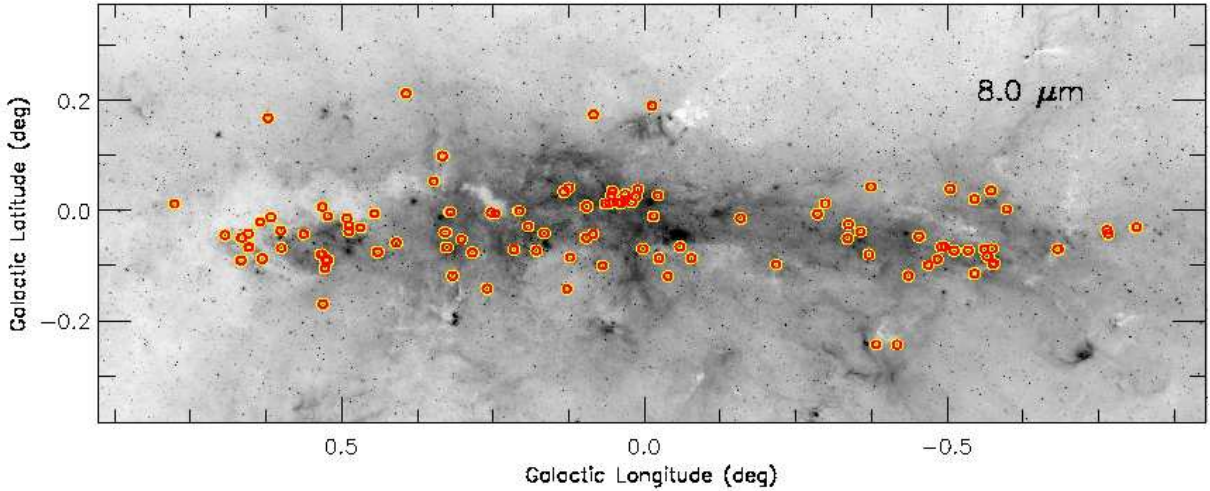}
\caption{Spatial distribution of 107 IRS targets on the IRAC $8.0\ \mu$m image
\citep{stolovy:06}.
The image shows the entire CMZ covering approximately $100\arcmin \times
40\arcmin$ centered on the GC.
Our IRS targets (shown in circles) were selected from the point sources
of this survey and they are uniformly distributed over the CMZ.
\label{fig:map}}
\end{figure*}

A literature search was performed on the 112 objects, yielding matches to 43
previously studied sources: 25 sources were previously-identified 
photometric YSO candidates,
4 were OH/IR stars, one was a Wolf-Rayet star, and the remaining 13 sources were
others (e.g., radio sources, X-ray sources, etc). 
Note that nearly 60\% ($\approx25/43$) of the objects had been selected
as YSO candidates by other methods. 
The 4 OH/IR stars and the Wolf-Rayet star were discarded from the final sample, 
giving a total of 107 massive YSO candidates.
The spatial location of the 107 massive YSO candidates of our sample is shown in 
Figure~\ref{fig:map}.

In spite of our efforts to exclude OH/IR stars from our YSO sample, 
we later realized that several of our targets appear to be coincident 
with a stellar maser source and/or a long-period variable
(SSTGC~284291, 425399, 564417, 
619964, 660708, 696367, and 711462).
SSTGC~517724 is now identified as an OB supergiant
by \citet{mauerhan:10}.
These sources known not to be YSOs have been helpful in
refining our spectroscopic YSO selection criteria.

Massive YSOs are our primary targets for the follow-up spectroscopic
observations (\S~\ref{sec:obs}) because our adopted color selection criteria set
the lower limit on the mass of the central object to be $M_* \ga \ 2.5\ M_\odot$
\citep{whitney:03,whitney:04}.
In addition, the source confusion limit in the input GC point 
source catalog 
\citep[see Figure~12 in][]{ramirez:08}, 
together with the 8~kpc distance to the GC \citep{reid:09} 
and $A_V\sim30$~mag of visual extinction, limits us to detecting 
YSOs with masses $\ga 6\ M_\odot$
(see \S~\ref{sec:analysis}).
Thus any YSO we identify in this paper is a {\it massive} YSO.

\subsection{Observations}\label{sec:obs}

Our IRS observations with a total integration time of $56$ hours were carried out 
in May and October 2008 (see Table~\ref{tab:tab1}) as part of {\it Spitzer}
Cycle~4 (Program ID: 40230, PI: S.\ Ram\'irez).
We observed our 107 targets with both high- and low-resolution IRS modules: 
short-high ({\tt SH}; $9.9\ \mu$m$-19.6\ \mu$m, $\lambda / \Delta \lambda \sim 600$),
long-high ({\tt LH}; $18.7\ \mu$m$-37.2\ \mu$m, $\lambda / \Delta \lambda \sim 600$),
short-low ({\tt SL}; $5.2\ \mu$m$-14.5\ \mu$m, $\lambda / \Delta \lambda \sim 60-127$), and
long-low ({\tt LL}; $14\ \mu$m$-38\ \mu$m, $\lambda / \Delta \lambda \sim 57-126$).

In Table~\ref{tab:tab2} we list IRS modules used in the current analysis for each
of the sources that are spectroscopically identified as a YSO 
or possible YSO in this paper
(see \S~\ref{sec:yso}). 
Most of these targets were observed with all of the four
IRS modules. 
We did not obtain spectra with
the {\tt SL} module for
some of the 107 YSO candidates,
including possible YSO SSTGC~610642, 
because of saturation 
in the IRS peak-up arrays (see below).
For a few sources we rejected
data in the first order of {\tt LL} ({\tt LL1};
$19.5\ \mu$m - $38.0\ \mu$m) because a large fraction of pixel
values were flagged as invalid. 

We divided our sample into four subsamples according to their IRAC [8.0]
magnitudes: [8.0] $\leq 6$~mag ($N=30$ objects), $6$~mag $ < $ [8.0] $ \leq 7$~mag
($N=28$), $7$~mag $<$ [8.0] $\leq 8$~mag ($N=28$), and [8.0] $>  8$~mag ($N=21$).
Exposure times were determined for each brightness subsample to achieve
a signal-to-noise (S/N) ratio of at least $50$ in {\tt SH} and {\tt SL}, and
a minimum S/N of $10$ in {\tt LH} and {\tt LL}. Our exposure times are
$6$~sec--$120$~sec in {\tt SH}, $6$~sec--$60$~sec in {\tt LH}, $6$~sec--$14$~sec 
in {\tt SL}, and $6$~sec in {\tt LL} modules. Subsamples were further grouped
based on spatial location. These groupings allowed us to observe 107 sources using
nine ``fixed cluster'' target observations, a strategy which proved to greatly
increase the observing efficiency by reducing  overheads due to telescope movement.
Each object was observed in the IRS staring mode with 4 exposures per source
(2 cycles) to properly correct for bad pixels.

Our observations were carried out without specific IRS peak-up sequences, since
target coordinates were accurate enough \citep[$<1\arcsec$;][]{ramirez:08} for our
science goals.
In addition, the background at the GC is too high for a peak-up sequence
to work even using a 2MASS source.
In fact, the background at the GC is so high that constraints were 
placed on the observing dates (which determine the telescope roll angle)
to avoid saturation of the IRS peak-up arrays.  Such saturation 
leads to incorrect droop corrections in the standard IRS pipelines and causes 
various defects on {\tt SL} frames (where the peak-up arrays are located).
The {\tt SL} observations were not carried out for 21 out of
107 targets because of
saturation in the peak-up arrays regardless of the date of observation.

Multiple off-source measurements at several different locations were carried out to
derive background spectra around each target, because strong and spatially variable
background at the GC can affect resulting line and/or continuum emission from the source.
Since the high-resolution slits are not long enough to take both source and background
measurements simultaneously, we located four background positions around each
target ($\sim \pm$1 \arcmin \ offsets in right ascension, $\sim \pm$1\arcmin \ 
offsets in declination).
Specific background positions for both {\tt SH} and {\tt LH} were determined to avoid background
sources and to properly interpolate background emission near the source position
over a $\sim1\arcmin$ scale.
The longer slit sizes of the low-resolution modules permit background
measurements along the on-source slit; we also identified two additional background
positions that are $\sim \pm$ 1 \arcmin\ away in the direction perpendicular to the {\tt SL} or {\tt LL} slit.
These dedicated background slits for {\tt SL} and {\tt LL} were centered on two of the high-resolution
background positions.

\begin{figure*}
\centering
\includegraphics[scale=0.26]{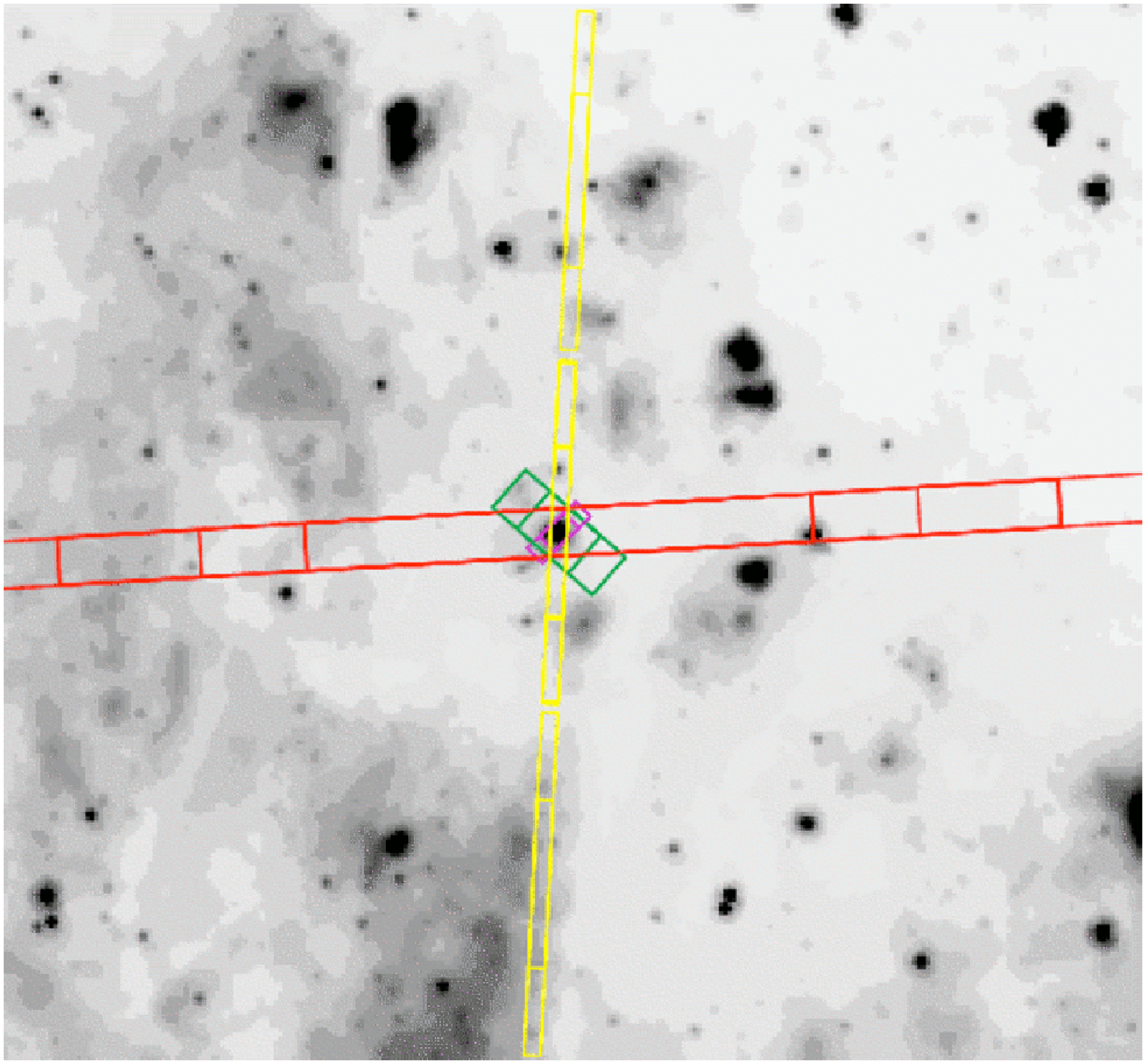}
\includegraphics[scale=0.26]{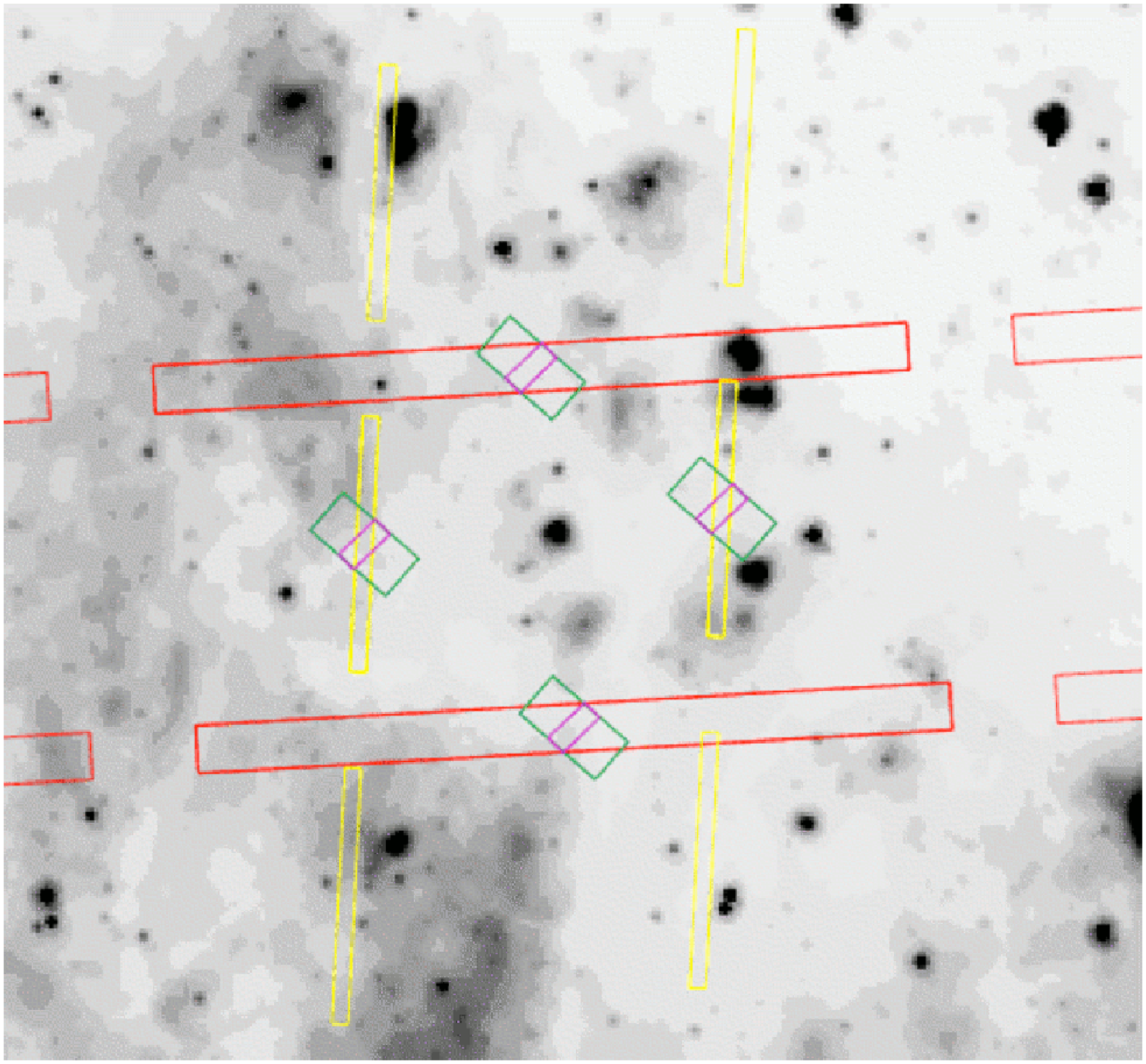}
\caption{IRAC $8.0\ \mu$m image showing a $4\arcmin \times 4\arcmin$ field
of view, centered on one of our IRS targets (SSTGC~797384).
{\it Left:} positions of all IRS slits ({\tt SL}: red, {\tt LL}: yellow,
{\tt SH}: magenta, and {\tt LH}: green)
for on-source measurements.
{\it Right:} Off-source measurements, showing background positions
around the source.
Similar slit formations were adopted for all of the spectroscopic targets.
The four off-source pointings were observed to derive a background spectrum
for each source, because of strong and spatially variable background towards
the GC.
\label{fig:slits}}
\end{figure*}

Figure~\ref{fig:slits} displays the IRS slit positions for on-source ({\it left})
and off-source ({\it right}) measurements for one of our sources (SSTGC~797384). 
Source and background spectra were taken consecutively to minimize zodiacal light
and instrumental variations. Each order of {\tt SL} or {\tt LL} was used to observe
a target, and different orders cover different parts of the sky near each target.
The low ecliptic latitude of the GC restricts the {\tt LL} slits to a position angle (PA)
of $\sim\pm90$\arcdeg\ and the {\tt SL} slits to a PA of $\sim0$\arcdeg\ or 180\arcdeg.

\begin{figure*}
\epsscale{0.8}
\plotone{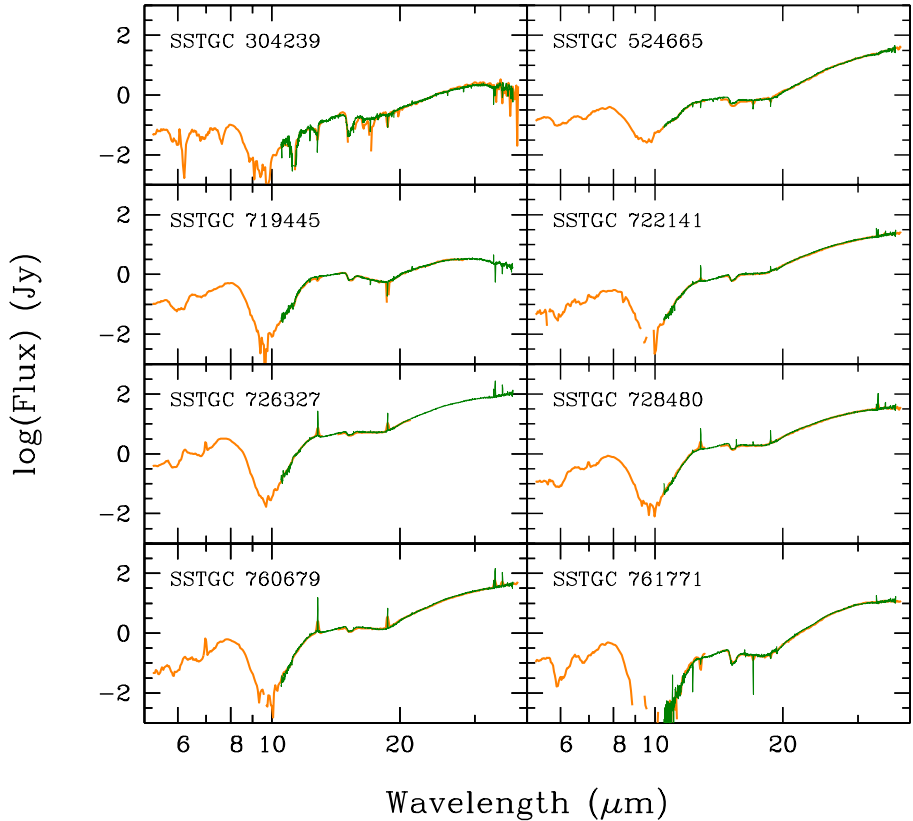}
\caption{IRS spectra of spectroscopically identified YSOs.
Orange lines are low-resolution ({\tt SL}, {\tt LL} modules) spectra,
and green lines are high-resolution ({\tt SH}, {\tt LH} modules) spectra.
The high-resolution spectra were scaled to match the flux in low-resolution
modules (see text).
\label{fig:allspec}}
\end{figure*}

\begin{figure*}
\epsscale{0.8}
\plotone{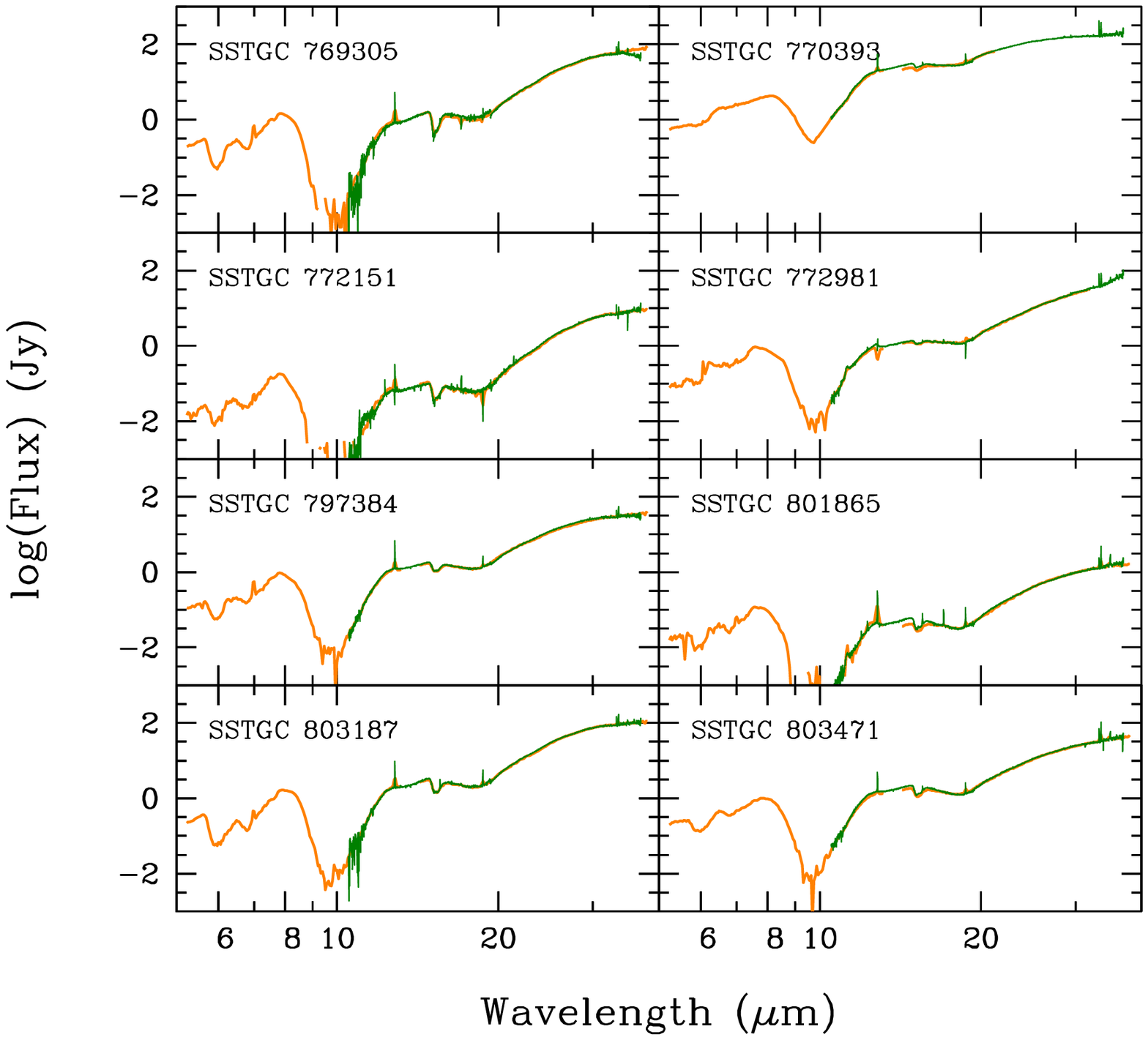}
\caption{Cont'd. Same as in Figure~\ref{fig:allspec}.
\label{fig:allspec2}}
\end{figure*}

\begin{figure*}
\epsscale{0.8}
\plotone{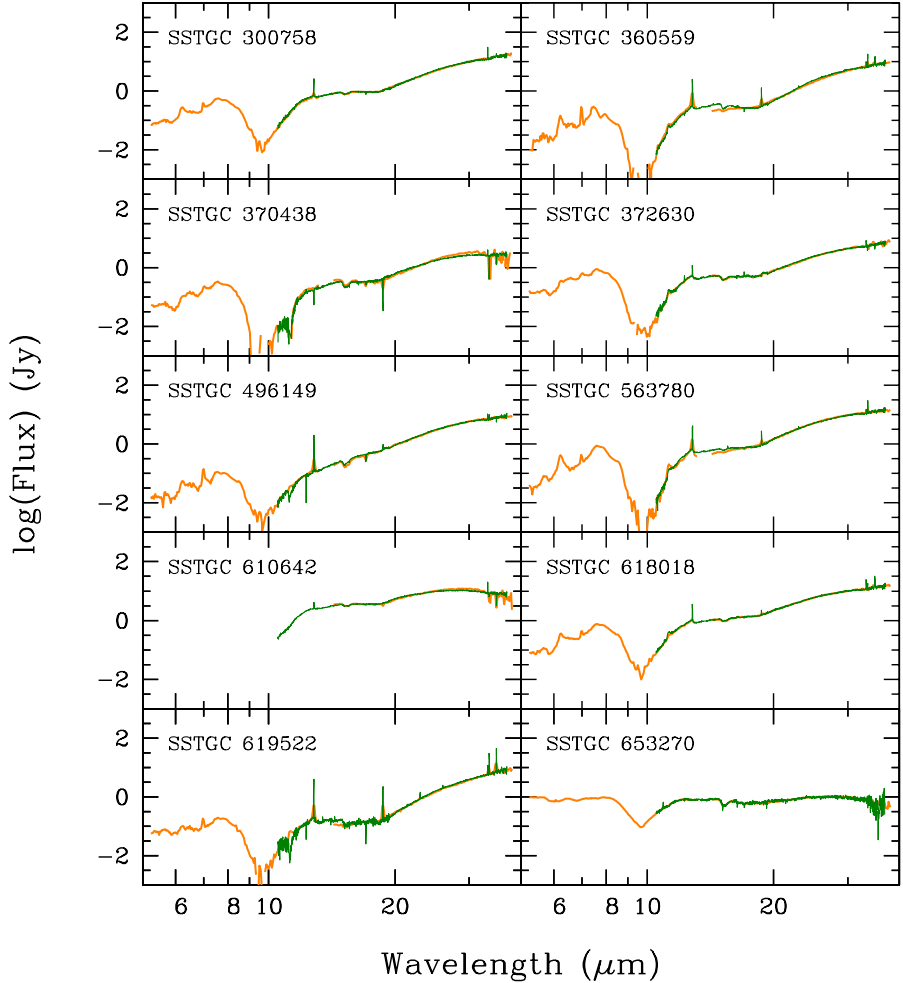}
\caption{IRS spectra of possible YSOs.
Line colors are the same as in Figure~\ref{fig:allspec}.
\label{fig:allspec3}}
\end{figure*}

\begin{figure*}
\epsscale{0.8}
\plotone{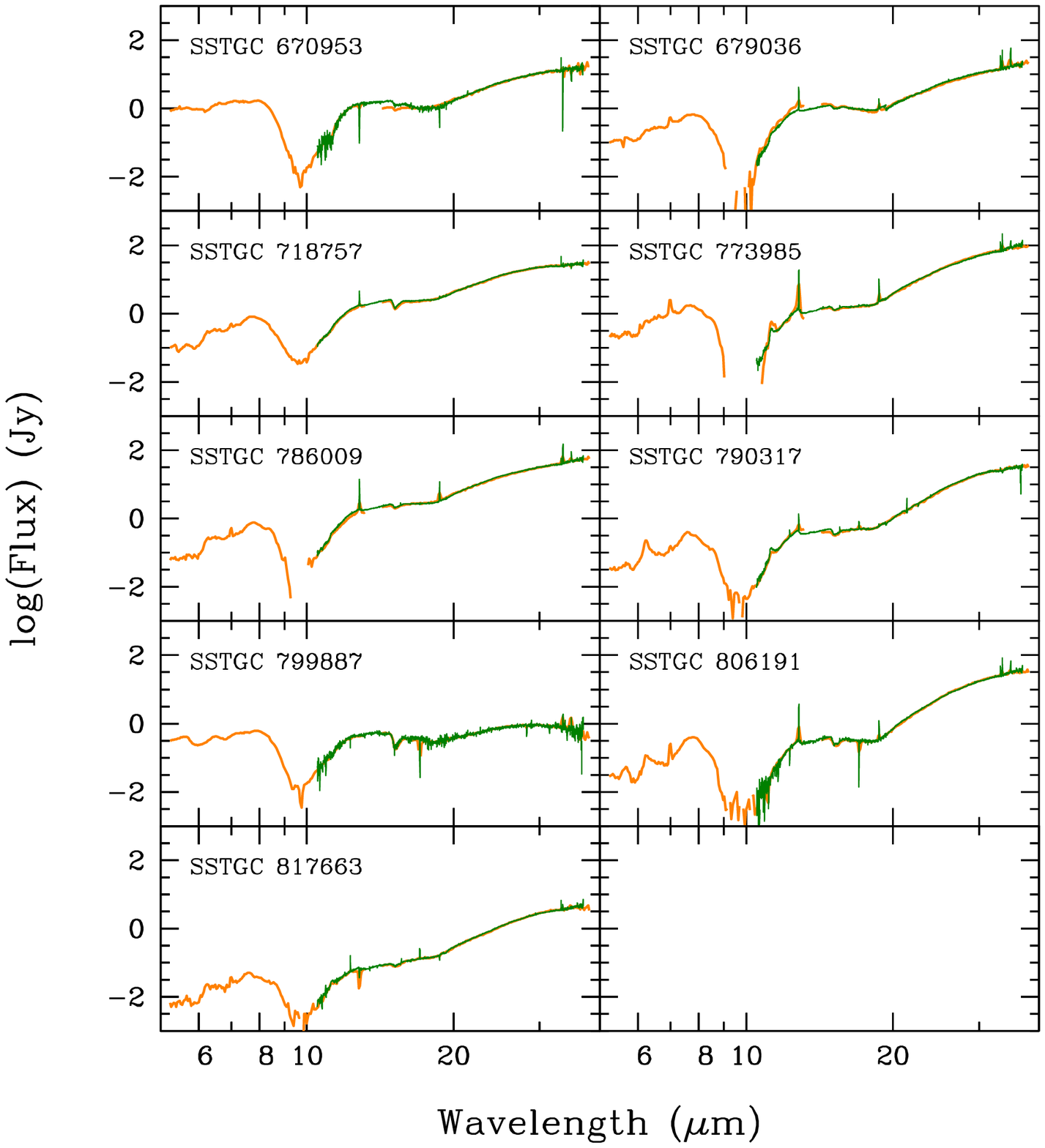}
\caption{Cont'd. Same as in Figure~\ref{fig:allspec3}.
\label{fig:allspec4}}
\end{figure*}

\begin{figure*}
\epsscale{0.8}
\plotone{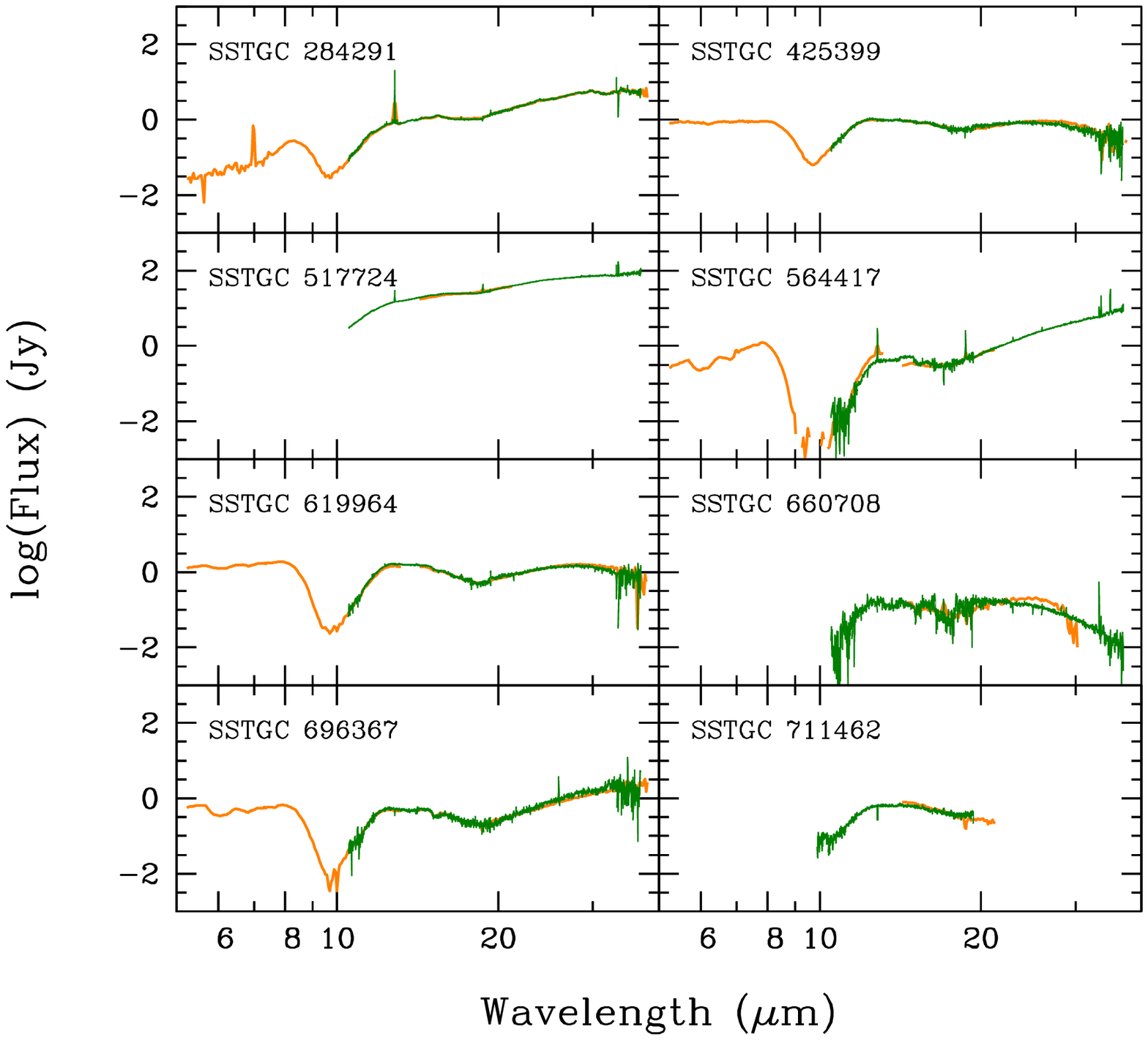}
\caption{IRS spectra of known stars (non-YSOs) in our target sample.
SSTGC~517724 is an OB supergiant star \citep{mauerhan:10} and the
other targets are OH/IR stars or long period variables.
Line colors are the same as in Figure~\ref{fig:allspec}.
\label{fig:allspec5}}
\end{figure*}

\subsection{Data Reduction}\label{sec:reduction}

We began reducing the high-resolution IRS spectra from the 
basic calibrated data (BCD), while we started with coadded products 
(post-BCD) for the low-resolution spectra.  We used the S18.7
version of the IRS pipeline for both. On the {\tt LH} frames we applied
the {\tt DARKSETTLE}\footnote{The SSC software packages can be found at\\
{\tt http://ssc.spitzer.caltech.edu/dataanalysistools/tools/}.} 
software package to correct for non-uniform dark currents.
We corrected for rogue pixel values using the SSC software package
{\tt IRSCLEAN}$^1$.
We only applied
campaign rogue masks (Campaigns 50 and 55 for the spring and autumn runs, respectively),
except in {\tt SL}, where we applied our own edited version to mask out hot pixels at
$\sim10\ \mu$m in addition to campaign rogue pixels. 
We then used {\tt SPICE}$^1$ to extract target and background spectra, 
and further corrected high-resolution spectra
({\tt SH}, {\tt LH}) for fringe patterns using the {\tt IRSFRINGE}$^1$ package.

Four background spectra per target were extracted for the high-resolution ({\tt SH}
and {\tt LH}) observations as four off-source pointings were obtained per target.
For the low-resolution ({\tt SL} and {\tt LL}) observations, two background spectra
were extracted from observations at the same positions as the high-resolution
background observations and two background spectra were extracted along 
the on-source slit observations.
For some objects, the high-resolution background slits were not
coincident with the on-source low-resolution slits.
For these, we inspected the slit positions on an IRAC/MIPS composite
image to determine the extraction position along the on-source
{\tt SL} or {\tt LL} slit that was closest in flux and position to
the high-resolution background slits.

For each module and for each target, we were able to extract four background 
spectra, that we used to estimate the target background by making a linear 
interpolation of the background flux at the source position at each wavelength.
Our interpolation scheme estimates the background flux 
gradient over a $\sim1\arcmin$ angular 
scale, since each background pointing is $\sim1\arcmin$ away from the science target. 
If the background emission is varying over a smaller angular scale, then 
background subtraction would be more uncertain.

We found that spectra extracted from various IRS modules usually do not match with 
each other at overlapping wavelengths, primarily due to the different 
sizes of the slit entrances. The {\tt SL} and {\tt LL} modules have $3.7\arcsec$ and
$10.7\arcsec$ slit widths, respectively, while the slit entrances of {\tt SH} and
{\tt LH} modules are $4.7\arcsec \times 11.3\arcsec$ and $11.1\arcsec \times 22.3\arcsec$,
respectively. Therefore, contamination by point sources and/or extended emission at the GC 
can easily lead to a flux mismatch among the various IRS modules.

To obtain internally consistent fluxes from all of the IRS modules, we scaled the
spectra to match the fluxes from the second order of {\tt LL} ({\tt LL2};
$14\ \mu$m - $21\ \mu$m). On the longer wavelength side, we scaled the {\tt LL1}
spectrum to the {\tt LL2} spectrum by estimating a median flux ratio for the two modules
in the overlapping wavelength region. We masked known emission features and rejected
points that were more than $3\ \sigma$ away from the median flux ratio. On the shorter
wavelength side, the scaling was done in a step-by-step fashion. We first scaled the
{\tt SH} spectrum to the base flux of the {\tt LL2} spectrum (before correcting for
order tilts; see below). We then matched the spectrum in the first order of {\tt SL}
({\tt SL1}; $7.4\ \mu$m - $14.5\ \mu$m) to the {\tt SH} spectrum, then scaled the flux
in {\tt SL3} (bonus order; $7.3\ \mu$m$-8.7\ \mu$m) to the {\tt SL1} spectrum, and
finally scaled the flux in the second order of {\tt SL} ({\tt SL2}; $5.2\ \mu$m
- $7.7\ \mu$m) to {\tt SL3}. We can describe this concisely as {\tt LL2} $\rightarrow$
{\tt LL1},  and {\tt LL2} $\rightarrow$ {\tt SH} $\rightarrow$ {\tt SL1} $\rightarrow$ 
{\tt SL3} $\rightarrow$ {\tt SL2}. If a source was not clearly separated from
extended background emission in {\tt LL2}, we opted to choose the {\tt SL1} spectrum
as the base flux for scaling. In this case, the flux calibration was done in the
following sequences: {\tt SL1} $\rightarrow$ {\tt SH} $\rightarrow$ {\tt LL2}
$\rightarrow$ {\tt LL1}, and {\tt SL1} $\rightarrow$ {\tt SL3} $\rightarrow$ {\tt SL2}.

Scaling factors applied to each module are shown in Table~\ref{tab:tab2}.
The second column in Table~\ref{tab:tab2} shows which module was used as the baseline
for the flux calibration for each target. Either large or small scaling factors
are found in {\tt LL} for SSTGC~360559, {\tt SL} for SSTGC~773985, and {\tt LL}
for SSTGC~801865. These objects are faint ([8.0]$=7.6$~mag and $8.8$~mag
for SSTGC~360559 and 801865, respectively, while SSTGC~773985 was not detected
in this bandpass) on top of bright or saturated background emission on
MIPS [24] images. As a result, their mid-IR fluxes are heavily contaminated by
background emissions at $\lambda \ga 20\ \mu$m, leading to an over-estimation of
flux from the target slit.

Order-tilt features remained in about 30 high-resolution spectra, after applying
{\tt IRSFRINGE} to {\tt SH} and {\tt LH} and {\tt DARKSETTLE} to {\tt LH}. To remove
this artifact, we applied a $1^{\rm st}$ order polynomial to each high-resolution
spectral order to force it to match the re-scaled low-resolution spectra. In addition,
three sources showed scalloping features in their high-resolution spectra; a $2^{\rm nd}$
order polynomial was applied to correct for this artifact. Individual spectra from
various orders were then merged together using a linear ramp.

Some data were excluded from the analysis due to problems in a particular spectrum,
such as saturation, excess bad pixels, or a poor match in background level. The last
column in Table~\ref{tab:tab2} lists any excluded data. Each target was observed at
two different nod positions in each module; any nod positions that were excluded for
a particular module are also given in the last column of Table~\ref{tab:tab2}.
A cardinal point (NSEW) given in parentheses in this column indicates that the background
spectrum offset in that direction from the source was not included in the background
determination for that module. 

Figures~\ref{fig:allspec}--\ref{fig:allspec5} display spectra resulting from
the above procedures; orange lines are low-resolution spectra, and green lines
represent high-resolution spectra. Only sources we spectroscopically identify
as a YSO or possible YSO are shown in Figures~\ref{fig:allspec}--\ref{fig:allspec4}.
Spectra of known OH/IR, long-period variable, or OB supergiant stars
in our sample are shown in Figure \ref{fig:allspec5} for comparison.

There are several sources of flux uncertainty in our spectra:
statistical, calibration (difference between different nods),
and the varying background (this last is usually largest).
This can cause spectral features observed in emission in the
background appear in absorption in some spectra,
such as the 11.3 $\mu$m polycyclic aromatic hydrocarbon (PAH) feature
(e.g. SSTGC 304239), H$_2$ emission at 17.0 $\mu$m
(e.g. SSTGC 761771), or forbidden lines such as
12.8 $\mu$m [Ne II] or 18.7 $\mu$m [S III]
(e.g. SSTGC 670953).
We have estimated the uncertainty due to background subtraction
by comparing results derived by excluding one of the four
background pointings from the interpolated background spectrum.
Throughout our analysis, we have added these uncertainties in quadrature
to derive final uncertainties in measured quantities
such as the CO$_2$ ice column density.

\section{Results}\label{sec:results}

\subsection{Spectroscopic Identification of Massive YSOs in the GC}\label{sec:yso}

\begin{figure*}
\epsscale{0.8}
\plotone{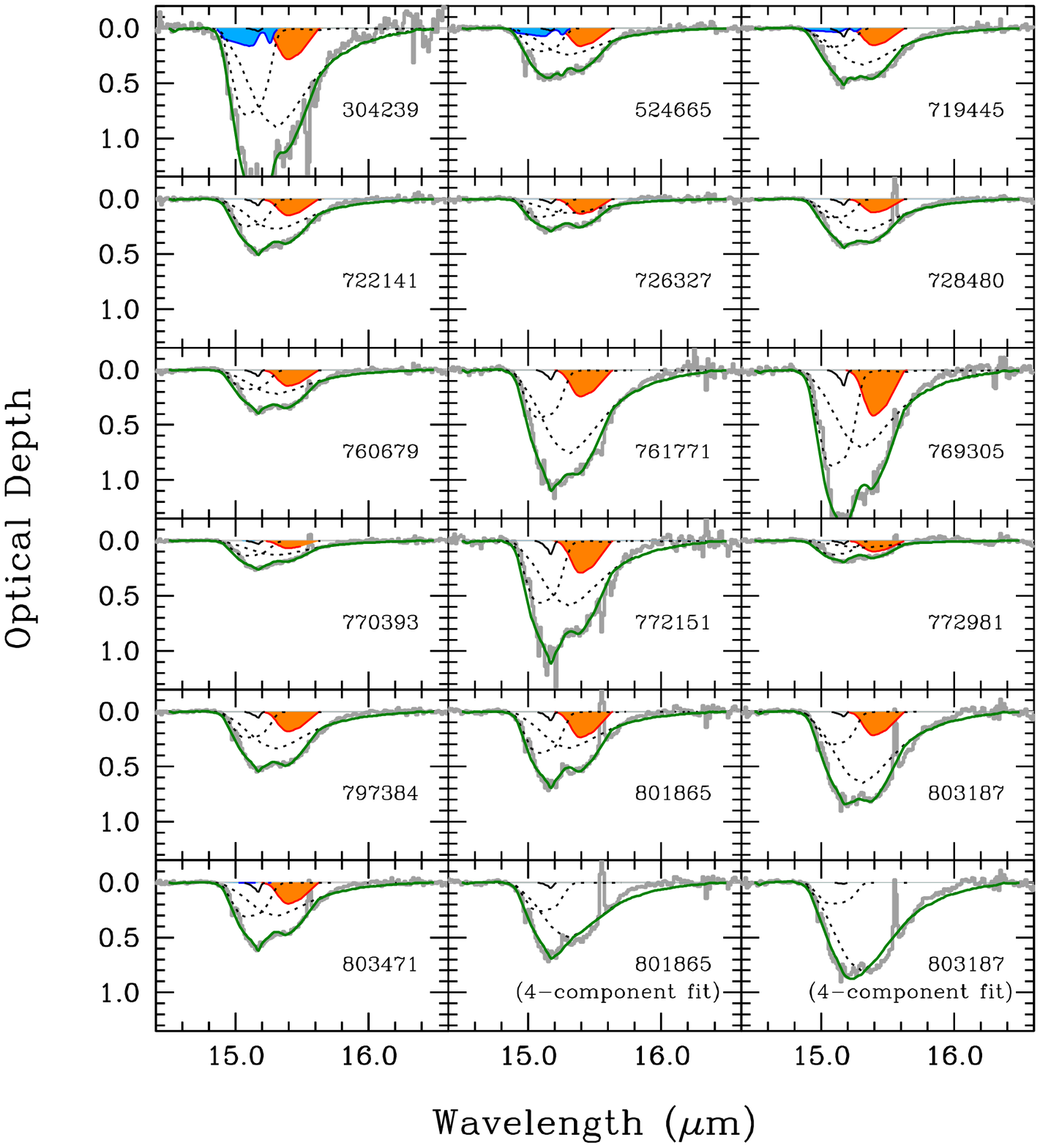}
\caption{Optical depth spectra of spectroscopically identified YSOs
centered on the $15\ \mu$m CO$_2$ ice absorption.
Best-fitting CO$_2$ ice models and individual fitting components are
displayed for each target: polar (dotted line, centered at $\sim15.3\ \mu$m),
apolar (dotted line, centered at $\sim15.1\ \mu$m), pure (blue shaded),
diluted (black solid line), and $15.4\ \mu$m shoulder (orange-shaded).
Sum of these components is shown as a green line. Bottom two panels show
examples of 4-component model fitting without the $15.4\ \mu$m shoulder ice.
\label{fig:co2}}
\end{figure*}

\begin{figure*}
\epsscale{0.8}
\plotone{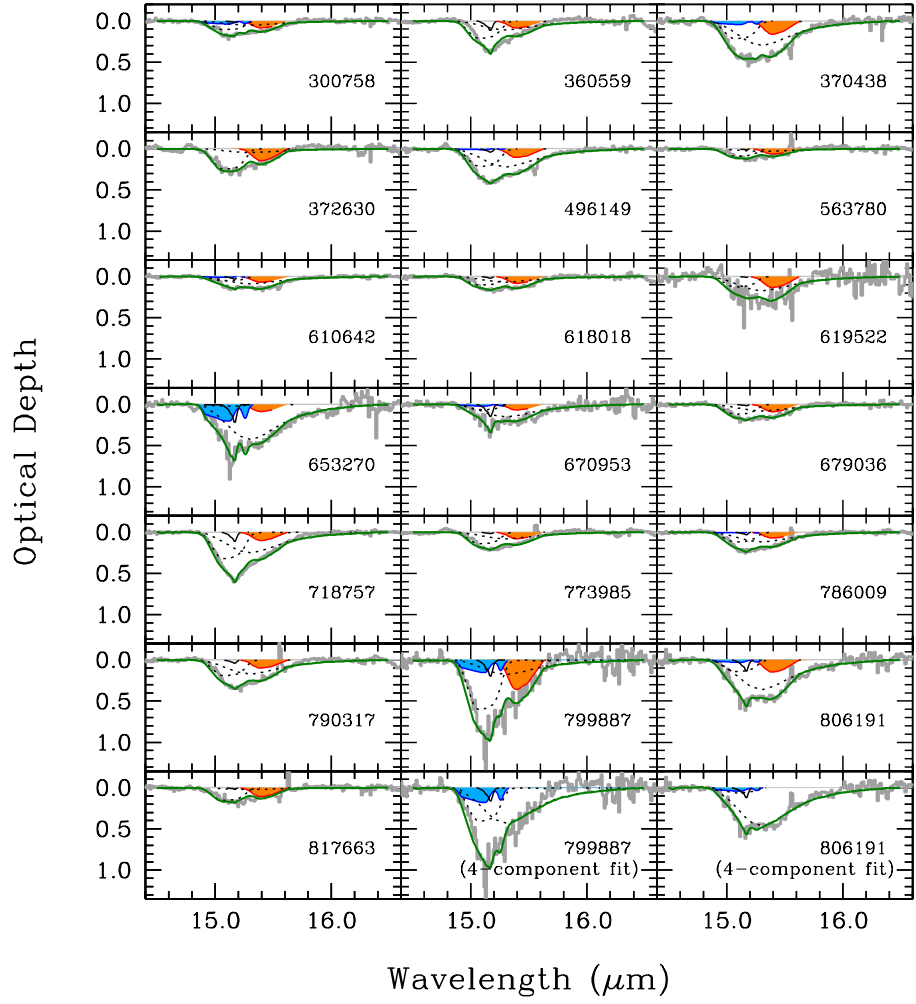}
\caption{Same as in Fig.~\ref{fig:co2}, but for possible YSOs.
\label{fig:co2.2}}
\end{figure*}

\begin{figure*}
\epsscale{0.8}
\plotone{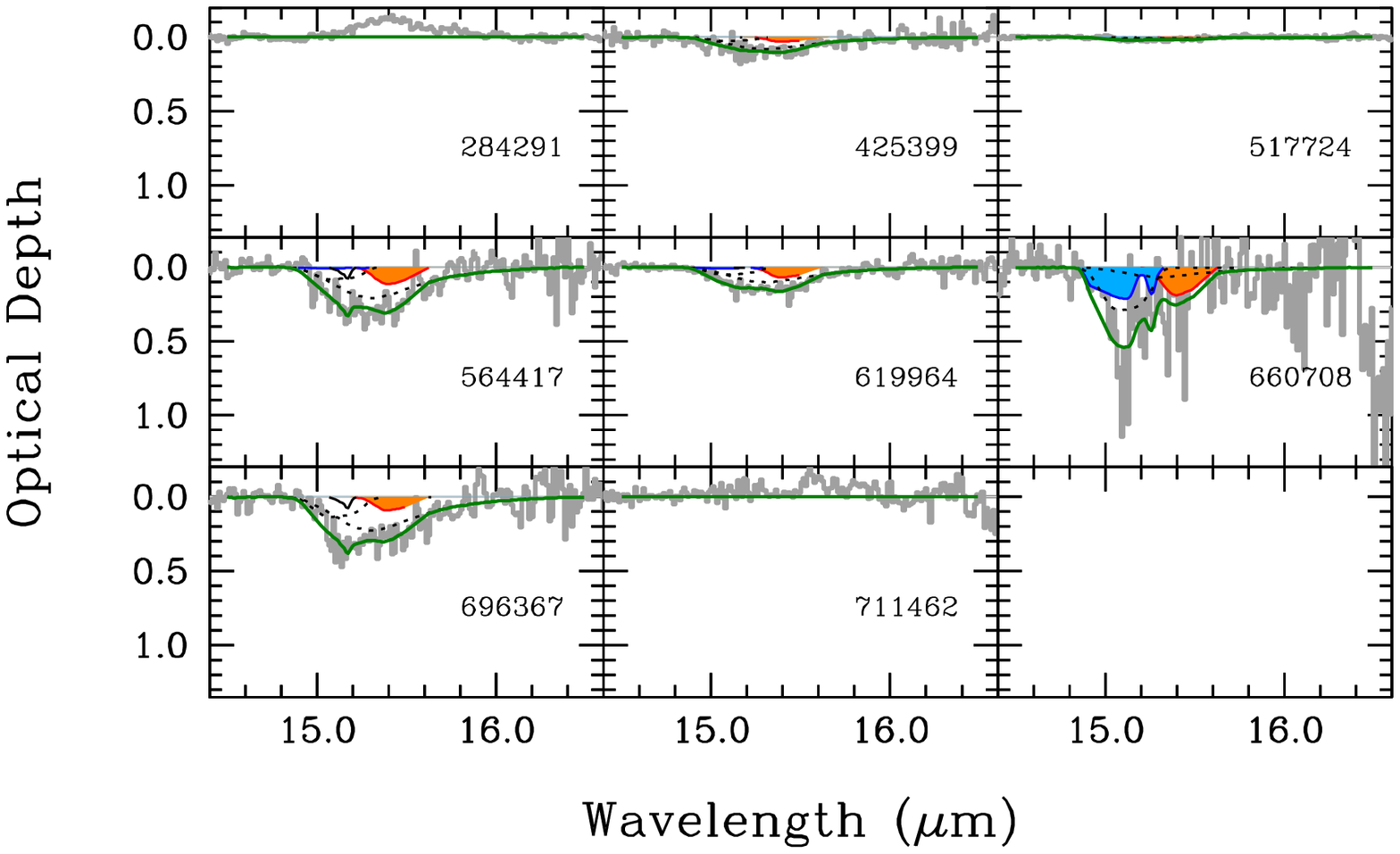}
\caption{Same as in Fig.~\ref{fig:co2}, but for known stars.
\label{fig:co2.3}}
\end{figure*}

To study the $15\ \mu$m CO$_2$ ice absorption profile, we fitted five
laboratory spectral components to the feature in all our 107 targets,
following the same procedure described in A09.
Figures~\ref{fig:co2}--\ref{fig:co2.3} show the CO$_2$ ice decomposition
for our YSOs, possible YSOs, and known stars for comparison, respectively.
We describe below how we selected YSOs and possible YSOs based on this
procedure.

First, we set a local continuum over $14.3\ \mu$m $\leq \lambda \leq 16.5\ \mu$m 
using a $3^{rd}$ order polynomial to derive the optical depth. Then we used
the modeling technique and laboratory data in \citet{pontoppidan:08} to
decompose the absorption profile with five laboratory spectral components;
these are polar CO$_2$ (CO$_2$:H$_2$O $= 14:100$ at 10~K; dotted line, centered
at $\sim15.3\ \mu$m), apolar CO$_2$ (CO:CO$_2 = 100:70$ at 10~K; dotted line,
centered at $\sim15.1\ \mu$m), pure CO$_2$ (15~K; blue shaded), diluted CO$_2$
(CO:CO$_2 = 100:4$ at 10~K; black solid line), and $15.4\ \mu$m shoulder CO$_2$
(modeled with two Gaussians in wavenumber space; orange shaded). We found
a best-fitting set of models from the non-linear least squares fitting routine 
MPFIT \citep{markwardt:09}.

Fitting results are shown in Table~\ref{tab:tab3}. The CO$_2$ ice column densities
were estimated from the integrated absorption, adopting the integrated line
strength $A = 1.1\times10^{-17}\ {\rm cm\ molecule^{-1}}$ \citep{gerakines:95}.
Background uncertainties were estimated by creating spectra with one of four
background positions excluded from the interpolated background spectrum
(\S\ref{sec:method}) and then comparing the column densities derived from
these spectra. We added these uncertainties in quadrature to the uncertainties
from comparing column densities from spectra at the two nod positions, and to
uncertainties in column densities due to the statistical uncertainties.
The $\chi^2_{\rm tot}$ and $N_{\rm tot}$ in Table~\ref{tab:tab3} represent
the total chi-square of the fit and the number of data points used in this fit.
The goodness of fit is generally poor, implying either underestimated flux errors
or our lack of knowledge of individual CO$_2$ ice models. Nevertheless,
the ice decomposition still provides useful information
on the nature of YSOs, as shown below. For comparison, fitting results for
some known stellar sources are included in Table~\ref{tab:tab3}.

Our primary method of identifying YSOs from our IRS observations is the CO$_2$
ice absorption profile at $15\ \mu$m, which is observed to have a different
spectral shape in and around YSOs \citep{ehrenfreund:99, dartois:99a}.
High-spectral resolution observations of many massive YSOs in our Galaxy
\citep{gerakines:99} and in the Large Magellanic Cloud \citep{seale:11}
found a ``shoulder'' at 15.4 $\mu$m on the CO$_2$ ice absorption profile.
This $15.4\ \mu$m shoulder is thought to be due to the presence of CH$_3$OH-rich
CO$_2$ ice grains \citep{ehrenfreund:99, dartois:99a}. Detailed fitting of
the 15 $\mu$m CO$_2$ ice profile shows that the 15.4 $\mu$m shoulder is weaker
in low-mass protostars \citep{pontoppidan:08, zasowski:09} and is not detected
towards field stars behind several molecular clouds
\citep{gerakines:99,bergin:05,knez:05,whittet:07,whittet:09}. Analysis of
the 15 $\mu$m CO$_2$
ice profile along the lines of sight to the Central Cluster and to two dusty WC9
stars in the Quintuplet Cluster demonstrates that none of these three GC spectra
shows a 15.4 $\mu$m shoulder on the 15 $\mu$m CO$_2$ ice absorption profile
\citep{gerakines:99}. Thus, the presence or absence of the 15.4 $\mu$m shoulder
is an empirical -- and quantitative -- way in the GC of distinguishing YSOs from
AGB stars behind molecular clouds.

The 15 $\mu$m CO$_2$ ice absorption profiles displayed in
Figures~\ref{fig:co2}--\ref{fig:co2.2} for 35 of our YSOs or possible YSOs
(see below) show two absorption peaks, at 15.15 $\mu$m and 15.4 $\mu$m. Many
previously studied YSOs show a double-peaked absorption profile, but with peaks
at shorter wavelengths of $15.10\ \mu$m and $15.25\ \mu$m
\citep[e.g.,][]{gerakines:99,pontoppidan:08,seale:11}. Double-peaked absorption at
15.10 $\mu$m and 15.25 $\mu$m is ascribed to pure CO$_2$ ices resulting from
the crystallization of heated H$_2$O-rich ices
\citep[e.g.,][]{gerakines:99,pontoppidan:08}. By contrast, \citet{ehrenfreund:99} and
\citet{dartois:99a} interpret CO$_2$ ice absorption peaking at 15.15 $\mu$m as due to
CO-rich CO$_2$ ices and absorption peaking at 15.4 $\mu$m as arising in CH$_3$OH-rich
CO$_2$ ices.

We selected YSOs by requiring that the model fit to the observed 15 $\mu$m CO$_2$ ice profile
significantly improves when the 15.4 $\mu$m shoulder is included in the model. We calculated
the reduced $\chi^2$ for fitting a four-component model (excluding the 15.4 $\mu$m shoulder)
to the 15 $\mu$m CO$_2$ profile, as shown in the bottom panels in Figures~\ref{fig:co2}--\ref{fig:co2.2}.
We then calculated the reduced $\chi^2$ for 
fitting the five-component model (including the 15.4 $\mu$m shoulder). Finally, we calculated
$\Delta \chi^2$, equal to the reduced $\chi^2$ for the four component model minus
the reduced $\chi^2$ for the five component model. We also required that the optical
depth from the 15.4 $\mu$m feature be more than 0.05, a limit set by the IRS flat field
uncertainty. This corresponds to a column density for the 15.4 $\mu$m CO$_2$ ice component of 
$N_{\rm col}$(shoulder) $\approx$ 0.5 $\times$ 10$^{17}$ cm$^{-2}$.

\begin{figure}
\epsscale{1.2}
\plotone{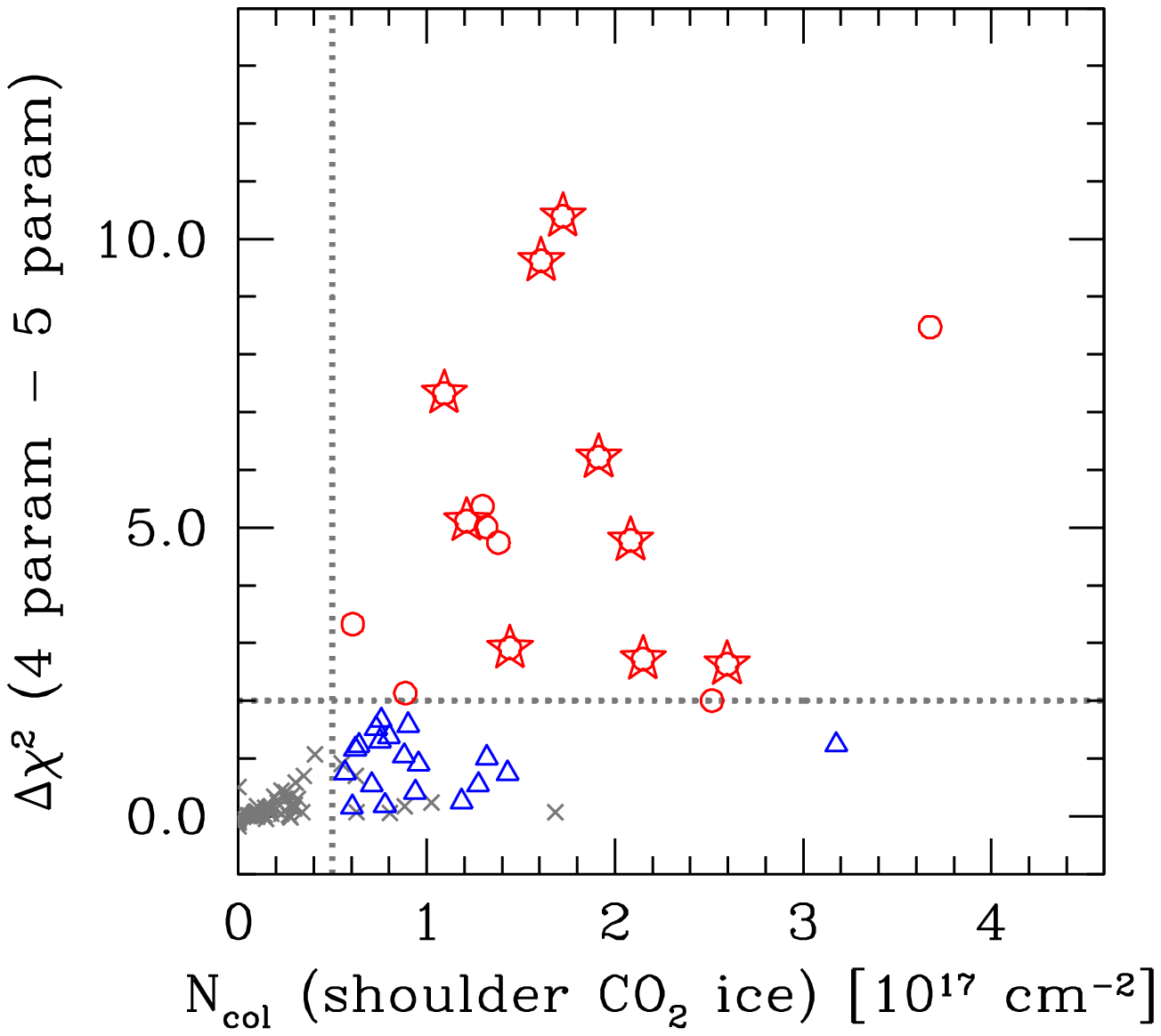} 
\caption{Difference ($\Delta \chi^2$) between the reduced $\chi^2$ of the four-component
CO$_2$ ice model (excluding the 15.4 $\mu$m shoulder) and the reduced $\chi^2$ of the
five-component CO$_2$ ice model (including the 15.4 $\mu$m shoulder) versus the column
density of the CO$_2$ 15.4 $\mu$m shoulder component. Two dotted lines represent the
criteria for our YSO identification. The YSOs (red circles) have $\Delta \chi^2 \geq 2$
and $N_{\rm col} {\rm (shoulder)} \geq 0.5 \times 10^{17} {\rm cm}^{-2}$ and possible
YSOs have $N_{\rm col} {\rm (shoulder)} \geq 0.5 \times 10^{17} {\rm cm}^{-2}$
and $0 < \Delta \chi^2 < 2$. Red stars mark YSOs with gas-phase absorption features.
Possible YSOs are shown as blue triangles and the remaining targets are
shown as grey crosses. The value of $\Delta \chi^2$ increases when adding the 15.4 $\mu$m 
shoulder component to the model significantly improves the fit.
\label{fig:chisqr}}
\end{figure}

We illustrate our YSO selection in Figure \ref{fig:chisqr} where we plot $\Delta \chi^2$ vs.  
$N_{\rm col}$(shoulder). We conclude that a GC source is a YSO if $\Delta \chi^2$ $\geq$ 2
and $N_{\rm col}$(shoulder) $\geq$ 0.5 $\times$ 10$^{17}$ cm$^{-2}$. We define a GC source
as a possible YSO if $0 < \Delta \chi^2 < 2$ and $N_{\rm col}$(shoulder) $\geq$ 0.5 $\times$
10$^{17}$ cm$^{-2}$. We visually inspected possible YSO spectra, and excluded some spectra as
clearly non-YSO: these are SSTGC~440424 (weak $15\ \mu$m CO$_2$ absorption),
SSTGC~564417 (OH/IR star), SSTGC~619964 (variable star), SSTGC~696367 (OH/IR star),
SSTGC~660708 (OH/IR star), SSTGC~732531 ($15\ \mu$m CO$_2$ absorption not significantly
different between source and background spectra), SSTGC~738126 (weak $15\ \mu$m CO$_2$
absorption). We consider all other GC sources not to be YSOs. These cutoff values of
$\Delta \chi ^ 2$ and $N_{\rm col}$(shoulder) closely agree with the YSO classification
that three of us (DA, SR, KS) did by visually inspecting the IRS spectra of all 107 targets.

Our spectroscopic classification of the 107 GC targets is shown in the fifth column of
Table~\ref{tab:tab1}. We conclude that 16 sources are YSOs (``yes'' in the fifth column
of Table~\ref{tab:tab1}) and 19 sources are possible YSOs (``maybe'' in the fifth column
of Table~\ref{tab:tab1}). The remaining columns in Table~\ref{tab:tab1} show
cross-identifications of our IRS sample with earlier photometry-based YSO selections
in \citet{felli:02}, \citet{schuller:06}, and \citet{yusefzadeh:09}. We describe these
cross-identifications in \S~\ref{sec:crossid}.

The strength of the $15.4\ \mu$m peak in our sources is similar to that of the
well-studied embedded massive YSO W33A \citep{gerakines:99}. It is ascribed to
a Lewis acid-base interaction of CO$_2$ (the Lewis acid) with CH$_3$OH \citep{dartois:99a}. 
Other species could be acting as a base as well, but CH$_3$OH is preferred due to its 
high abundance toward W33A, which is $5\%$--$22\%$ relative to solid H$_2$O 
\citep{dartois:99b}. Two other massive YSOs (AFGL~7009S, AFGL~2136) show a prominent
$15.4\ \mu$m peak, and indeed these sources have high CH$_3$OH abundances as well
\citep{dartois:99b,gibb:04}. We therefore suggest that the GC YSOs and possible YSOs
may also have high solid CH$_3$OH abundances. Although the origin of the large
quantities of CH$_3$OH in the previously studied massive YSOs is not fully understood
\citep{dartois:99a}, all lines of sight with high solid CH$_3$OH abundances are
associated with star formation, strengthening the argument that the sources studied
in this paper are indeed YSOs.

\subsection{Gas-phase Absorption}\label{sec:gas}

Many of our YSOs have gas-phase absorption from C$_2$H$_2$ ($13.71\ \mu$m, $\nu_5 = 1-0$),
HCN ($14.05\ \mu$m, $\nu_2 = 1-0$), and/or CO$_2$ ($14.97\ \mu$m, $\nu_2 = 1-0$). These
gaseous bandheads have been detected toward other massive YSOs, tracing warm and dense gas
\citep[e.g.,][]{lahuis:00,boonman:03,knez:09}. All GC sources with these gas absorption
bands have been already identified as YSOs through the strength of the 15.4 $\mu$m CO$_2$
ice shoulder, thus strengthening our identification. An AGB star can show either CO$_2$
gas or C$_2$H$_2$ gas but not both, because while O-rich AGB stars sometimes show CO$_2$
gas in emission or absorption \citep{justtanont:98}, C$_2$H$_2$ gas absorption is found
only in C-rich AGB stars \citep{aoki:99}.

\begin{figure*}
\epsscale{0.8}
\plotone{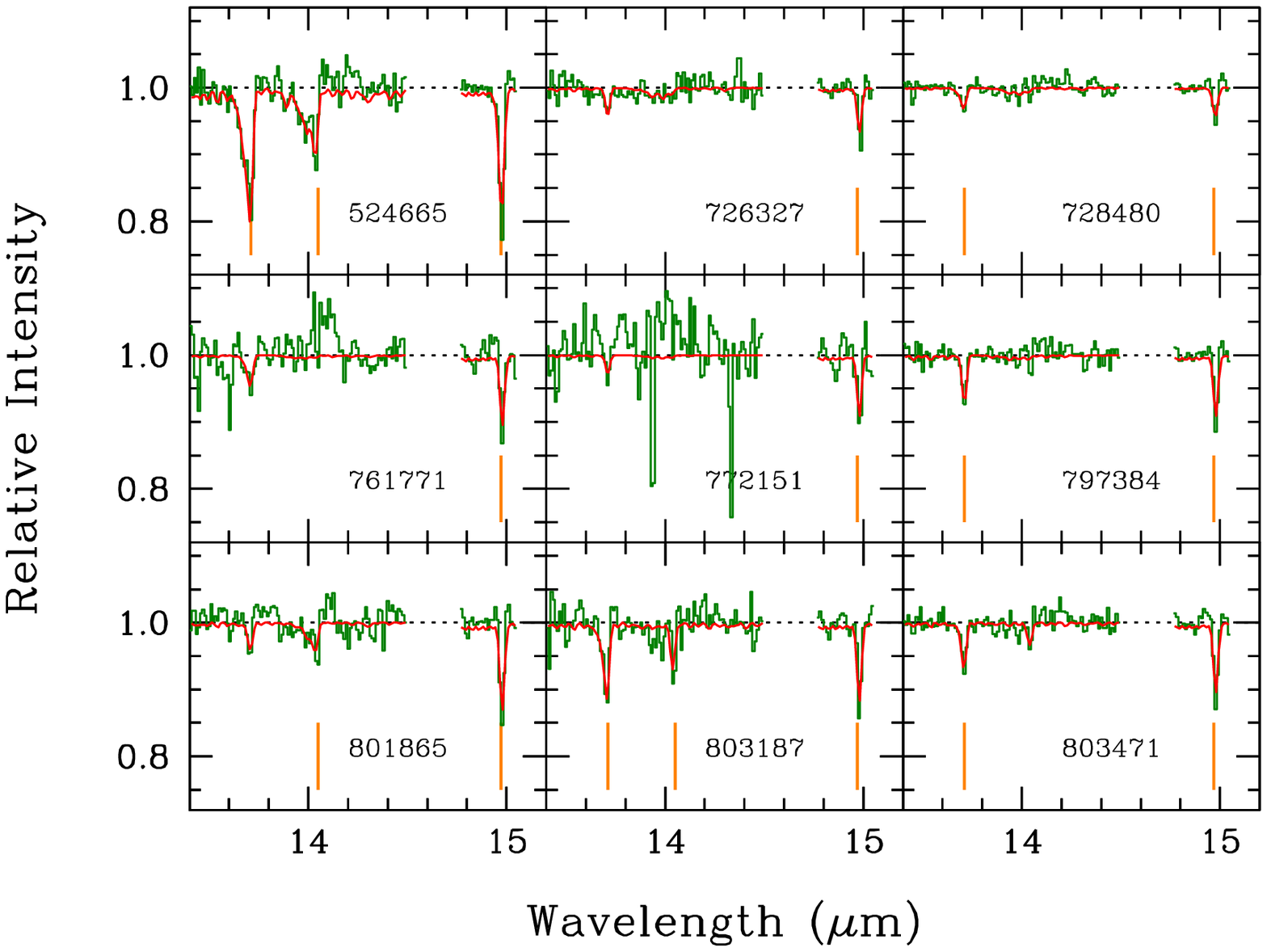}
\caption{Gas-phase molecular absorptions from C$_2$H$_2$ $\nu_5 = 1-0$
($13.71\ \mu$m), HCN $\nu_2 = 1-0$ ($14.05\ \mu$m), and
CO$_2$ $\nu_2 = 1-0$ ($14.97\ \mu$m).
These gas-phase molecular features trace warm and dense gas detected
towards galactic massive YSOs.
Red lines represent models with best-fitting $T_{\rm ex}$ and $N_{\rm col}$
(see text).
Individual absorption lines are marked with vertical bars if they
were identified, independent of the model fitting.
All objects with identified gas-phase absorptions are
selected as YSOs through the detection of the 15.4 $\mu$m
shoulder component of the 15 $\mu$m CO$_2$ ice absorption feature.
\label{fig:gas}}
\end{figure*}

Figure~\ref{fig:gas} shows relative intensity spectra for the nine YSOs
in our sample that show gas-phase absorption from at least one of these
species. Three YSOs (SSTGC~524665, SSTGC~797384, SSTGC~803187) presented
in A09 are shown in Figure~\ref{fig:gas} together with six additional YSOs
with significant gas-phase absorption. The relative intensity was determined
by using a second order polynomial to set a local continuum at
$13.30\ \mu$m $\leq \lambda \leq 14.55\ \mu$m for C$_2$H$_2$ and HCN, and
at $14.77\ \mu$m $\leq \lambda \leq 15.06\ \mu$m for CO$_2$.

As in A09, we used model spectra from \citet{spectrafactory}. These models are
based on the {\tt HITRAN04} linelist \citep{hitran} for C$_2$H$_2$ and HCN, and
based on {\tt HITEMP} \citep{hitemp} for CO$_2$. We did not include isotopes in
the computation because of the limited parameter span in the model grids.
However, even a relatively high isotopic fraction in the GC
\citep[$^{12}{\rm C}/^{13}{\rm C}\sim25$;][]{wannier:80,gusten:85} has a negligible
impact on the model fitting. Best-fitting model values of the excitation
temperature, $T_{\rm ex}$, and the gas-phase column density, $N_{\rm col}$, were
found by searching for the minimum
$\chi^2$ of the fits over 100~K $\leq T_{\rm ex} \leq$ 1000~K in steps of
$\Delta T_{\rm ex} = 100$~K, and $15 \leq \log{N_{\rm col}} \leq 18$ for C$_2$H$_2$,
$16 \leq \log{N_{\rm col}} \leq 18$ for HCN,
and $16 \leq \log{N_{\rm col}} \leq 22$ for CO$_2$ with intervals of $0.1$~dex.
Errors in these parameters were estimated from $\Delta \chi^2$, where $1\sigma$ 
measurement errors were taken from the scatter in flux. Systematic errors from
background subtraction and nodding differences were then added in quadrature.
We tested varying covering factors (the fraction of the background continuum
source covered by the component in question), but found that the best-fitting
value is equal to or close to unity. We adopted a Doppler parameter of 3 km s$^{-1}$.

Figure~\ref{fig:gas} shows the best-fitting models for each molecular species
in red lines. Individual absorption lines are marked with vertical bars if they
were identified by three of us (DA, SR, KS) by visually inspecting the IRS spectra
of all 107 targets, independent of the model fitting. Some of the lines were
marked undetected (e.g., C$_2$H$_2$ of SSTGC~761771) because of a low signal-to-noise
ratio of its spectrum. The best-fitting model excitation temperatures ($T_{\rm ex}$)
and column densities ($N_{\rm col}$) of identified lines are listed in
Table~\ref{tab:tab4}. All objects with identified gas-phase absorptions are
selected as YSOs through the detection of the 15.4 $\mu$m
shoulder component of the 15 $\mu$m CO$_2$ ice absorption feature.

\subsection{Extinction}\label{sec:extinction}

The extinction for our sources can be derived from the optical depths of $9.7\ \mu$m
and $18\ \mu$m silicate absorption features in the IRS spectra. We derived two
estimates of the dust extinction: one using the low-resolution modules {\tt SL}+{\tt LL} 
[hereafter $A_V$ ({\tt SL}+{\tt LL})], and one using the high-resolution modules
{\tt SH}+{\tt LH} [hereafter $A_V$ ({\tt SH}+{\tt LH})]. The determination of 
$A_V$ ({\tt SL}+{\tt LL}) takes both the $9.7\ \mu$m and $18\ \mu$m silicate
features into account. The high-resolution data do not include the short wavelength
side of the $9.7\ \mu$m silicate feature, and so $A_V$ ({\tt SH}+{\tt LH}) is mainly
constrained by the $18\ \mu$m silicate feature. The $18\ \mu$m feature is broader and
shallower than the $9.7\ \mu$m absorption, so it provides a weaker constraint on $A_V$.
$A_V$ ({\tt SH}+{\tt LH}), however, provides a useful diagnostic when {\tt SL} is not
available, as many sources near Sgr~A do not have {\tt SL} data due to saturation 
in the peak-up arrays (e.g., SSTGC~610642). Since the high-resolution spectra were
scaled to the flux in the low-resolution modules, $A_V$ ({\tt SL}+{\tt LL}) and
$A_V$ ({\tt SH}+{\tt LH}) are not independent from each other.

To determine the dust extinction, we model the $5\ \mu$m\ --\ $32\ \mu$m spectrum by
simultaneously fitting the underlying continuum, the silicate dust features
centered at 9.7 $\mu$m and 18 $\mu$m, and the 13 $\mu$m librational H$_2$O ice absorption
(see Figure~4 in A09). The entire silicate extinction curve, derived using
the GCS~3 spectrum from the Infrared Space Observatory (ISO) Short Wavelength
Spectrometer (SWS) \citep{kemper:04}, is characterized by the optical depth
at 9.7 $\mu$m, $\tau_{9.7}$. We adopted the laboratory spectrum of pure amorphous
H$_2$O ice at $T = 10$~K \citep{hudgins:93} to model the $13\ \mu$m librational
H$_2$O absorption. This shallow absorption is not well-constrained, however,
so that the resulting column density of H$_2$O ice, $N_{{\rm col}} (13\ \mu {\rm m})$,
is uncertain. We used a second-order polynomial to simulate the overall shape of
the SED plus grey extinction in the absence of silicate and H$_2$O absorption.
Before performing this non-linear least squares fit \citep{markwardt:09}
we masked molecular absorption features at $5.5\ \mu$m $< \lambda \la 7.5\ \mu$m,
PAH emission at $\sim11.3\ \mu$m, CO$_2$ ice absorption at $\sim15\ \mu$m, strong
emission lines, as well as the noisy bottom part of the $9.7\ \mu$m silicate feature
($9.3\ \mu$m $< \lambda < 10.1\ \mu$m). We derived $A_V$ ({\tt SL}+{\tt LL}) from 
$\tau_{9.7}$ by adopting $A_V / \tau_{9.7} = 9$ \citep{roche:85}, the value
measured for lines of sight towards the GC. We derived $A_V$ ({\tt SH}+{\tt LH}) in
the same way, except that we modeled the $10\ \mu$m\ --\ $32\ \mu$m high-resolution
spectra instead.

The uncertainties in $A_V$ ({\tt SL}+{\tt LL}) and $A_V$ ({\tt SH}+{\tt LH})
are dominated by the uncertainty in choosing the continuum. We estimated these
uncertainties by comparing results where the continuum was derived from the same
wavelength regions in all spectra to results where each continuum was set interactively. 
Applying a second-order polynomial for a continuum generally results in a good fit
over $10\ \mu$m $\la \lambda \la 32\ \mu$m, but underestimates fluxes at $<8\ \mu$m,
which may be due to under-subtraction of background PAH emission at $\sim7.7\ \mu$m.
We followed the prescription in \citet{boogert:08} to force the continuum
to match (by eye) the observed flux at $\sim5.5\ \mu$m and $\sim7.5\ \mu$m and 
to set an approximate flux at $\sim30\ \mu$m. We tried a number of interactive continuum
settings, but this approach generally results in a much worse agreement of the
model fits with observed flux over $10\ \mu$m $\la \lambda \la 32\ \mu$m.
We took this as an upper $1\sigma$ boundary of $A_V$ ({\tt SL}+{\tt LL}). Errors in
$A_V$ ({\tt SH}+{\tt LH}) include statistical uncertainties,
where we took the scatter of points at $20\ \mu$m $\leq \lambda \leq 30\ \mu$m with
respect to a second-order polynomial as the flux errors over the entire wavelength
range, added in quadrature to uncertainties from varying the background subtraction
and uncertainties between the two nod positions.

\begin{figure}
\epsscale{1.05}
\plotone{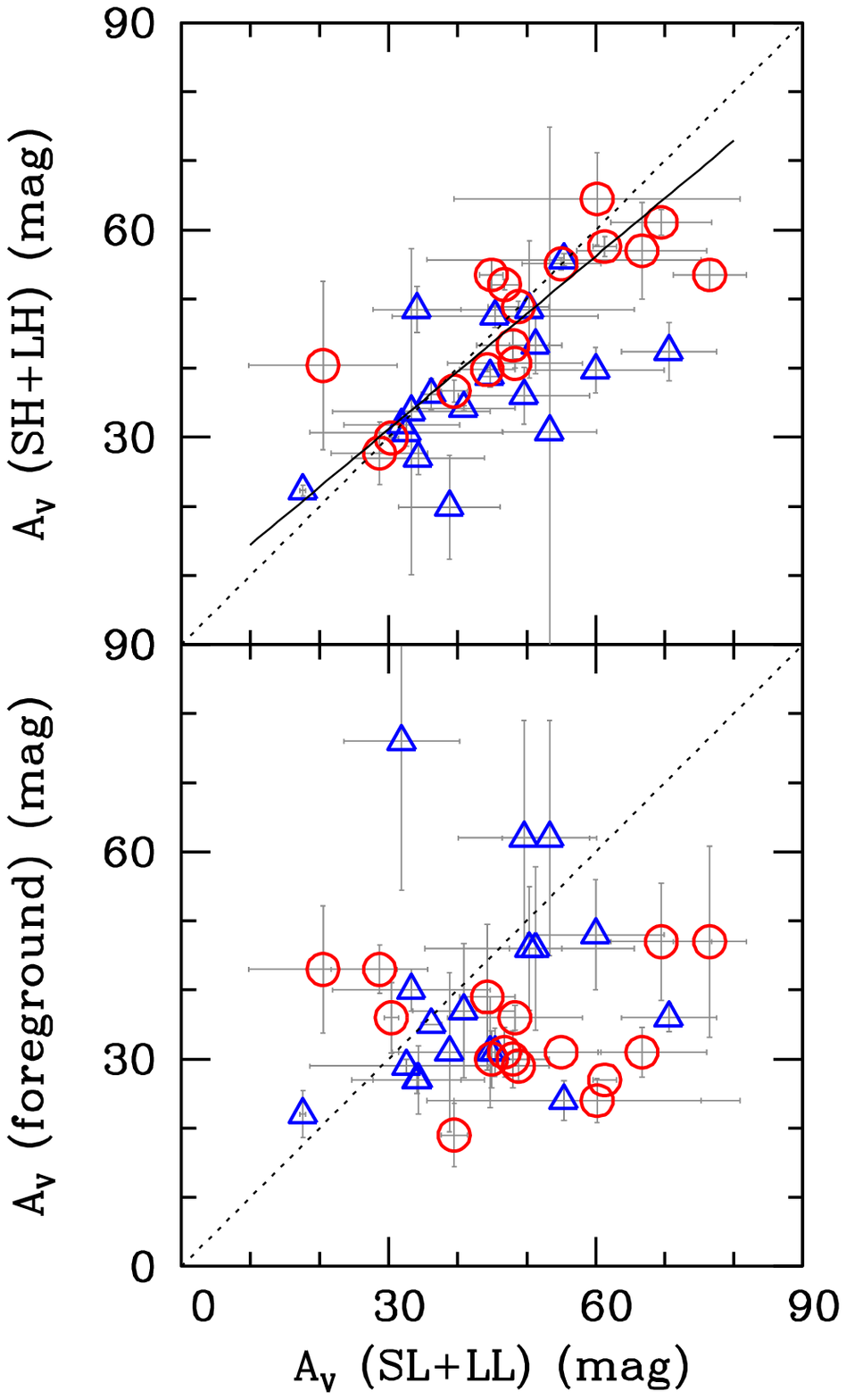}
\caption{Comparisons of $A_V$ ({\tt SL} +{\tt LL}) with $A_V$ ({\tt SH}+{\tt LH})
({\it top}) and with those based on the colors of giant stars in the field
\citep*{schultheis:09} ({\it bottom}). Red circles are YSOs, and blue triangles
are possible YSOs selected in this work. Dotted line represents equal values for
$A_V$ ({\tt SL}+{\tt LL}) and $A_V$ ({\tt SH}+{\tt LH}). Solid line in the {\it top}
panel shows a linear fit to the data using errors in both axes.
$A_V$ ({\tt SL}+{\tt LL}) is systematically larger than $A_V$ (foreground)
as $A_V$ ({\tt SL}+{\tt LL}) is the sum of the line-of-sight extinction
and the localized extinction from the dusty envelope of a YSO while
$A_V$ (foreground) is a spatially averaged line-of-sight extinction.
\label{fig:av}}
\end{figure}

Table~\ref{tab:tab5} shows $A_V$ ({\tt SL}+{\tt LL}) and $A_V$ ({\tt SH}+{\tt LH})
estimates for YSOs and possible YSOs. We compare these two extinction estimates
in the upper panel of Figure~\ref{fig:av}. The extinction for each YSO is
a combination of extinction along the line of sight to the GC and extinction
intrinsic to the YSO; sources which are not YSOs will not always have intrinsic
extinction. There is a good correlation between  $A_V$ ({\tt SL}+{\tt LL}) and
$A_V$ ({\tt SH}+{\tt LH}) as illustrated in Figure~\ref{fig:av}. This is expected,
because the high-resolution spectra are scaled to the low-resolution spectra, and so
the two methods are not completely independent from each other. The weighted mean
difference is $\langle A_V ({\tt SH}+{\tt LH}) - A_V ({\tt SL}+{\tt LL}) \rangle=
+0.38\pm0.65$~mag for both YSOs and possible YSOs. The (unweighted) rms difference
is $11$~mag, compared to the formal uncertainties of $\sim 9$ mag from both axes.

The lower panel in Figure~\ref{fig:av} shows a comparison of $A_V$ ({\tt SL}+{\tt LL}) to
$A_V$(foreground) derived from the extinction map in \citet{schultheis:09}. The latter is based
on the 2MASS and IRAC color-magnitude diagrams of GC red giant branch stars
within $2\arcmin$ from each source. The errors are the rms difference of $A_V$(foreground)
derived at the positions of four background pointings. As seen in the figure,
$A_V$ ({\tt SL}+{\tt LL}) is systematically larger than $A_V$ (foreground) for
YSOs and possible YSOs. Such overall behavior is expected for YSOs, since
$A_V$(foreground) from  \citet{schultheis:09} is a spatially averaged 
line-of-sight extinction to the GC, while $A_V$ ({\tt SL}+{\tt LL}) is the sum of the 
line-of-sight extinction to the GC and the localized extinction from the dusty envelope
of the YSO.

\subsection{Molecular Abundances}

By using the dust extinction values derived in the previous section, we derived
abundances for gas-phase molecular absorbers with respect to hydrogen. We obtained
a total hydrogen column density from the optical depth of the $9.7\ \mu$m
silicate absorption, assuming $A_V / \tau_{9.7} = 9$ \citep{roche:85} and
$N_{\rm H} / A_V \approx 1.87 \times 10^{21}$ cm$^{-2}\ {\rm mag}^{-1}$
\citep{bohlin:78} at $R_V = 3.1$. We used $A_V$({\tt SL}+{\tt LL}) to derive the
H$_2$ column density, assuming $N_{\rm H_2} = N_{\rm H}/2$. Here we implicitly
assumed that the H$_2$ column density along the full 8~kpc line of sight is
comparable to the local value near the YSO. A factor of two difference would exist,
if the local and the full H$_2$ column densities are the same, but we neglected this
difference.

The gas-phase molecular abundances relative to H$_2$ (i.e., ratios of column
densities) are shown in Table \ref{tab:tab4}. Our derived abundances for
C$_2$H$_2$ and HCN are $10^{-6.9}$--$10^{-5.3}$, and our gas-phase CO$_2$
abundances are $10^{-6.4}$--$10^{-5.1}$. Intervening molecular clouds in
the line of sight to the GC are less likely the main cause of these absorptions
because the average HCN abundance of $2.5\times10^{-8}$ towards Sgr~B2(M)
\citep{greaves:96}, where half of our YSOs and possible YSOs are found
(\S~\ref{sec:spatial}), is an order of magnitude lower than our measurements.

Individual gas-phase abundances are comparable to or generally higher
than those in earlier studies. \citet{lahuis:00} found abundances of
$10^{-8}$--$10^{-6}$ for C$_2$H$_2$ and HCN in the warm gas for several massive
YSOs, and \citet{knez:09} found $10^{-6.1}$ for C$_2$H$_2$ and $10^{-8.3}$ for
HCN towards IRS~1 in NGC~7538. \citet{boonman:03} estimated CO$_2$ abundances
of $10^{-7.2}$--$10^{-6.5}$ towards lines of sight to several YSOs.
However, these differences could be due to the uncertainties of comparing
different techniques of deriving $N$(H$_2$). If we consider the column
densities of warm gas towards massive YSOs, our values are in good agreement with
those found in the previous work \citep{lahuis:00,boonman:03,knez:09}. In addition,
our gas to solid abundance ratios for CO$_2$ ($10^{-1}$--$10^{-2}$), which do not
require knowledge of the foreground extinction, are consistent with \citet{boonman:03}.

\section{Properties of Massive YSOs in the GC}\label{sec:analysis}

In the above section, we spectroscopically identified 16 YSOs 
and 19 possible YSOs from among 107 IRS targets in the GC.
Although our selection of massive YSOs is primarily based on the $15\ \mu$m CO$_2$
ice absorption profile, absorption from hot and dense molecular
gases further supports our selection procedures. 
In this section, we derive and inspect properties of these YSOs and
possible YSOs using SED 
model fits, and look for a spatial correlation of these sources in the CMZ.

\subsection{YSO Parameters from SED Fitting}\label{sec:model}

For our 35 YSOs and possible YSOs, we performed SED fitting using a set
of models in \citet{robitaille:06}. For this purpose, we used the
{\tt Online SED Fitter}\footnote{\tt http://caravan.astro.wisc.edu/protostars}
\citep{robitaille:07} to derive YSO parameters, such as the mass of the
central object, the bolometric luminosity, and the accretion rate from
the envelope.

\begin{figure*}
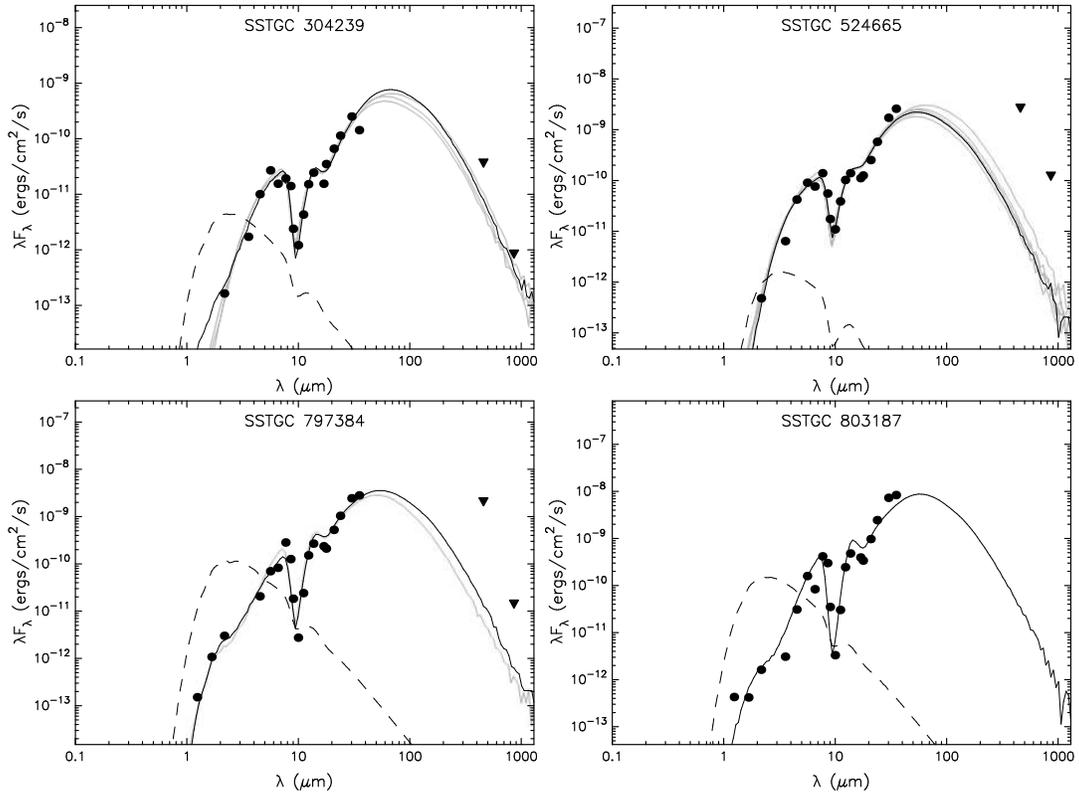

\centering
\includegraphics[scale=0.6]{fig14-02.eps}
\includegraphics[scale=0.6]{fig14-07.eps}
\includegraphics[scale=0.6]{fig14-29.eps}
\includegraphics[scale=0.6]{fig14-32.eps}
\caption{SED fitting results for four GC YSOs using a set of models in
\citet{robitaille:06} and based on near-infrared photometry, synthetic
values derived from the IRS spectra, and SCUBA observations (Table~\ref{tab:tab6}).
Observed points are shown as filled circles and upper limits are shown
as downward pointing triangles. The black line is a best-fitting model,
and grey lines represent acceptable fits. Dashed line is the emission from
the central object in the absence of the dusty envelope. The SED model fitting
suggests that our spectroscopically selected YSOs are massive Stage-I YSOs.
(An extended version of this figure is available in the online journal,
showing all 35 GC YSOs and possible YSOs.)
\label{fig:sed}}
\end{figure*}

As an input to the {\tt SED Fitter}, we used available near- and mid-IR photometry
as listed in Table~\ref{tab:tab6}. The near-IR $JHK$ observations are
{\tt Aperture3} magnitudes from UKIDSS DR2 \citep{warren:07}. Many of
our YSOs and possible YSOs are found on
saturated pixels on the MIPS $24\ \mu$m images \citep{carey:09}. Therefore,
we derived synthetic photometry at $24\ \mu$m by convolving the MIPS [24]
filter response function on the IRS spectra, following the prescriptions on
the {\it Spitzer} website\footnote{See 
{\tt http://ssc.spitzer.caltech.edu/dataanalysistools/cookbook/10/}.}.
These values are listed in Table~\ref{tab:tab6}. Synthetic values for our
IRS targets are $1.05\pm0.17$~mag ($N_{\rm comp}=29$) systematically smaller
(brighter) than MIPS [24] photometry (S.\ Carey, 2008, private communication).
Similarly, we found a mean difference of $0.72\pm0.09$~mag ($N_{\rm comp}=77$)
between IRAC [8.0] photometry \citep{ramirez:08} and synthetic values. Again
the sense of the difference is that synthetic values are brighter than
Ramirez et al. values. This is likely due to extended emission around YSOs.
We also utilized $450\ \mu$m and $850\ \mu$m observations
from the Submillimetre Common User Bolometer Array \citep[SCUBA;][]{difrancesco:08},
measured with a 23$''$ diameter beam.

In addition to the above photometry, we derived monochromatic fluxes at $14$
wavelength points: 5.58 $\mu$m, 6.4 $\mu$m, 7.65 $\mu$m, 8.5 $\mu$m,
9.0 $\mu$m, 9.7 $\mu$m, 11.0 $\mu$m, 12.0 $\mu$m, 13.5 $\mu$m,
17.0 $\mu$m, 18.0 $\mu$m, 21.0 $\mu$m, 30.0 $\mu$m, and 35.0 $\mu$m.
These points were selected to characterize the overall shape of a SED with
as little ice features as possible, because the models do not include ices.
We computed a monochromatic flux with a 2\%-wide Gaussian filter in these wavelength
points, except at 9.7 $\mu$m where we used a 3\%-wide filter, 
to better characterize the bottom of the silicate absorption band. Note that
we did not use the IRAC [5.8] and [8.0] photometry and instead used the above
synthetic values to avoid strong $6\ \mu$m and $7\ \mu$m absorption bands, which
are not included in the models.

For each source, we ran the {\tt Online SED Fitter} using the above set of photometry
and collected results that satisfy $(\chi^2 - \chi^2_{\rm min})/N_{\rm tot} < 5$,
where $\chi^2_{\rm min}$ is the minimum $\chi^2$ value from the available model sets,
and $N_{\rm tot}$ is the total number of data points, which are between 12 and 22
for our sources. We note that the fitting is not strictly statistical, given the limited parameter 
space of models for all 14 YSO parameters \citep[see][]{robitaille:07}.
We chose the above cut to include reasonable fitting results, and then estimated a 
mean and a standard deviation for each YSO parameter.

Figure~\ref{fig:sed} displays SED fitting results overplotted on the input photometry
for four YSOs. All 35 YSOs and possible YSOs are shown in the online journal. The black
solid line shows a best-fitting SED, and grey lines show acceptable fits from the
on-line SED fitter. The dashed line represents the emission from the central source in
the absence of extinction from the dusty envelope. In this fitting exercise, we restricted
the source distance, $d$, to 7~kpc $\leq d \leq$ 9~kpc from the Sun, and interstellar
extinction along the line of sight to the GC to 20~mag $\leq A_V \leq$ 40~mag.

\begin{figure*}
\epsscale{0.85}
\plotone{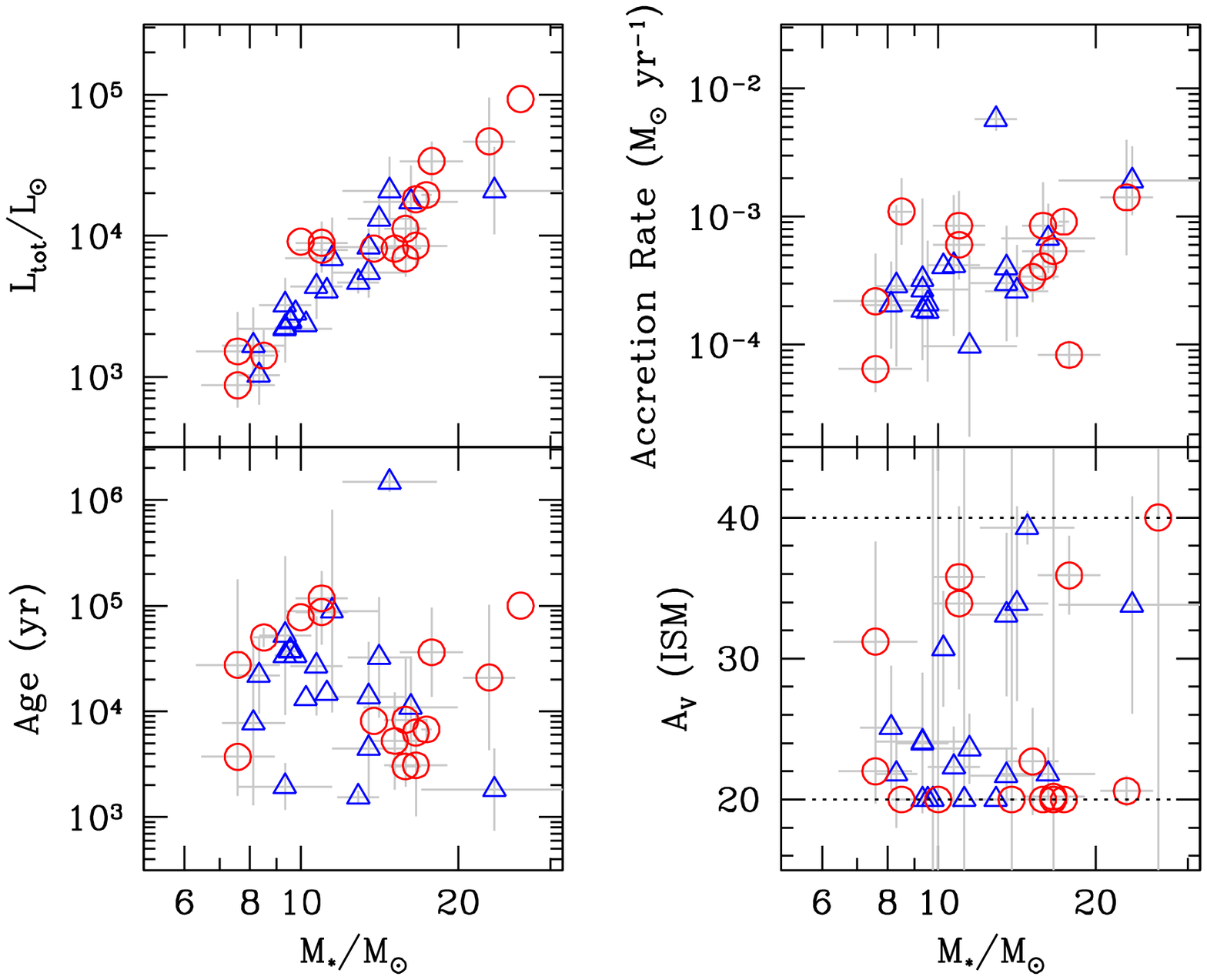}
\caption{SED fitting results for YSOs (red circles)
and possible YSOs (blue triangles) in our sample.
The total luminosity, age, envelope accretion rate, and foreground
extinction (from the top left to the bottom right panels) are shown
as a function of the mass for a central object.
\label{fig:modelvalue}}
\end{figure*}

Figure~\ref{fig:modelvalue} shows results for derived YSO parameters, and
Table~\ref{tab:tab7} summarizes the results. Entries with no error bars indicate
that a single solution is found within $(\chi^2 - \chi^2_{\rm min})/N_{\rm tot} < 5$.
Our derived masses of central objects span $8\ M_\odot \la M_* \la 23\ M_\odot$, 
and the total luminosities range over $10^3 L_\odot \la L_{\rm tot} \la 10^5 L_\odot$. 
Note that the mass is not directly determined from the SED; rather, it is the 
bolometric luminosity and the temperature we are determining, and the mass is 
implicitly constrained by these from the evolutionary tracks built-in to the model 
grid. The mass accretion rate from the envelope is another indicator for the evolutionary 
stage of YSOs. For our YSOs and possible YSOs, we found a heavy infall rate,
$10^{-4} M_\odot\ yr^{-1} \la$ \.{M}$_{\rm env} \la 10^{-3} M_\odot\ yr^{-1}$,
which is consistent with those for Stage-I YSOs \citep{robitaille:06}. The range
of these parameters recovered from the {\tt SED fitting tool} remained essentially
unchanged if we instead imposed a $A_V$ limit using $A_V$ (foreground) measurements
in Table~\ref{tab:tab5} with its $\pm2\sigma$ error bounds \citep{schultheis:09}.
Our SED fitting suggests that our sources are massive YSOs in their early stages of
protostar evolution.

\begin{figure}
\epsscale{1.1}
\plotone{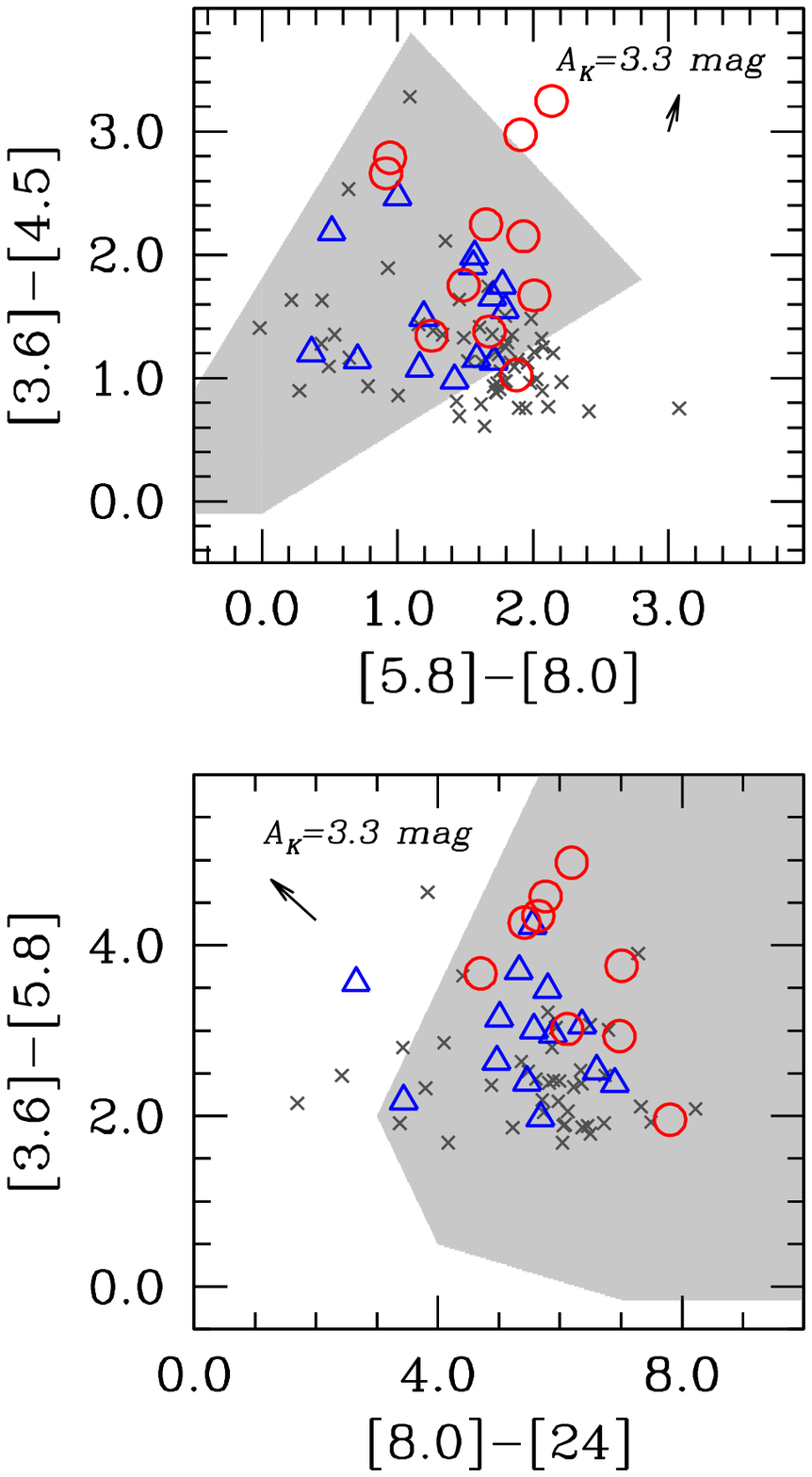}
\caption{Distribution of YSOs (red circles) and possible YSOs (blue
triangles) on IRAC/MIPS color-color diagrams.
Grey crosses represent our remaining IRS targets.
Grey regions are theoretically predicted color ranges
for Stage-I YSOs \citep[adapted from][]{robitaille:06}.
Photometry is not corrected for extinction.
The arrows indicate the reddening vector from the extinction law in
\citet{chiar:06} at $A_{K}=3.28$~mag or $A_V=29$~mag \citep{figer:99}.
There is significant overlap between YSOs and non-YSOs in our sample
within the predicted colors of Stage-I YSOs.
\label{fig:cmd}}
\end{figure}

Figure~\ref{fig:cmd} shows the color distribution of YSOs and possible YSOs
in the mid-IR color-color diagrams, overlaid with regions occupied by theoretical
Stage-I objects \citep{robitaille:06}. The colors of YSOs and possible YSOs,
relative to non-YSOs, are a bit bluer for [5.8]-[8.0]. On the other hand, YSOs
and possible YSOs are redder in [3.6]-[4.5] and [3.6]-[5.8]. All of our 107
sources have similar [8.0]-[24] colors. Although YSOs and possible YSOs in the
GC have colors that are similar to the theoretically predicted colors, non-YSOs
are also found in the same color space. This confirms earlier theoretical work
\citep[e.g.,][]{robitaille:07}, concluding that broad-band colors are not sufficient
to separate YSOs from non-YSOs.

\subsection{Mass Estimates from Radio Continuum}

Eight YSOs and possible YSOs are coincident with radio continuum sources, and
are thus likely to be compact \ion{H}{2} regions. These are listed in the last
two columns of Table~\ref{tab:tab7}. We used radio continuum data
\citep{zoonematkermani:90,mehringer:92,mehringer:93,becker:94,mehringer:95,
lazio:98,yusefzadeh:04,white:05,lazio:08,yusefzadeh:09} to derive the number of ionizing
photons for each \ion{H}{2} region, assuming a distance of 8~kpc. We then
converted the number of ionizing photons to stellar mass by using the results
of \citet{panagia:73} and assuming a surface gravity of $\log{g} = 4.2$.
Our derived masses, listed in Table \ref{tab:tab7}, agree on average with
those estimated from SED fits in the previous section (\S~\ref{sec:model}).

\subsection{Spatial Distribution of YSOs in the GC}\label{sec:spatial}

We confined our spectroscopic sample to those within $|b| < 15\arcmin$ to avoid
likely foreground objects (\S~\ref{sec:sample}). Nevertheless, this spatial cut
is generous enough that our spectroscopic survey is almost free from a spatially
dependent sample bias, and enables us to map out active star-forming regions in
the GC and to study their relation to the interstellar medium (ISM).

\begin{figure*}
\epsscale{0.85}
\plotone{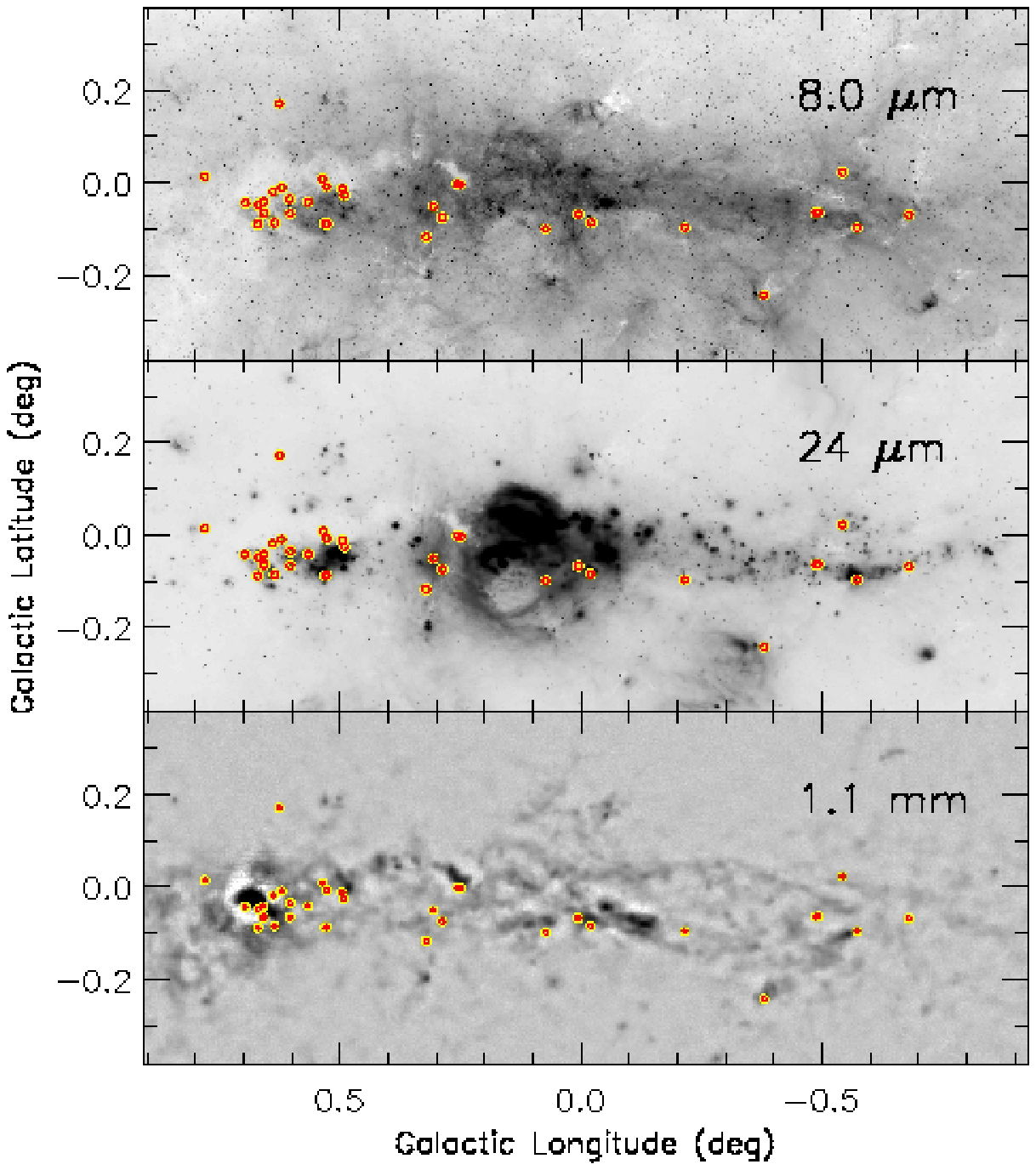}
\caption{Spatial distribution of 35 YSOs and possible YSOs on images from
{\it Spitzer}/IRAC $8\ \mu$m \citep[][top]{stolovy:06}, MIPS $24\ \mu$m
combined with images from Midcourse Space Experiment (MSX) $21.34\ \mu$m
E band \citep[][middle]{yusefzadeh:09}, and $1.1\ $mm continuum survey image
from the Bolocam Galactic Plane Survey \citep[BGPS;][bottom]{BGPS}. Although
YSOs and possible YSOs are found throughout the CMZ, about half of them are
found in and around the Sgr B molecular complex ($l \sim +0.6\deg, b \sim 0.0\deg$).
\label{fig:map.yso}}
\end{figure*}

\begin{figure*}
\epsscale{0.85}
\plotone{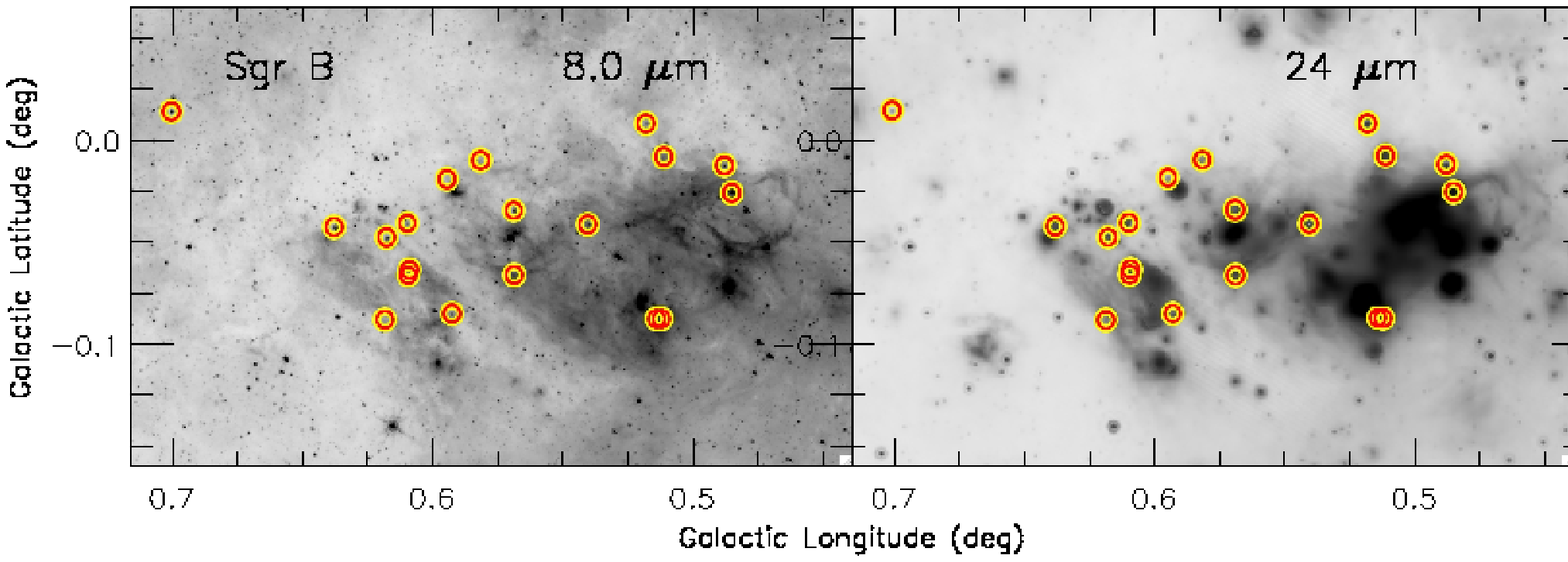}
\caption{Same as in Figure~\ref{fig:map.yso}, but in the Sgr~B region.
The YSOs and possible YSOs identified in our work are found on the
edge of the strong 24 $\mu$m emission regions.
\label{fig:map.yso.sgrB}}
\end{figure*}

Figure~\ref{fig:map.yso} displays the locations of 35 YSOs and possible YSOs in
the CMZ (see Figure~\ref{fig:map} for the locations of all of our spectroscopic
targets). Although YSOs and possible YSOs are found throughout the CMZ, it is
striking to see that half of these sources (18 out of 35) are found in and around
Sgr~B. Sgr~B is known as the most active star-forming region in the Galaxy
\citep{BGPS}, but this is the first direct evidence of the presence
of YSOs in this region at the earliest stage of star formation ($\la 1$ Myr).
Figure~\ref{fig:map.yso.sgrB} shows the Sgr~B region with locations of our
YSOs and possible YSOs. As seen on the $24\ \mu$m map, our sources are
preferentially found on the edge of strong $24\ \mu$m emission regions.

\subsection{Star Formation Rate at the GC}\label{sec:crossid}

YSOs are direct tracers of early star formation, and can be used to estimate the
{\it in situ} star formation rate (SFR) in the GC. Previous identifications of
YSOs based on broad-band photometry were used to infer the SFR in the GC, but
the heavy extinction towards the GC limits any estimate of the SFR based on
photometrically selected YSOs. This is because reddened AGB stars have similar
colors (e.g., Figure~\ref{fig:cmd}). Our IRS spectra provide a unique opportunity to
check how well earlier studies selected their YSO candidates, and can be used to
refine SFR estimates at the GC.

Table~\ref{tab:tab1} includes cross-identifications of our IRS sample with earlier
photometry-based YSO selections in \citet{felli:02}, \citet{schuller:06}, and
\citet{yusefzadeh:09}. YSO selections in both \citet{felli:02} and \citet{schuller:06}
are based on ISOGAL photometry \citep{omont:03,schuller:03}, while that of
\citet{yusefzadeh:09} is based on the {\it Spitzer} IRAC \citep{ramirez:08} and
MIPS photometry \citep{hinz:09}. The source catalogs \citep{omont:03,schuller:03,hinz:09}
do not cover the entire CMZ, in particular near Sgr~A. Sources with missing data
(``\nodata'') in Table~\ref{tab:tab1} represent our spectroscopic targets that
were not detected in these catalogs in a $3\arcsec$ search radius.

\citet{felli:02} used ISOGAL photometry at 7 $\mu$m and 15 $\mu$m to select bright
YSO candidates, using the mid-infrared color-magnitude diagram for ultra-compact
\ion{H}{2} regions. In total, 28 sources identified by \citet{felli:02} as
photometric YSOs (``yes'' in column 6 of Table~\ref{tab:tab1}) were cross-matched
with our IRS targets (Table~\ref{tab:tab1}, column 5) in a $3\arcsec$ search radius,
but we identified only 36\% of them (10/28) as YSOs (4) or possible YSOs (6) in
our study.

YSO candidates were also selected by \citet{schuller:06} based on ISOGAL photometry
at $7\ \mu$m and $15\ \mu$m and spatial extent of ISOGAL sources. Their study
focused on a small $20\arcmin\times20\arcmin$ field between Sgr~A and Sgr~C.
We have obtained IRS spectra of only eight ISOGAL sources in this field.
\citet{schuller:06} photometrically identified five GC sources as YSOs
(Table~\ref{tab:tab1}, column 7). However, none of them are identified by
us as YSOs or possible YSOs (Table~\ref{tab:tab1}, column 5).
The low rate in the YSO identification could be due to their selection criteria based
on the spatial extent of sources, while our spectroscopic targets were selected from
point sources in the IRAC bandpasses (\S~\ref{sec:sample}).

A comparable hit rate to that from \citet{felli:02} was found for YSO candidates
from the most recent photometric study by \citet{yusefzadeh:09}, whose YSO candidates
were identified based on the {\it Spitzer} IRAC and MIPS images. In total 17
photometric YSOs (``yes'' in Table~\ref{tab:tab1}, column 8) in their list were
cross-matched with our IRS targets in a $3\arcsec$ search radius, but only 47\%
(8/17; ``yes'' in Table~\ref{tab:tab1}, column 8) of them were found to be either
YSOs (3) or possible YSOs (5) in our study.

A complete analysis on the SFR estimate requires a better understanding of the sample
bias in our spectroscopic target selection, which is the subject of the next
papers of this series. Nonetheless, we can make a preliminary estimate on the SFR
based on the result in this paper: since the hit rate of the photometric YSO selection
in \citet{yusefzadeh:09} is $\sim50\%$, their SFR estimate for Stage I YSOs would have
been overestimated by a factor of $\sim2$. They have concluded that the Stage I SFR is
$\sim0.14\ M_\odot\ yr^{-1}$, so this implies a revised SFR $\sim0.07\ M_\odot\ yr^{-1}$
at the GC. If we assume a gas surface density of the GC from the total mass of
$5.3\times10^7\ M_\odot$ \citep{pierceprice:00} over the entire CMZ, both values of
the star formation rate are roughly consistent with the Kennicutt-Schmidt law
\citep{kennicutt:98}.

\section{Summary}

We obtained {\it Spitzer}/IRS spectra for 107 sources in the GC, which were selected
based on near- and mid-IR photometry including those obtained from {\it Spitzer}/IRAC.
Based on the shape of the $15\ \mu$m CO$_2$ spectral feature
and the strength of the 15.4 $\mu$m shoulder CO$_2$ ice component, we
selected 35 YSOs and possible YSOs. Our identifications are further supported by the presence of hot and
dense gas-phase molecular absorptions such as C$_2$H$_2$, HCN, and CO$_2$
for some YSOs. This is the first spectroscopic identification of a large
YSO population, tracing an early stage of star formation in the GC.
Spectroscopic confirmation of candidate YSOs in the GC is
essential because the older stellar population in the GC, when reddened
by $A_V$ $\sim$ 30, has infrared colors similar to those of YSOs.

From the SED model fitting, we inferred that the masses of these objects
are typically $\sim 8 - 23 \ M_\odot$, and that the high infall rate from
the envelope suggests that they are on Stage I, an early evolutionary stage
of protostars \citep[e.g.,][]{robitaille:06}. We found that these YSOs and
possible YSOs are found throughout the whole CMZ, but half of them are
located in and around the Sgr~B. We found that about 50\% of photometrically
selected YSOs are spectroscopically confirmed by our study. We estimated
a preliminary star formation rate, based on an earlier photometric study by
\citet{yusefzadeh:09}, to be $\sim0.07\ M_\odot\ yr^{-1}$.

Our {\it Spitzer}/IRS survey is limited to YSOs of at least $\sim3~M_\odot$
(masses of central objects). However, next generation telescopes, such as
the Giant Magellan Telescope (GMT) or the James Webb Space Telescope (JWST),
will overcome this limit, exploring significantly less massive stars with
high-resolution imaging and moderate/high-resolution spectroscopic capabilities
in the near- and mid-IR range, allowing detailed studies of the initial mass
function in these crowded fields. Until then, our {\it Spitzer}/IRS data will
remain as a unique database for studying the star formation process in the GC.

\acknowledgements

We thank the referee for careful and detailed comments.
This work is based on observations made with the {\it Spitzer Space Telescope}, which is
operated by the Jet Propulsion Laboratory, California Institute of Technology under
a contract with NASA. Support for this work was provided by NASA through an award
issued by JPL/Caltech.
This research has made use of the SIMBAD database, operated at CDS, Strasbourg, France.

\begin{deluxetable*}{crrlcccc}
\tablewidth{0pt}
\tabletypesize{\scriptsize}
\tablecaption{IRS Sample of Candidate YSOs\label{tab:tab1}}
\tablehead{
  \colhead{Source ID} &
  \colhead{R.A.} &
  \colhead{Decl.} &
  \colhead{Date of} &
  \multicolumn{4}{c}{YSO Status} \nl
  \cline{5-8}
  \colhead{SSTGC} &
  \colhead{(J2000.0)} &
  \colhead{(J2000.0)} &
  \colhead{Observation} &
  \colhead{This work\tablenotemark{a}} &
  \colhead{Felli et al.} &
  \colhead{Schuller et al.} &
  \colhead{Yusef-Zadeh et al.}
}
\startdata
244532 & 17 43 47.97 & $-29$ 38 41.2  & Oct. 2008 & no    & no      & \nodata & no      \nl 
260956 & 17 43 55.98 & $-29$ 36 22.4  & Oct. 2008 & no    & yes     & \nodata & \nodata \nl 
263857 & 17 43 57.32 & $-29$ 36 40.6  & Oct. 2008 & no    & \nodata & \nodata & yes     \nl 
284291 & 17 44 06.91 & $-29$ 24 17.4  & May  2008 & no    & yes     & \nodata & yes     \nl 
293528 & 17 44 11.20 & $-29$ 26 37.9  & May  2008 & no    & yes     & \nodata & \nodata \nl 
300758 & 17 44 14.49 & $-29$ 23 22.2  & May  2008 & maybe & yes     & \nodata & yes     \nl 
303865 & 17 44 15.85 & $-29$ 20 43.7  & May  2008 & no    & no      & \nodata & \nodata \nl 
304239 & 17 44 16.03 & $-29$ 33 16.6  & Oct. 2008 & yes   & no      & \nodata & \nodata \nl 
343554 & 17 44 31.54 & $-29$ 27 39.0  & Oct. 2008 & no    & yes     & \nodata & yes     \nl 
348392 & 17 44 33.41 & $-29$ 27 02.0  & Oct. 2008 & no    & yes     & \nodata & \nodata \nl 
349071 & 17 44 33.68 & $-29$ 13 55.7  & May  2008 & no    & yes     & yes     & \nodata \nl 
354683 & 17 44 35.87 & $-29$ 27 44.8  & Oct. 2008 & no    & \nodata & \nodata & \nodata \nl 
358370 & 17 44 37.26 & $-29$ 28 41.7  & Oct. 2008 & no    & \nodata & \nodata & \nodata \nl 
360055 & 17 44 37.90 & $-29$ 25 46.5  & Oct. 2008 & no    & yes     & \nodata & \nodata \nl 
360559 & 17 44 38.09 & $-29$ 28 38.9  & Oct. 2008 & maybe & \nodata & \nodata & \nodata \nl 
368854 & 17 44 41.29 & $-29$ 24 35.4  & May  2008 & no    & no      & \nodata & \nodata \nl 
370438 & 17 44 41.90 & $-29$ 23 32.2  & May  2008 & maybe & no      & \nodata & yes     \nl 
372630 & 17 44 42.79 & $-29$ 23 16.3  & May  2008 & maybe & no      & \nodata & yes     \nl 
374813 & 17 44 43.59 & $-29$ 20 48.8  & May  2008 & no    & no      & \nodata & yes     \nl 
381931 & 17 44 46.32 & $-29$ 27 39.3  & Oct. 2008 & no    & \nodata & \nodata & yes     \nl 
388790 & 17 44 48.94 & $-29$ 23 42.8  & Oct. 2008 & no    & no      & \nodata & \nodata \nl 
394248 & 17 44 51.02 & $-28$ 50 46.6  & May  2008 & no    & no      & \nodata & no      \nl 
395805 & 17 44 51.68 & $-29$ 11 00.2  & May  2008 & no    & \nodata & \nodata & \nodata \nl 
401264 & 17 44 53.73 & $-29$ 23 12.5  & Oct. 2008 & no    & \nodata & \nodata & yes     \nl 
404312 & 17 44 54.89 & $-29$ 14 13.1  & Oct. 2008 & no    & yes     & yes     & \nodata \nl 
405235 & 17 44 55.25 & $-29$ 15 37.8  & Oct. 2008 & no    & yes     & yes     & \nodata \nl 
412509 & 17 44 58.01 & $-29$ 10 56.6  & May  2008 & no    & \nodata & no      & yes     \nl 
421092 & 17 45 01.27 & $-29$ 14 55.7  & Oct. 2008 & no    & \nodata & no      & yes     \nl 
425399 & 17 45 02.91 & $-29$ 22 11.2  & Oct. 2008 & no    & no      & \nodata & no      \nl 
426214 & 17 45 03.21 & $-29$ 17 38.3  & Oct. 2008 & no    & yes     & yes     & \nodata \nl 
440424 & 17 45 08.58 & $-28$ 46 17.7  & May  2008 & no    & yes     & \nodata & no      \nl 
465659 & 17 45 18.10 & $-29$ 04 40.6  & Oct. 2008 & no    & yes     & yes     & \nodata \nl 
492222 & 17 45 27.95 & $-28$ 56 22.7  & May  2008 & no    & \nodata & \nodata & \nodata \nl 
496149 & 17 45 29.42 & $-29$ 10 21.8  & Oct. 2008 & maybe & no      & no      & yes     \nl 
497500 & 17 45 29.91 & $-28$ 54 22.8  & May  2008 & no    & \nodata & \nodata & \nodata \nl 
507261 & 17 45 33.50 & $-28$ 54 37.2  & May  2008 & no    & \nodata & \nodata & \nodata \nl 
511261 & 17 45 34.94 & $-29$ 25 10.3  & Oct. 2008 & no    & \nodata & \nodata & \nodata \nl 
511666 & 17 45 35.08 & $-28$ 53 34.2  & May  2008 & no    & \nodata & \nodata & \nodata \nl 
516435 & 17 45 36.84 & $-28$ 52 21.2  & May  2008 & no    & \nodata & \nodata & \nodata \nl 
516756 & 17 45 36.94 & $-28$ 54 33.4  & Oct. 2008 & no    & \nodata & \nodata & \nodata \nl 
517724 & 17 45 37.30 & $-28$ 53 53.7  & May  2008 & no    & \nodata & \nodata & \nodata \nl 
519103 & 17 45 37.80 & $-28$ 57 16.2  & May  2008 & no    & \nodata & \nodata & \nodata \nl 
521894 & 17 45 38.82 & $-28$ 52 31.9  & May  2008 & no    & \nodata & \nodata & \nodata \nl 
524419 & 17 45 39.80 & $-28$ 53 44.4  & May  2008 & no    & \nodata & \nodata & \nodata \nl 
524665 & 17 45 39.86 & $-29$ 23 23.4  & Oct. 2008 & yes   & \nodata & \nodata & \nodata \nl 
525666 & 17 45 40.22 & $-28$ 53 28.2  & May  2008 & no    & \nodata & \nodata & \nodata \nl 
\enddata
\tablenotetext{a}{YSOs are marked as ``yes'', possible YSOs are marked as
``maybe'', and the remaining targets are marked as ``no''.}
\tablenotetext{b}{$4.5\ \mu$m excess source without a $24\ \mu$m counterpart.}
\end{deluxetable*}

\begin{deluxetable*}{crrlcccc}
\tablewidth{0pt}
\tabletypesize{\scriptsize}
\tablecaption{IRS Sample of Candidate YSOs\label{tab:tab1}}
\tablehead{
  \colhead{Source ID} &
  \colhead{R.A.} &
  \colhead{Decl.} &
  \colhead{Date of} &
  \multicolumn{4}{c}{YSO Status} \nl
  \cline{5-8}
  \colhead{SSTGC} &
  \colhead{(J2000.0)} &
  \colhead{(J2000.0)} &
  \colhead{Observation} &
  \colhead{This work\tablenotemark{a}} &
  \colhead{Felli et al.} &
  \colhead{Schuller et al.} &
  \colhead{Yusef-Zadeh et al.}
}
\startdata
531300 & 17 45 42.32 & $-28$ 52 47.3  & May  2008 & no    & \nodata & \nodata & \nodata \nl
534806 & 17 45 43.57 & $-28$ 29 16.9  & Oct. 2008 & no    & \nodata & \nodata & no      \nl
535007 & 17 45 43.64 & $-28$ 52 24.9  & May  2008 & no    & \nodata & \nodata & \nodata \nl
536969 & 17 45 44.35 & $-29$ 01 13.8  & Oct. 2008 & no    & \nodata & \nodata & \nodata \nl
540840 & 17 45 45.74 & $-28$ 48 29.7  & May  2008 & no    & \nodata & \nodata & \nodata \nl
543691 & 17 45 46.76 & $-29$ 02 48.0  & Oct. 2008 & no    & \nodata & \nodata & \nodata \nl
547817 & 17 45 48.24 & $-28$ 48 16.6  & May  2008 & no    & \nodata & \nodata & \nodata \nl
550608 & 17 45 49.30 & $-28$ 50 58.8  & Oct. 2008 & no    & \nodata & \nodata & \nodata \nl
563780 & 17 45 54.11 & $-28$ 58 12.1  & Oct. 2008 & maybe & \nodata & \nodata & \nodata \nl
564417 & 17 45 54.33 & $-29$ 00 03.2  & Oct. 2008 & no    & \nodata & \nodata & \nodata \nl
579667 & 17 45 59.90 & $-28$ 53 07.2  & Oct. 2008 & no    & \nodata & \nodata & \nodata \nl
580183 & 17 46 00.07 & $-29$ 01 49.3  & Oct. 2008 & no    & \nodata & \nodata & yes     \nl
584613 & 17 46 01.67 & $-28$ 35 53.9  & May  2008 & no    & \nodata & \nodata & no      \nl
588220 & 17 46 02.98 & $-28$ 52 45.0  & Oct. 2008 & no    & \nodata & \nodata & \nodata \nl
600274 & 17 46 07.39 & $-28$ 45 32.0  & May  2008 & no    & \nodata & \nodata & \nodata \nl 
609613 & 17 46 10.71 & $-28$ 48 55.0  & Oct. 2008 & no    & \nodata & \nodata & \nodata \nl 
610642 & 17 46 11.08 & $-28$ 55 40.9  & May  2008 & maybe & \nodata & \nodata & \nodata \nl 
612688 & 17 46 11.83 & $-28$ 47 12.0  & May  2008 & no    & \nodata & \nodata & \nodata \nl 
618018 & 17 46 13.81 & $-28$ 43 44.5  & May  2008 & maybe & yes     & \nodata & \nodata \nl 
619522 & 17 46 14.33 & $-28$ 43 18.4  & May  2008 & maybe & \nodata & \nodata & \nodata \nl 
619964 & 17 46 14.48 & $-28$ 36 39.7  & May  2008 & no    & no      & \nodata & no      \nl 
621858 & 17 46 15.18 & $-28$ 52 31.4  & Oct. 2008 & no    & \nodata & \nodata & \nodata \nl 
635358 & 17 46 20.01 & $-28$ 49 18.3  & Oct. 2008 & no    & \nodata & \nodata & \nodata \nl 
646021 & 17 46 23.89 & $-28$ 39 48.1  & May  2008 & no    & no      & \nodata & \nodata \nl 
648790 & 17 46 24.93 & $-28$ 47 18.2  & Oct. 2008 & no    & \nodata & \nodata & \nodata \nl 
653270 & 17 46 26.55 & $-28$ 18 59.9  & Oct. 2008 & maybe & yes     & \nodata & no      \nl 
660708 & 17 46 29.27 & $-28$ 54 03.9  & May  2008 & no    & no      & \nodata & \nodata \nl 
670953 & 17 46 32.95 & $-28$ 42 16.3  & May  2008 & maybe & no      & \nodata & \nodata \nl 
673151 & 17 46 33.76 & $-28$ 40 32.9  & May  2008 & no    & no      & \nodata & no      \nl 
679036 & 17 46 35.98 & $-28$ 43 58.2  & May  2008 & maybe & yes     & \nodata & \nodata \nl 
689397 & 17 46 39.67 & $-28$ 41 27.8  & May  2008 & no    & yes     & \nodata & \nodata \nl 
696367 & 17 46 42.28 & $-28$ 33 26.3  & May  2008 & no    & no      & \nodata & no      \nl 
711462 & 17 46 47.82 & $-28$ 47 15.4  & May  2008 & no    & yes     & \nodata & no      \nl 
716531 & 17 46 49.64 & $-28$ 36 57.4  & Oct. 2008 & no    & yes     & \nodata & \nodata \nl 
718757 & 17 46 50.50 & $-28$ 43 33.4  & May  2008 & maybe & yes     & \nodata & \nodata \nl 
719445 & 17 46 50.72 & $-28$ 31 24.7  & May  2008 & yes   & yes     & \nodata & yes     \nl 
721436 & 17 46 51.49 & $-28$ 33 06.2  & May  2008 & no    & no      & \nodata & \nodata \nl 
722141 & 17 46 51.68 & $-28$ 28 41.6  & May  2008 & yes   & yes     & \nodata & yes     \nl 
726327 & 17 46 53.29 & $-28$ 32 01.2  & Oct. 2008 & yes   & yes     & \nodata & \nodata \nl 
728480 & 17 46 54.13 & $-28$ 29 39.5  & May  2008 & yes   & yes     & \nodata & \nodata \nl 
732531 & 17 46 55.74 & $-28$ 32 20.2  & Oct. 2008 & no    & \nodata & \nodata & \nodata \nl 
738126 & 17 46 57.95 & $-28$ 35 54.5  & May  2008 & no    & \nodata & \nodata & \nodata \nl 
760679 & 17 47 07.45 & $-28$ 28 41.9  & May  2008 & yes   & \nodata & \nodata & \nodata \nl 
761771 & 17 47 07.94 & $-28$ 24 53.2  & May  2008 & yes   & \nodata & \nodata & \nodata \nl 
769305 & 17 47 11.27 & $-28$ 26 31.7  & May  2008 & yes   & no      & \nodata & \nodata \nl 
770393 & 17 47 11.75 & $-28$ 31 21.9  & Oct. 2008 & yes   & \nodata & \nodata & \nodata \nl 
771791 & 17 47 12.35 & $-28$ 31 10.8  & Oct. 2008 & no    & \nodata & \nodata & \nodata \nl 
772151 & 17 47 12.50 & $-28$ 24 15.6  & May  2008 & yes   & \nodata & \nodata & \nodata \nl 
772981 & 17 47 12.90 & $-28$ 32 05.5  & Oct. 2008 & yes   & \nodata & \nodata & \nodata \nl 
773985 & 17 47 13.34 & $-28$ 31 56.9  & Oct. 2008 & maybe & \nodata & \nodata & \nodata \nl 
782872 & 17 47 17.31 & $-28$ 32 20.2  & Oct. 2008 & no    & yes     & \nodata & \nodata \nl 
786009 & 17 47 18.69 & $-28$ 27 31.7  & May  2008 & maybe & \nodata & \nodata & \nodata \nl 
790317 & 17 47 20.55 & $-28$ 23 54.8  & May  2008 & maybe & \nodata & \nodata & \nodata \nl 
797384 & 17 47 23.68 & $-28$ 23 34.6  & May  2008 & yes   & \nodata & \nodata & \nodata \nl 
799887 & 17 47 24.80 & $-28$ 15 56.8  & May  2008 & maybe & yes     & \nodata & no      \nl 
801865 & 17 47 25.69 & $-28$ 24 40.2  & May  2008 & yes   & \nodata & \nodata & \nodata \nl 
803187 & 17 47 26.29 & $-28$ 22 01.5  & May  2008 & yes   & \nodata & \nodata & yes\tablenotemark{b} \nl 
803471 & 17 47 26.40 & $-28$ 24 43.7  & May  2008 & yes   & \nodata & \nodata & \nodata              \nl 
806191 & 17 47 27.66 & $-28$ 26 28.4  & Oct. 2008 & maybe & no      & \nodata & yes                  \nl 
817031 & 17 47 32.97 & $-28$ 34 12.0  & Oct. 2008 & no    & yes     & \nodata & no                   \nl 
817663 & 17 47 33.28 & $-28$ 24 47.4  & May  2008 & maybe & \nodata & \nodata & \nodata              \nl 
\enddata
\tablenotetext{a}{YSOs are marked as ``yes'', possible YSOs are marked as
``maybe'', and the remaining targets are marked as ``no''.}
\tablenotetext{b}{$4.5\ \mu$m excess source without a $24\ \mu$m counterpart.}
\end{deluxetable*}

\begin{deluxetable*}{ccrrrrrl}
\tablewidth{0pt}
\tabletypesize{\scriptsize}
\tablecaption{Data Reduction Summary for YSOs and possible YSOs\label{tab:tab2}}
\tablehead{
  \colhead{SSTGC} &
  \colhead{Baseline} &
  \multicolumn{5}{c}{Flux Scaling Factor\tablenotemark{b}} &
  \colhead{Excluded} \nl
  \cline{3-7}
  \colhead{ID} &
  \colhead{Module\tablenotemark{a}} &
  \colhead{{\tt SL1}} &
  \colhead{{\tt SL2}} &
  \colhead{{\tt SL3}} &
  \colhead{{\tt LL1}} &
  \colhead{{\tt LL2}} &
  \colhead{Data Set\tablenotemark{c}}
}
\startdata
300758 & {\tt LL2} & $1.25\pm0.02$ & $1.18\pm0.06$ & $1.15\pm0.01$ & $1.03\pm0.03$ & \nodata       & \nl
304239 & {\tt SL1} & \nodata       & $0.89\pm0.08$ & $1.31\pm0.02$ & $1.80\pm0.14$ & $1.81\pm0.14$ & {\tt SL2} (1st nod, N), LL (W) \nl
360559 & {\tt SL1} & \nodata       & $2.34\pm0.06$ & $2.70\pm0.14$ & $0.10\pm0.01$ & $0.10\pm0.01$ & \nl
370438 & {\tt LL2} & $1.30\pm0.10$ & $1.17\pm0.08$ & $1.23\pm0.03$ & $0.98\pm0.04$ & \nodata       & \nl
372630 & {\tt LL2} & $1.85\pm0.07$ & $1.60\pm0.15$ & $1.61\pm0.02$ & $0.96\pm0.03$ & \nodata       & {\tt SL1} (E), {\tt SL2} (2nd nod, N, E) \nl
496149 & {\tt LL2} & $0.87\pm0.03$ & $0.85\pm0.04$ & $0.92\pm0.01$ & $1.00\pm0.02$ & \nodata       & \nl
524665 & {\tt SL1} & \nodata       & $1.03\pm0.06$ & $1.06\pm0.01$ & $0.86\pm0.01$ & $0.82\pm0.02$ & \nl
563780 & {\tt SL1} & \nodata       & $1.04\pm0.05$ & $1.06\pm0.01$ & $0.27\pm0.01$ & $0.25\pm0.01$ & {\tt SL1} (W) \nl
610642 & {\tt LL2} & \nodata       & \nodata       & \nodata       & $0.97\pm0.02$ & \nodata       & SL not available \nl
618018 & {\tt LL2} & $1.32\pm0.02$ & $1.34\pm0.06$ & $1.34\pm0.01$ & $1.02\pm0.02$ & \nodata & \nl
619522 & {\tt SL1} & \nodata       & $1.05\pm0.03$ & $1.03\pm0.02$ & $0.54\pm0.02$ & $0.58\pm0.02$ & \nl
653270 & {\tt LL2} & $0.99\pm0.01$ & $0.99\pm0.07$ & $0.99\pm0.01$ & $1.02\pm0.03$ & \nodata       & \nl
670953 & {\tt LL2} & $2.06\pm0.11$ & $2.09\pm0.06$ & $2.10\pm0.02$ & $1.00\pm0.03$ & \nodata       & {\tt SL2} (1st nod) \nl
679036 & {\tt LL2} & $1.36\pm0.10$ & $1.33\pm0.08$ & $1.36\pm0.01$ & $0.94\pm0.04$ & \nodata       & {\tt SL1} (W) \nl
718757 & {\tt LL2} & $1.66\pm0.02$ & $1.56\pm0.09$ & $1.57\pm0.02$ & $0.98\pm0.02$ & \nodata       & {\tt SL2} (E) \nl
719445 & {\tt LL2} & $1.29\pm0.04$ & $1.25\pm0.09$ & $1.28\pm0.01$ & $0.97\pm0.02$ & \nodata       & \nl
722141 & {\tt LL2} & $1.81\pm0.11$ & $1.77\pm0.11$ & $1.85\pm0.02$ & $0.96\pm0.02$ & \nodata       & {\tt SL1} (W) \nl
726327 & {\tt LL2} & $1.81\pm0.11$ & $1.77\pm0.11$ & $1.85\pm0.02$ & \nodata       & \nodata       & {\tt SL1} (W), {\tt SL2} (1st nod, N), {\tt LL1} (both nods) \nl
728480 & {\tt LL2} & $2.71\pm0.04$ & $2.60\pm0.18$ & $2.69\pm0.02$ & $1.01\pm0.02$ & \nodata       & {\tt SL1} (W) \nl
760679 & {\tt LL2} & $1.53\pm0.04$ & $1.50\pm0.05$ & $1.51\pm0.02$ & $0.95\pm0.01$ & \nodata       & {\tt SL1} (E) \nl
761771 & {\tt LL2} & $3.21\pm0.50$ & $2.93\pm0.16$ & $3.25\pm0.13$ & $0.98\pm0.02$ & \nodata       & \nl
769305 & {\tt LL2} & $1.42\pm0.06$ & $1.39\pm0.09$ & $1.42\pm0.02$ & $1.00\pm0.04$ & \nodata       & \nl
770393 & {\tt LL2} & $3.29\pm0.04$ & $3.12\pm0.18$ & $3.28\pm0.02$ & \nodata       & \nodata       & {\tt LL1} (both nods) \nl
772151 & {\tt LL2} & $1.37\pm0.08$ & $1.35\pm0.05$ & $1.54\pm0.03$ & $1.01\pm0.06$ & \nodata       & {\tt SL1} (S), {\tt SL2} (W) \nl
772981 & {\tt LL2} & $2.20\pm0.08$ & $1.88\pm0.01$ & $1.99\pm0.01$ & $0.64\pm0.02$ & \nodata       & {\tt SL2} (1st nod, N), {\tt LL1} (N) \nl
773985 & {\tt LL2} & $7.87\pm1.59$ & $6.02\pm0.10$ & $6.51\pm0.14$ & $0.90\pm0.03$ & \nodata       & {\tt LL1} (N, S) \nl
786009 & {\tt LL2} & $1.45\pm0.03$ & $1.81\pm0.13$ & $1.89\pm0.06$ & $1.00\pm0.02$ & \nodata       & {\tt SL1} (1st nod, S) \nl
790317 & {\tt LL2} & $1.45\pm0.05$ & $1.45\pm0.09$ & $1.49\pm0.01$ & $0.93\pm0.02$ & \nodata       & {\tt SH} (S) \nl
797384 & {\tt LL2} & $1.33\pm0.05$ & $1.34\pm0.09$ & $1.39\pm0.01$ & $0.99\pm0.03$ & \nodata       & \nl
799887 & {\tt LL2} & $1.02\pm0.06$ & $1.02\pm0.07$ & $1.00\pm0.01$ & $0.96\pm0.02$ & \nodata       & \nl
801865 & {\tt SL1} & \nodata       & $0.93\pm0.07$ & $1.00\pm0.01$ & $0.05\pm0.01$ & $0.04\pm0.01$ & \nl
803187 & {\tt LL2} & $1.33\pm0.13$ & $1.31\pm0.08$ & $1.33\pm0.01$ & $0.98\pm0.03$ & \nodata       & \nl
803471 & {\tt LL2} & $1.83\pm0.04$ & $1.85\pm0.13$ & $1.90\pm0.01$ & $1.03\pm0.03$ & \nodata       & \nl
806191 & {\tt LL2} & $1.37\pm0.14$ & \nodata       & \nodata       & $1.10\pm0.03$ & \nodata       & N from all sky positions, {\tt SL2} (W), {\tt SL3} (both nods, S) \nl
817663 & {\tt SL1} & \nodata       & $0.69\pm0.04$ & $0.72\pm0.05$ & $0.42\pm0.01$ & $0.48\pm0.01$ & {\tt SL2} (E)
\enddata
\tablecomments{IRS modules:
short-high ({\tt SH}; $9.9\ \mu$m$-19.6\ \mu$m, 
$\lambda / \Delta \lambda \sim 600$),
long-high ({\tt LH}; $18.7\ \mu$m$-37.2\ \mu$m, 
$\lambda / \Delta \lambda \sim 600$),
short-low ({\tt SL} [1st order {\tt SL1} $7.4\ \mu$m$-14.5\ \mu$m, 
2nd order {\tt SL2} $5.2\ \mu$m$-7.7\ \mu$m,
3rd order {\tt SL3} $7.3\ \mu$m$-8.7\ \mu$m, 
$\lambda / \Delta \lambda \sim 60-127$]), and
long-low ({\tt LL} [1st order {\tt LL1} $19.5\ \mu$m$-38.0\ \mu$m, 
2nd order {\tt LL2} $14.0\ \mu$m$-21.3\ \mu$m,
3rd order {\tt LL3} $19.4\ \mu$m$-21.7\ \mu$m, 
$\lambda / \Delta \lambda \sim 57-126$]).}
\tablenotetext{a}{IRS module selected as a baseline for the flux calibration. See text.}
\tablenotetext{b}{Adopted scaling factor in each module. The 
value represents the scaled flux divided by the original flux.}
\tablenotetext{c}{Specific modules/orders that contain defective data. 
These were excluded in the spectral analysis. 
``NSEW'' denote background observations northern/southern/eastern/western
from the source target. ``Nod'' represents a specific nod
position for a set of target spectra.}
\end{deluxetable*}

\begin{deluxetable*}{lccccccccl}
\tablewidth{0pt}
\tabletypesize{\scriptsize}
\tablecaption{CO$_2$ Ice Decomposition for 
YSOs and possible YSOs \label{tab:tab3}}
\tablehead{
  \colhead{SSTGC} &
  \colhead{$\log{N_{\rm col}}$(polar)} &
  \colhead{$\log{N_{\rm col}}$(apolar)} &
  \colhead{$\log{N_{\rm col}}$(shoulder)} &
  \colhead{$\log{N_{\rm col}}$(diluted)} &
  \colhead{$\log{N_{\rm col}}$(pure)} &
  \colhead{$\log{N_{\rm col}}$(total)} &
  \colhead{$\chi^2_{\rm tot}$} &
  \colhead{$N_{\rm tot}$} &
  \colhead{YSO} \nl
  \colhead{ID} &
  \colhead{(${\rm cm}^{-2}$)} &
  \colhead{(${\rm cm}^{-2}$)} &
  \colhead{(${\rm cm}^{-2}$)} &
  \colhead{(${\rm cm}^{-2}$)} &
  \colhead{(${\rm cm}^{-2}$)} &
  \colhead{(${\rm cm}^{-2}$)} &
  \colhead{} &
  \colhead{} &
  \colhead{Status}
}
\startdata
300758 & $17.08\pm0.08$ & $17.11\pm0.06$ & $16.88\pm0.06$ & $<16.17      $ & $<16.78      $ & $17.57\pm0.03$ & $ 488$ & $152$ & maybe \nl
304239 & $18.31\pm0.08$ & $17.98\pm0.06$ & $17.40\pm0.16$ & $<16.17      $ & $17.29\pm0.25$ & $18.54\pm0.04$ & $1615$ & $152$ & yes   \nl
360559 & $17.40\pm0.17$ & $17.41\pm0.15$ & $16.88\pm0.21$ & $16.54\pm0.31$ & $<16.78      $ & $17.79\pm0.01$ & $2181$ & $152$ & maybe \nl
370438 & $17.83\pm0.19$ & $17.42\pm0.05$ & $17.16\pm0.08$ & $<16.17      $ & $<16.78      $ & $18.06\pm0.12$ & $ 280$ & $152$ & maybe \nl
372630 & $17.06\pm0.16$ & $17.48\pm0.01$ & $17.11\pm0.05$ & $<16.17      $ & $<16.78      $ & $17.73\pm0.06$ & $ 237$ & $152$ & maybe \nl
496149 & $17.70\pm0.10$ & $17.42\pm0.04$ & $16.97\pm0.11$ & $<16.17      $ & $<16.78      $ & $17.95\pm0.04$ & $  99$ & $152$ & maybe \nl
524665 & $17.75\pm0.01$ & $17.41\pm0.05$ & $17.16\pm0.02$ & $<16.17      $ & $16.94\pm0.06$ & $18.02\pm0.01$ & $ 424$ & $152$ & yes   \nl
563780 & $17.06\pm0.01$ & $17.05\pm0.07$ & $16.75\pm0.06$ & $<16.17      $ & $<16.78      $ & $17.39\pm0.11$ & $ 594$ & $152$ & maybe \nl
610642 & $17.31\pm0.06$ & $16.80\pm0.09$ & $16.79\pm0.04$ & $<16.17      $ & $<16.78      $ & $17.54\pm0.03$ & $ 219$ & $118$ & maybe \nl
618018 & $17.18\pm0.03$ & $17.10\pm0.03$ & $16.86\pm0.03$ & $<16.17      $ & $<16.78      $ & $17.55\pm0.02$ & $ 535$ & $152$ & maybe \nl
619522 & $17.60\pm0.69$ & $17.19\pm0.30$ & $17.07\pm0.23$ & $<16.17      $ & $<16.78      $ & $17.83\pm0.62$ & $1027$ & $152$ & maybe \nl
653270 & $17.99\pm0.05$ & $16.79\pm0.21$ & $16.89\pm0.11$ & $16.82\pm0.23$ & $17.39\pm0.17$ & $18.14\pm0.04$ & $ 980$ & $152$ & maybe \nl
670953 & $17.54\pm0.39$ & $16.90\pm0.30$ & $16.78\pm0.20$ & $16.64\pm0.31$ & $<16.78      $ & $17.75\pm0.14$ & $ 710$ & $152$ & maybe \nl
679036 & $17.26\pm0.15$ & $17.17\pm0.04$ & $16.90\pm0.06$ & $<16.17      $ & $<16.78      $ & $17.62\pm0.04$ & $ 473$ & $152$ & maybe \nl
718757 & $17.76\pm0.03$ & $17.59\pm0.01$ & $16.96\pm0.01$ & $16.52\pm0.04$ & $<16.78      $ & $18.04\pm0.02$ & $ 658$ & $152$ & maybe \nl
719445 & $17.88\pm0.03$ & $17.32\pm0.04$ & $17.14\pm0.03$ & $16.36\pm0.15$ & $<16.78      $ & $18.07\pm0.03$ & $ 889$ & $152$ & yes   \nl
722141 & $17.80\pm0.02$ & $17.48\pm0.01$ & $17.12\pm0.02$ & $16.24\pm0.12$ & $<16.78      $ & $18.03\pm0.01$ & $ 990$ & $152$ & yes   \nl
726327 & $17.47\pm0.10$ & $17.30\pm0.07$ & $17.08\pm0.07$ & $<16.17      $ & $<16.78      $ & $17.80\pm0.01$ & $ 867$ & $152$ & yes   \nl
728480 & $17.83\pm0.01$ & $17.32\pm0.01$ & $17.04\pm0.01$ & $16.20\pm0.03$ & $<16.78      $ & $18.00\pm0.01$ & $1175$ & $152$ & yes   \nl
760679 & $17.70\pm0.01$ & $17.34\pm0.02$ & $17.11\pm0.03$ & $16.23\pm0.06$ & $<16.78      $ & $17.94\pm0.01$ & $ 834$ & $152$ & yes   \nl
761771 & $18.24\pm0.03$ & $17.73\pm0.01$ & $17.33\pm0.02$ & $16.42\pm0.05$ & $<16.78      $ & $18.40\pm0.02$ & $ 963$ & $152$ & yes   \nl
769305 & $18.21\pm0.12$ & $18.03\pm0.06$ & $17.57\pm0.05$ & $16.68\pm0.04$ & $<16.78      $ & $18.49\pm0.04$ & $2347$ & $152$ & yes   \nl
770393 & $17.48\pm0.03$ & $17.23\pm0.01$ & $16.78\pm0.01$ & $<16.17      $ & $<16.78      $ & $17.74\pm0.02$ & $1077$ & $152$ & yes   \nl
772151 & $18.13\pm0.10$ & $17.84\pm0.06$ & $17.41\pm0.07$ & $16.56\pm0.31$ & $<16.78      $ & $18.37\pm0.05$ & $1081$ & $152$ & yes   \nl
772981 & $17.15\pm0.15$ & $17.19\pm0.07$ & $16.95\pm0.05$ & $<16.17      $ & $<16.78      $ & $17.60\pm0.13$ & $ 630$ & $152$ & yes   \nl
773985 & $17.26\pm0.14$ & $17.24\pm0.03$ & $16.85\pm0.05$ & $<16.17      $ & $<16.78      $ & $17.64\pm0.04$ & $ 340$ & $152$ & maybe \nl
786009 & $17.43\pm0.02$ & $17.16\pm0.02$ & $16.81\pm0.02$ & $<16.17      $ & $<16.78      $ & $17.70\pm0.01$ & $ 465$ & $152$ & maybe \nl
790317 & $17.59\pm0.19$ & $17.37\pm0.02$ & $16.94\pm0.11$ & $<16.17      $ & $<16.78      $ & $17.86\pm0.10$ & $ 573$ & $152$ & maybe \nl
797384 & $17.88\pm0.02$ & $17.47\pm0.03$ & $17.21\pm0.01$ & $16.23\pm0.08$ & $<16.78      $ & $18.09\pm0.01$ & $1109$ & $152$ & yes   \nl
799887 & $17.60\pm0.18$ & $17.87\pm0.10$ & $17.50\pm0.06$ & $16.77\pm0.32$ & $17.27\pm0.42$ & $18.23\pm0.03$ & $ 817$ & $152$ & maybe \nl
801865 & $17.88\pm0.01$ & $17.66\pm0.06$ & $17.32\pm0.02$ & $16.38\pm0.10$ & $<16.78      $ & $18.16\pm0.03$ & $ 698$ & $152$ & yes   \nl
803187 & $18.17\pm0.03$ & $17.56\pm0.05$ & $17.28\pm0.04$ & $<16.17      $ & $<16.78      $ & $18.31\pm0.01$ & $1648$ & $152$ & yes   \nl
803471 & $17.84\pm0.02$ & $17.58\pm0.03$ & $17.24\pm0.03$ & $16.41\pm0.04$ & $<16.78      $ & $18.11\pm0.01$ & $1253$ & $152$ & yes   \nl
806191 & $17.92\pm0.04$ & $17.24\pm0.07$ & $17.12\pm0.05$ & $16.60\pm0.13$ & $<16.78      $ & $18.08\pm0.02$ & $ 657$ & $152$ & maybe \nl
817663 & $17.06\pm0.14$ & $17.28\pm0.13$ & $16.98\pm0.21$ & $<16.17      $ & $<16.78      $ & $17.54\pm0.08$ & $ 789$ & $152$ & maybe \nl
\hline
\multicolumn{8}{c}{Known Stars} \nl
\hline
425399 & $17.26\pm0.21$ & $<16.79      $ & $<16.65      $ & $<16.17      $ & $<16.78      $ & $17.38\pm3.38$ & $ 330$ & $152$ & known stars \nl
564417 & $17.68\pm0.63$ & $17.00\pm0.29$ & $17.01\pm0.40$ & $16.37\pm0.21$ & $<16.78      $ & $17.86\pm3.86$ & $1205$ & $152$ & known stars \nl
619964 & $17.37\pm0.32$ & $16.80\pm0.02$ & $16.80\pm0.17$ & $<16.17      $ & $<16.78      $ & $17.58\pm3.58$ & $1076$ & $152$ & known stars \nl
660708 & $17.18\pm0.40$ & $17.55\pm0.88$ & $17.23\pm0.68$ & $<16.17      $ & $17.41\pm0.63$ & $17.97\pm3.97$ & $1856$ & $151$ & known stars \nl
696367 & $17.72\pm0.66$ & $17.21\pm0.43$ & $16.91\pm0.27$ & $16.35\pm0.18$ & $<16.78      $ & $17.90\pm3.90$ & $ 499$ & $152$ & known stars \nl
\enddata
\end{deluxetable*}

\begin{deluxetable*}{lcccccccccccc}
\tablewidth{0pt}
\tabletypesize{\scriptsize}
\tablecaption{Gas-Phase Absorption Features of YSOs and possible YSOs\label{tab:tab4}}
\tablehead{
  \colhead{SSTGC} &
  \multicolumn{3}{c}{C$_2$H$_2$} &
  \colhead{} &
  \multicolumn{3}{c}{HCN} &
  \colhead{} &
  \multicolumn{3}{c}{CO$_2$} &
  \colhead{CO$_2$ gas}\nl
  \cline{2-4}
  \cline{6-8}
  \cline{10-12}
  \colhead{ID} &
  \colhead{$T_{\rm ex}$} &
  \colhead{$\log{N_{\rm col}}$} &
  \colhead{Abundance\tablenotemark{a}} &
  \colhead{} &
  \colhead{$T_{\rm ex}$} &
  \colhead{$\log{N_{\rm col}}$} &
  \colhead{Abundance\tablenotemark{a}} &
  \colhead{} &
  \colhead{$T_{\rm ex}$} &
  \colhead{$\log{N_{\rm col}}$} &
  \colhead{Abundance\tablenotemark{a}} &
  \colhead{to solid ratio} \nl
  \colhead{} &
  \colhead{(K)} &
  \colhead{(${\rm cm}^{-2}$)} &
  \colhead{} &
  \colhead{} &
  \colhead{(K)} &
  \colhead{(${\rm cm}^{-2}$)} &
  \colhead{} &
  \colhead{} &
  \colhead{(K)} &
  \colhead{(${\rm cm}^{-2}$)} &
  \colhead{}
}
\startdata
524665 & $400\pm190$ & $16.9\pm0.1$ & $-5.4\pm0.1$ && $400\pm70$  & $17.0\pm0.2$ & $-5.3\pm0.2$ &&  $200\pm100$ & $17.2\pm0.3$ & $-5.1\pm0.3$ & $0.15\pm0.10$ \nl
726327 & \nodata     & \nodata      & \nodata      && \nodata     & \nodata      & \nodata      &&  $100\pm 50$ & $16.4\pm0.3$ & $-6.3\pm0.3$ & $0.04\pm0.03$ \nl
728480 & $300\pm170$ & $15.7\pm0.3$ & $-6.9\pm0.3$ && \nodata     & \nodata      & \nodata      &&  $200\pm170$ & $16.2\pm0.2$ & $-6.4\pm0.2$ & $0.02\pm0.01$ \nl
761771 & \nodata     & \nodata      & \nodata      && \nodata     & \nodata      & \nodata      &&  $100\pm 50$ & $16.7\pm0.3$ & $-6.1\pm0.3$ & $0.02\pm0.01$ \nl
772151 & \nodata     & \nodata      & \nodata      && \nodata     & \nodata      & \nodata      &&  $100\pm173$ & $16.6\pm0.3$ & $-6.2\pm0.3$ & $0.02\pm0.01$ \nl
797384 & $100\pm160$ & $16.0\pm0.2$ & $-6.7\pm0.2$ && \nodata     & \nodata      & \nodata      &&  $100\pm 50$ & $16.6\pm0.1$ & $-6.1\pm0.1$ & $0.03\pm0.01$ \nl
801865 & \nodata     & \nodata      & \nodata      && $400\pm500$ & $16.5\pm0.4$ & $-6.3\pm0.4$ &&  $100\pm158$ & $16.9\pm0.2$ & $-5.9\pm0.2$ & $0.05\pm0.03$ \nl
803187 & $300\pm170$ & $16.4\pm0.2$ & $-6.4\pm0.2$ && $100\pm580$\tablenotemark{b} & $16.4\pm0.6$\tablenotemark{b} & $-6.4\pm0.6$ &&  $100\pm 50$ & $16.8\pm0.2$ & $-6.0\pm0.2$ & $0.03\pm0.01$ \nl
803471 & $200\pm160$ & $16.0\pm0.2$ & $-6.6\pm0.2$ && \nodata     & \nodata      & \nodata      &&  $100\pm150$ & $16.7\pm0.2$ & $-5.9\pm0.2$ & $0.04\pm0.02$ \nl
\enddata
\tablenotetext{a}{Abundance relative to molecular hydrogren, $\log{N/N({\rm H_2})}$.}
\tablenotetext{b}{Two local $\chi^2$ minima were found at $T_{\rm ex}=100~K$, $\log{N_{\rm col}}=16.4$
and $T_{\rm ex}\approx700~K$, $\log{N_{\rm col}}\approx16.7$.}
\end{deluxetable*}

\begin{deluxetable}{crrr}
\tablecaption{$A_V$ for GC YSOs and possible YSOs\label{tab:tab5}}
\tablehead{
  \colhead{SSTGC} &
  \colhead{$A_V\ ({\tt SL+LL})$} &
  \colhead{$A_V\ ({\tt SH+LH})$} &
  \colhead{$A_V$ (foreground)\tablenotemark{a}} \nl
  \colhead{ID} &
  \colhead{(mag)} &
  \colhead{(mag)} &
  \colhead{(mag)}
}
\startdata
300758 & $33.3\pm11.4$ & $33.7\pm23.6$ & $40\pm 3$ \nl 
304239 & $28.7\pm 7.0$ & $27.7\pm 4.5$ & $43\pm 4$ \nl 
360559 & $50.4\pm15.2$ & $48.5\pm 9.9$ & $46\pm 9$ \nl 
370438 & $53.3\pm 6.8$ & $30.8\pm44.1$ & $62\pm17$ \nl 
372630 & $49.6\pm 9.5$ & $36.0\pm 4.1$ & $62\pm17$ \nl 
496149 & $38.8\pm 7.4$ & $19.9\pm 7.5$ & $31\pm12$ \nl 
524665 & $20.5\pm10.7$ & $40.4\pm12.2$ & $43\pm 9$ \nl 
563780 & $60.0\pm 9.9$ & $39.7\pm 3.3$ & $48\pm 8$ \nl 
610642 & \nodata       & $24.6\pm 9.5$ & $28\pm 2$ \nl 
618018 & $34.3\pm 9.6$ & $27.0\pm 2.4$ & $27\pm 5$ \nl 
619522 & $34.1\pm 6.4$ & $48.5\pm 3.3$ & $27\pm 2$ \nl 
653270 & $17.6\pm 0.4$ & $22.3\pm 0.8$ & $22\pm 3$ \nl 
670953 & $44.7\pm 1.5$ & $38.8\pm 1.6$ & $31\pm 8$ \nl 
679036 & $51.3\pm 3.8$ & $43.3\pm 4.1$ & $46\pm12$ \nl 
718757 & $31.9\pm 8.4$ & $31.8\pm 0.3$ & $76\pm22$ \nl 
719445 & $44.9\pm 1.7$ & $53.5\pm 2.1$ & $30\pm 4$ \nl 
722141 & $39.5\pm 1.9$ & $36.7\pm 1.6$ & $19\pm 5$ \nl 
726327 & $48.0\pm 5.2$ & $43.3\pm 2.1$ & $30\pm 4$ \nl 
728480 & $44.3\pm 4.0$ & $39.8\pm 0.3$ & $39\pm11$ \nl 
760679 & $48.8\pm 4.4$ & $48.9\pm 0.8$ & $29\pm 2$ \nl 
761771 & $69.5\pm 7.3$ & $61.1\pm 1.9$ & $47\pm 8$ \nl 
769305 & $76.5\pm 5.3$ & $53.5\pm 2.3$ & $47\pm14$ \nl 
770393 & $30.4\pm 1.0$ & $29.9\pm 0.5$ & $36\pm 5$ \nl 
772151 & $60.2\pm20.7$ & $64.5\pm 6.7$ & $24\pm 3$ \nl 
772981 & $48.3\pm 9.8$ & $40.7\pm 0.7$ & $36\pm 2$ \nl 
773985 & $70.6\pm 6.9$ & $42.4\pm 4.2$ & $36\pm 2$ \nl 
786009 & $40.9\pm 7.4$ & $34.2\pm 0.4$ & $37\pm10$ \nl 
790317 & $45.4\pm14.9$ & $47.5\pm 1.6$ & $31\pm 4$ \nl 
797384 & $55.0\pm 5.7$ & $55.1\pm 0.9$ & $31\pm 1$ \nl 
799887 & $36.2\pm 1.5$ & $36.2\pm 2.3$ & $35\pm 1$ \nl 
801865 & $66.7\pm 9.4$ & $57.0\pm 7.0$ & $31\pm 4$ \nl 
803187 & $61.3\pm 1.7$ & $57.6\pm 1.4$ & $27\pm 1$ \nl 
803471 & $46.8\pm 2.1$ & $52.1\pm 0.8$ & $31\pm 4$ \nl 
806191 & $55.4\pm19.8$ & $55.7\pm 0.9$ & $24\pm 3$ \nl 
817663 & $32.6\pm14.0$ & $30.7\pm 2.0$ & $29\pm 1$ \nl 
\enddata
\tablenotetext{a}{Based on the 2MASS and IRAC color-magnitude diagrams of GC red
giant branch stars within $2\arcmin$ of the source \citep{schultheis:09}.}
\end{deluxetable}

\begin{deluxetable*}{lrrrrrrrrrrr}
\tablewidth{0pt}
\tabletypesize{\scriptsize}
\tablecaption{Photometric Properties of YSOs and possible YSOs\label{tab:tab6}}
\tablehead{
  \colhead{SSTGC} &
  \multicolumn{3}{c}{UKIDSS\tablenotemark{a}} &
  \colhead{} &
  \multicolumn{4}{c}{IRAC\tablenotemark{b}}&
  \colhead{Synthetic} &
  \multicolumn{2}{c}{SCUBA\tablenotemark{d}} \nl
  \cline{2-4}
  \cline{6-9}
  \cline{11-12}
  \colhead{ID} &
  \colhead{$J$} &
  \colhead{$H$} &
  \colhead{$K$} &
  \colhead{} &
  \colhead{[3.6]} &
  \colhead{[4.5]} &
  \colhead{[5.8]} &
  \colhead{[8.0]} &
  \colhead{[24]\tablenotemark{c}} &
  \colhead{450 $\mu$m} &
  \colhead{850 $\mu$m} \nl
  \colhead{} &
  \colhead{(mag)} &
  \colhead{(mag)} &
  \colhead{(mag)} &
  \colhead{} &
  \colhead{(mag)} &
  \colhead{(mag)} &
  \colhead{(mag)} &
  \colhead{(mag)} &
  \colhead{(mag)} &
  \colhead{(Jy)} &
  \colhead{(Jy)} 
}
\startdata
300758 & $19.06\pm0.11$ & $14.77\pm0.01$ & $12.36\pm0.01$ &  & $10.2 $ & $ 9.0 $ & $ 7.8 $ & $ 6.2 $ & $ 0.7 $ & $   12.2\pm1.4$ & $    0.3\pm0.1$\nl
304239 & \nodata        & \nodata        & $16.88\pm0.18$ &  & $12.8 $ & $10.2 $ & $ 8.6 $ & $ 7.7 $ & $ 2.2 $ & $<   5.0      $ & $    0.2\pm0.1$\nl
360559 & \nodata        & $18.74\pm0.29$ & $14.60\pm0.03$ &  & $12.2 $ & $11.1 $ & $ 9.3 $ & $ 7.6 $ & $ 1.7 $ & \nodata        & \nodata   \nl
370438 & \nodata        & \nodata        & $14.93\pm0.04$ &  & $11.9 $ & $10.0 $ & $ 8.9 $ & $ 7.4 $ & $ 1.8 $ & \nodata        & \nodata   \nl
372630 & \nodata        & $16.39\pm0.03$ & $13.63\pm0.01$ &  & $10.3 $ & $ 8.8 $ & $ 7.7 $ & $ 6.5 $ & $ 1.5 $ & $    1.9\pm0.8$ & $    0.1\pm0.1$\nl
496149 & \nodata        & $16.93\pm0.11$ & $13.89\pm0.03$ &  & $12.0 $ & $11.0 $ & $ 9.6 $ & $ 8.2 $ & $ 1.3 $ & \nodata         & \nodata   \nl
524665 & \nodata        & \nodata        & $15.71\pm0.10$ &  & $11.4 $ & $ 8.6 $ & $ 7.1 $ & $ 6.1 $ & $ 0.5 $ & $  428.5\pm1.1$ & $   36.9\pm0.1$\nl
563780 & $18.56\pm0.10$ & $15.93\pm0.04$ & $14.09\pm0.04$ &  & $11.6 $ & $10.8 $ & $ 8.2 $ & \nodata & $ 0.7 $ & \nodata         & \nodata   \nl
610642 & $18.97\pm0.15$ & \nodata        & $12.56\pm0.01$ &  & $ 9.8 $ & $ 8.0 $ & $ 6.6 $ & $ 4.8 $ & $-0.2 $ & \nodata         & \nodata   \nl
618018 & $15.33\pm0.01$ & $14.52\pm0.01$ & $13.79\pm0.02$ &  & \nodata & $ 9.6 $ & $ 7.9 $ & $ 6.5 $ & $ 0.5 $ & \nodata         & \nodata   \nl
619522 & \nodata        & $14.50\pm0.01$ & $11.97\pm0.01$ &  & $10.4 $ & $ 9.3 $ & $ 8.4 $ & $ 7.7 $ & $ 2.0 $ & \nodata         & \nodata   \nl
653270 & \nodata        & $16.43\pm0.06$ & $10.98\pm0.01$ &  & $ 8.3 $ & $ 7.1 $ & $ 6.1 $ & $ 5.7 $ & $ 2.3 $ & $   26.2\pm1.6$ & $    1.1\pm0.1$\nl
670953 & \nodata        & $17.38\pm0.15$ & $14.92\pm0.06$ &  & $11.4 $ & $ 8.9 $ & $ 7.1 $ & $ 6.1 $ & $ 0.6 $ & \nodata         & \nodata   \nl
679036 & $17.05\pm0.02$ & $16.50\pm0.04$ & $14.66\pm0.04$ &  & $11.3 $ & $ 9.4 $ & $ 7.6 $ & $ 6.1 $ & $ 0.7 $ & $    8.1\pm0.7$ & $    1.3\pm0.1$\nl
718757 & \nodata        & \nodata        & $15.41\pm0.07$ &  & $10.9 $ & $ 9.3 $ & $ 7.8 $ & $ 6.0 $ & $-0.3 $ & \nodata         & \nodata   \nl
719445 & \nodata        & $16.50\pm0.04$ & $13.51\pm0.01$ &  & $11.4 $ & $ 9.1 $ & $ 7.7 $ & $ 6.0 $ & $ 1.3 $ & \nodata         & \nodata   \nl
722141 & \nodata        & \nodata        & $15.35\pm0.05$ &  & $13.0 $ & $10.9 $ & $ 9.3 $ & $ 7.3 $ & $ 0.3 $ & \nodata         & \nodata   \nl
726327 & $18.29\pm0.07$ & $13.55\pm0.01$ & $11.55\pm0.01$ &  & $ 9.3 $ & $ 7.9 $ & $ 6.6 $ & $ 4.9 $ & \nodata & \nodata         & \nodata   \nl
728480 & \nodata        & \nodata        & $13.20\pm0.01$ &  & $11.3 $ & $10.3 $ & $ 9.3 $ & $ 7.5 $ & $-0.3 $ & $   47.9\pm0.9$ & $    8.0\pm0.1$\nl
760679 & \nodata        & \nodata        & $15.67\pm0.07$ &  & \nodata & $10.3 $ & $ 8.3 $ & $ 6.5 $ & $-0.2 $ & $   21.0\pm0.9$ & $    0.9\pm0.1$\nl
761771 & \nodata        & \nodata        & $15.43\pm0.06$ &  & $13.2 $ & $10.6 $ & \nodata & $ 8.0 $ & $ 1.3 $ & \nodata         & \nodata   \nl
769305 & $18.50\pm0.08$ & $18.23\pm0.19$ & \nodata        &  & $11.8 $ & $ 8.8 $ & $ 7.2 $ & $ 5.3 $ & $-0.5 $ & $  492.6\pm1.2$ & $   28.1\pm0.1$\nl
770393 & \nodata        & $17.17\pm0.07$ & $13.81\pm0.01$ &  & $10.4 $ & $ 8.7 $ & $ 7.2 $ & $ 5.1 $ & \nodata & \nodata         & \nodata   \nl
772151 & $15.88\pm0.01$ & $14.85\pm0.01$ & $14.22\pm0.02$ &  & $13.1 $ & $11.7 $ & $10.2 $ & \nodata & $ 2.1 $ & $  568.3\pm1.4$ & $   21.5\pm0.1$\nl
772981 & \nodata        & $16.11\pm0.03$ & $13.54\pm0.01$ &  & $11.2 $ & $ 9.8 $ & $ 8.2 $ & $ 7.0 $ & $ 0.0 $ & \nodata         & \nodata   \nl
773985 & \nodata        & \nodata        & $14.30\pm0.02$ &  & $12.0 $ & $10.7 $ & \nodata & \nodata & $-0.7 $ & \nodata         & \nodata   \nl
786009 & \nodata        & \nodata        & \nodata        &  & \nodata & $10.7 $ & \nodata & \nodata & $-0.6 $ & $   62.4\pm1.0$ & $    2.0\pm0.1$\nl
790317 & \nodata        & $17.21\pm0.07$ & $14.40\pm0.02$ &  & \nodata & $10.8 $ & $ 9.2 $ & $ 7.5 $ & $ 0.5 $ & $ 7214.4\pm3.5$ & $  326.2\pm0.3$\nl
797384 & $18.52\pm0.08$ & $15.59\pm0.02$ & $13.72\pm0.01$ &  & \nodata & $ 9.4 $ & $ 7.7 $ & $ 5.6 $ & $-0.2 $ & $  340.4\pm3.6$ & $    4.4\pm0.4$\nl
799887 & \nodata        & \nodata        & $14.24\pm0.02$ &  & $ 9.4 $ & $ 7.2 $ & $ 5.8 $ & $ 5.3 $ & $ 2.7 $ & \nodata         & \nodata   \nl
801865 & \nodata        & $17.07\pm0.07$ & \nodata        &  & \nodata & $11.3 $ & $10.3 $ & $ 8.8 $ & $ 3.3 $ & \nodata         & \nodata   \nl
803187 & $17.39\pm0.03$ & $16.61\pm0.05$ & $14.39\pm0.02$ &  & $12.2 $ & $ 9.0 $ & $ 7.2 $ & $ 5.1 $ & $-1.1 $ & \nodata         & \nodata   \nl
803471 & \nodata        & \nodata        & $13.50\pm0.01$ &  & $10.5 $ & $ 8.8 $ & $ 7.5 $ & $ 6.0 $ & $-0.1 $ & \nodata         & \nodata   \nl
806191 & $16.52\pm0.01$ & $15.43\pm0.01$ & $14.61\pm0.03$ &  & $12.6 $ & $11.0 $ & $ 9.1 $ & $ 7.4 $ & $ 1.6 $ & $  243.5\pm1.2$ & $   11.3\pm0.1$\nl
817663 & \nodata        & $16.35\pm0.03$ & $14.34\pm0.02$ &  & $12.6 $ & $11.5 $ & $10.1 $ & $ 8.9 $ & $ 2.3 $ & $  133.9\pm1.1$ & $    8.1\pm0.1$\nl
\enddata
\tablenotetext{a}{Aperture3 magnitudes from UKIDSS DR2 \citep{warren:07}.}
\tablenotetext{b}{Systematic errors of IRAC photometry were determined to be 0.1~mag,
0.1~mag, 0.15~mag, and 0.2 mags for channels 1, 2, 3, and 4, respectively, by comparing
the \citet{ramirez:08} values 
with the measurements from the GLIMPSE II catalog \citep{churchwell:09}.}
\tablenotetext{c}{Synthetic photometry based on IRS spectra.}
\tablenotetext{d}{Fluxes from SCUBA Legacy Catalogues \citep{difrancesco:08}.}
\end{deluxetable*}

\begin{deluxetable*}{lcccccrc}
\tablewidth{0pt}
\tabletypesize{\scriptsize}
\tablecaption{SED Fitting Results and Mass Estimates from Radio Continuum\label{tab:tab7}}
\tablehead{
  \colhead{SSTGC} &
  \colhead{$\log{M_*}$} &
  \colhead{$\log{L_{tot}}$} &
  \colhead{$A_V$ (ISM)} &
  \colhead{$\log{}$\.{M}$_{env}$} &
  \colhead{$\log{\rm Age}$} &
  \multicolumn{2}{c}{Radio Modeling} \nl
  \cline{7-8}
  \colhead{ID} &
  \colhead{($M_\odot$)} &
  \colhead{($L_\odot$)} &
  \colhead{(mag)} &
  \colhead{($M_\odot\ yr^{-1}$)} &
  \colhead{(yr)} &
  \colhead{$\log{M_*}$ ($M_\odot$)} &
  \colhead{References}
}
\startdata
300758 & $1.03\pm0.05$ & $3.64\pm0.23$ & $22.3\pm2.9$ & $-3.38\pm0.55$ & $4.43\pm0.47$ & $1.18$ & 1 \nl
304239 & $0.88\pm0.08$ & $3.18\pm0.28$ & $31.2\pm7.1$ & $-3.66\pm0.37$ & $4.44\pm0.81$ & & \nl
360559 & $1.01       $ & $3.37       $ & $30.7\pm4.1$ & $-3.39       $ & $4.12       $ & & \nl
370438 & $0.91\pm0.06$ & $3.22\pm0.27$ & $25.1\pm4.4$ & $-3.69\pm0.34$ & $3.89\pm0.78$ & & \nl
372630 & $0.97\pm0.09$ & $3.34\pm0.24$ & $24.1\pm3.4$ & $-3.57\pm0.24$ & $3.29\pm0.22$ & & \nl
496149 & $0.97\pm0.05$ & $3.51\pm0.19$ & $24.0\pm5.0$ & $-3.73\pm0.37$ & $4.72\pm0.75$ & & \nl
524665 & $1.04\pm0.05$ & $3.90\pm0.16$ & $33.9\pm6.1$ & $-3.07\pm0.27$ & $4.94\pm0.31$ & & \nl
563780 & $0.99$        & $3.45$        & $20.0\pm9.0$ & \nodata        & $4.53$        & & \nl
610642 & $1.06\pm0.09$ & $3.84\pm0.29$ & $23.6\pm2.5$ & $-4.01\pm0.71$ & $4.95\pm0.96$ & & \nl
618018 & $0.97\pm0.02$ & $3.35\pm0.10$ & $20.0      $ & $-3.49\pm0.63$ & $4.53\pm0.20$ & & \nl
619522 & $0.98       $ & $3.40       $ & $20.0      $ & $-3.68       $ & $4.58       $ & & \nl
653270 & $1.17\pm0.09$ & $4.32\pm0.24$ & $39.3\pm1.2$ & \nodata        & $6.17\pm0.09$ & & \nl
670953 & $1.37\pm0.14$ & $4.32\pm0.31$ & $33.8\pm7.7$ & $-2.72\pm0.27$ & $3.26\pm0.39$ & & \nl
679036 & $0.98\pm0.01$ & $3.39\pm0.08$ & $20.0      $ & $-3.74\pm0.55$ & $4.57\pm0.06$ & & \nl
718757 & $1.13\pm0.07$ & $3.92\pm0.13$ & $33.1\pm5.8$ & $-3.52\pm0.45$ & $4.14\pm0.52$ & & \nl
719445 & $0.88\pm0.07$ & $2.94\pm0.16$ & $22.0\pm2.3$ & $-4.19\pm0.18$ & $3.57\pm0.37$ & & \nl
722141 & $1.04\pm0.05$ & $3.95\pm0.15$ & $35.8\pm5.0$ & $-3.22\pm0.27$ & $5.07\pm0.26$ & & \nl
726327 & $1.36\pm0.05$ & $4.67\pm0.31$ & $20.6\pm1.1$ & $-2.85\pm0.45$ & $4.32\pm0.69$ & $1.27$ & 2 \nl
728480 & $1.18\pm0.05$ & $3.91\pm0.11$ & $22.7\pm3.8$ & $-3.47\pm0.20$ & $3.72\pm0.46$ & $1.26$ & 3 \nl
760679 & $1.42$        & $4.97$        & $40.0$       & \nodata        & $5.00$        & $1.26$ & 4 \nl
761771 & $1.00$        & $3.96$        & $20.0$       & \nodata        & $4.89$        & & \nl
769305 & $1.24\pm0.01$ & $4.29\pm0.04$ & $20.0      $ & $-3.04\pm0.09$ & $3.83\pm0.10$ & $1.27$ & 5 \nl
770393 & $1.25\pm0.06$ & $4.53\pm0.14$ & $35.9\pm2.8$ & $-4.08\pm0.11$ & $4.56\pm0.42$ & & \nl
772151 & $1.14$        & $3.91$        & $20.0$       & \nodata        & $3.91$        & $1.26$ & 6 \nl
772981 & $1.20\pm0.03$ & $3.84\pm0.13$ & $20.0\pm0.1$ & $-3.39\pm0.18$ & $3.48\pm0.19$ & & \nl
773985 & $1.21\pm0.09$ & $4.24\pm0.26$ & $21.8\pm1.9$ & $-3.17\pm0.27$ & $4.04\pm0.48$ & & \nl
786009 & $1.15\pm0.06$ & $4.12\pm0.14$ & $33.9\pm6.9$ & $-3.58\pm0.36$ & $4.51\pm0.57$ & & \nl
790317 & $1.13\pm0.07$ & $3.74\pm0.18$ & $21.7\pm2.0$ & $-3.40\pm0.25$ & $3.65\pm0.53$ & & \nl
797384 & $1.20\pm0.04$ & $4.05\pm0.24$ & $20.0      $ & $-3.07\pm0.34$ & $3.92\pm0.58$ & $1.29$ & 7 \nl
799887 & $1.11\pm0.04$ & $3.67\pm0.08$ & $20.0      $ & $-2.24\pm0.09$ & $3.19\pm0.06$ & & \nl
801865 & $0.93\pm0.02$ & $3.15\pm0.18$ & $20.0      $ & $-2.96\pm0.26$ & $4.70\pm0.09$ & & \nl
803187 & $1.22$        & $4.26$        & $20.0\pm9.0$ & \nodata        & $3.80$        & $1.39$ & 8 \nl
803471 & $1.22\pm0.06$ & $3.93\pm0.16$ & $20.2\pm0.4$ & $-3.27\pm0.22$ & $3.49\pm0.48$ & & \nl
806191 & $1.05$        & $3.61$        & $20.0\pm9.0$ & \nodata        & $4.17$        & & \nl 
817663 & $0.92\pm0.04$ & $3.01\pm0.21$ & $21.8\pm3.8$ & $-3.54\pm0.63$ & $4.34\pm0.36$ & & \nl 
\enddata
\tablerefs{References for radio observations:
(1) \citet{yusefzadeh:09};
(2) GPSR5 0.488-0.028, \citet{mehringer:92,becker:94};
(3) \#8, \citet{mehringer:92,mehringer:95};
(4) 2LC 000.563-0.044, \citet{mehringer:92,yusefzadeh:04,lazio:08};
(5) GPSR5 0.602-0.037, \citet{becker:94,lazio:98,yusefzadeh:04,lazio:08};
(6) 1LC 000.635-0.020, \citet{lazio:98,white:05};
(7) SGR B2 HII P, \citet{mehringer:93,yusefzadeh:04};
(8) GPSR5 0.693-0.046, \citet{zoonematkermani:90,becker:94,lazio:98,yusefzadeh:04,white:05,lazio:08}.}
\end{deluxetable*}


\begin{thebibliography}

\bibitem[An et al.(2009)]{an:09}
An, D., et al.\ 2009, \apjl, 702, L128 (A09)

\bibitem[Aoki et al.(1999)]{aoki:99}
Aoki, W., Tsuji, T., \& Ohnaka, K.\ 1999, \aap, 350, 945

\bibitem[Bally et al.(1987)]{bally:87}
Bally, J., et al.\ 1987, \apjs, 65, 13

\bibitem[Bally et al.(1988)]{bally:88}
Bally, J., Stark, A.~A., Wilson, R.~W., \& Henkel, C.\ 1988, \apj, 324, 223

\bibitem[Bally et al.(2010)]{BGPS}
Bally, J., et al.\ 2010, \apj, 721, 137

\bibitem[Becker et al.(1994)]{becker:94}
Becker, R.~H., White, R.~L., Helfand, D.~J., \& Zoonematkermani, S.\ 1994, \apjs, 91, 347 

\bibitem[Bergin et al.(2005)]{bergin:05}
Bergin, E.~A., Melnick, G.~J., Gerakines, P.~A., Neufeld, D.~A., 
\& Whittet, D.~C.~B.\ 2005, \apjl, 627, L33 

\bibitem[Bohlin et al.(1978)]{bohlin:78}
Bohlin, R.~C., Savage, B.~D., \& Drake, J.~F.\ 1978, \apj, 224, 132

\bibitem[Boogert et al.(2008)]{boogert:08}
Boogert, A.~C.~A., et al.\ 2008, \apj, 678, 985 

\bibitem[Boonman et al.(2003)]{boonman:03}
Boonman, A.~M.~S., van Dishoeck, E.~F., Lahuis, F., \& Doty, S.~D.\ 2003,
\aap, 399, 1063

\bibitem[Cami et al.(2010)]{spectrafactory}
Cami, J., van Malderen, R., \& Markwick, A.~J.\ 2010, \apjs, 187, 409

\bibitem[Carey et al.(2009)]{carey:09}
Carey, S.~J., et al.\ 2009, \pasp, 121, 76

\bibitem[Chiar \& Tielens(2006)]{chiar:06}
Chiar, J.~E., \& Tielens, A.~G.~G.~M.\ 2006, \apj, 637, 774 

\bibitem[Churchwell et al.(2009)]{churchwell:09}
Churchwell, E., Babler, B.~L., Meade, M.~R., Whitney, B.~A.,
Benjamin, R., Indebetouw, R., Cyganowski, C., Robitaille, T.~P., 
Povich, M., Watson, C.,
Bracker, S.\ 2009, \pasp, 121, 213

\bibitem[Dartois et al.(1999a)]{dartois:99a}
Dartois, E., Demyk, K., d'Hendecourt, L., \& Ehrenfreund, P.\ 1999a, \aap, 351, 1066

\bibitem[Dartois et al.(1999b)]{dartois:99b}
Dartois, E., Schutte, W., Geballe, T.~R., Demyk, K., Ehrenfreund, P.,
\& D'Hendecourt, L.\ 1999b, \aap, 342, L32

\bibitem[Di Francesco et al.(2008)]{difrancesco:08}
Di Francesco, J., Johnstone, D., Kirk, H., MacKenzie, T., 
\& Ledwosinska, E.\ 2008, \apjs, 175, 277 

\bibitem[Ehrenfreund et al.(1999)]{ehrenfreund:99} 
Ehrenfreund, P., et al.\ 1999, \aap, 350, 240 

\bibitem[Fatuzzo \& Melia(2009)]{fatuzzo:09}
Fatuzzo, M., \& Melia, F.\ 2009, \pasp, 121, 585 

\bibitem[Fazio et al.(2004)]{fazio:04}
Fazio, G.~G., et al.\ 2004, \apjs, 154, 10

\bibitem[Felli et al.(2002)]{felli:02}
Felli, M., Testi, L., Schuller, F., \& Omont, A.\ 2002, \aap, 392, 971

\bibitem[Figer et al.(1999)]{figer:99}
Figer, D. F., et al., 1999, \apj, 525, 750

\bibitem[Frogel \& Whitford(1987)]{frogel:87}
Frogel, J. A., \& Whitford, A. E., 1987, \apj, 320, 199

\bibitem[Gerakines et al.(1995)]{gerakines:95}
Gerakines, P.~A., Schutte, W.~A., Greenberg, J.~M.,
\& van Dishoeck, E.~F.\ 1995, \aap, 296, 810

\bibitem[Gerakines et al.(1999)]{gerakines:99}
Gerakines, P.~A., et al.\ 1999, \apj, 522, 357

\bibitem[Gibb et al.(2004)]{gibb:04}
Gibb, E.~L., Whittet, D.~C.~B., Boogert, A.~C.~A., \& Tielens,
A.~G.~G.~M.\ 2004, \apjs, 151, 35

\bibitem[Greaves \& Nyman(1996)]{greaves:96}
Greaves, J.~S., \& Nyman, L.-A.\ 1996, \aap, 305, 950

\bibitem[G\"usten et al.(1985)]{gusten:85}
G\"usten, R., Henkel, C., \& Batrla, W.\ 1985, \aap, 149, 195

\bibitem[Hinz et al.(2009)]{hinz:09}
Hinz, J.~L., Rieke, G.~H., Yusef-Zadeh, F., Hewitt, J., Balog, Z., \&
Block, M.\ 2009, \apjs, 181, 227 

\bibitem[Houck et al.(2004)]{houck:04}
Houck, J.~R., et al.\ 2004, \apjs, 154, 18

\bibitem[Hudgins et al.(1993)]{hudgins:93}
Hudgins, D.~M., Sandford, S.~A., Allamandola, L.~J., 
\& Tielens, A.~G.~G.~M.\ 1993, \apjs, 86, 713

\bibitem[Justtanont et al.(1998)]{justtanont:98} 
Justtanont, K., Feuchtgruber, H., de Jong, T., Cami, J., 
Waters, L.~B.~F.~M., Yamamura, I., \& Onaka, T.\ 1998, \aap, 330, L17 

\bibitem[Kemper et al.(2004)]{kemper:04}
Kemper, F., Vriend, W.~J., \& Tielens, A.~G.~G.~M.\ 2004, \apj, 609, 826

\bibitem[Kennicutt(1998)]{kennicutt:98}
Kennicutt, R.~C., Jr.\ 1998, \apj, 498, 541

\bibitem[Knez et al.(2005)]{knez:05}
Knez, C., et al.\ 2005, \apjl, 635, L145

\bibitem[Krabbe et al.(1991)]{krabbe:91} Krabbe, A., Genzel, R., 
Drapatz, S., \& Rotaciuc, V.\ 1991, \apjl, 382, L19 

\bibitem[Knez et al.(2009)]{knez:09}
Knez, C., Lacy, J.~H., Evans, N.~J., van Dishoeck, E.~F., \&
Richter, M.~J.\ 2009, \apj, 696, 471

\bibitem[Lahuis \& van Dishoeck(2000)]{lahuis:00}
Lahuis, F., \& van Dishoeck, E.~F.\ 2000, \aap, 355, 699

\bibitem[Lazio \& Cordes(1998)]{lazio:98}
Lazio, T.~J.~W., \& Cordes, J.~M.\ 1998, \apjs, 118, 201 

\bibitem[Lazio \& Cordes(2008)]{lazio:08}
Lazio, T.~J.~W., \& Cordes, J.~M.\ 2008, \apjs, 174, 481 

\bibitem[Markwardt(2009)]{markwardt:09}
Markwardt, C.~B.\ 2009, Astronomical Society of the Pacific Conference Series, 411, 251 

\bibitem[Mauerhan et al.(2010)]{mauerhan:10}
Mauerhan, J.~C., Muno, 
M.~P., Morris, M.~R., Stolovy, S.~R., \& Cotera, A.\ 2010, \apj, 710, 706

\bibitem[Mehringer(1995)]{mehringer:95}
Mehringer, D.~M.\ 1995, \apj, 454, 782

\bibitem[Mehringer et al.(1993)]{mehringer:93}
Mehringer, D.~M., Palmer, P., Goss, W.~M., \& Yusef-Zadeh, F.\ 1993, \apj, 412, 684

\bibitem[Mehringer et al.(1992)]{mehringer:92}
Mehringer, D.~M., Yusef-Zadeh, F., Palmer, P., \& Goss, W.~M.\ 1992, \apj, 401, 168 

\bibitem[Morris \& Serabyn(1996)]{morris:96}
Morris, M., \& Serabyn, E.\ 1996, \araa, 34, 645

\bibitem[Nishimura et al.(1980)]{nishimura:80}
Nishimura, T., Low, F.~J., \& Kurtz, R.~F.\ 1980, \apjl, 239, L101

\bibitem[Omont et al.(2003)]{omont:03}
Omont, A., et al.\ 2003, \aap, 403, 975

\bibitem[Panagia(1973)]{panagia:73} Panagia, N.\ 1973, \aj, 78, 929 

\bibitem[Pierce-Price et al.(2000)]{pierceprice:00}
Pierce-Price, D., et al.\ 2000, \apjl, 545, L121 

\bibitem[Pontoppidan et al.(2008)]{pontoppidan:08}
Pontoppidan, K.~M., et al.\ 2008, \apj, 678, 1005

\bibitem[Price et al.(2001)]{price:01}
Price, S.~D., Egan, M.~P., Carey, S.~J., Mizuno, D.~R.,
\& Kuchar, T.~A.\ 2001, \aj, 121, 2819

\bibitem[Ram{\'{\i}}rez et al.(2008)]{ramirez:08}
Ram{\'{\i}}rez, S.~V., Arendt, R.~G., Sellgren, K., Stolovy, S.~R., Cotera, A.,
Smith, H.~A., \& Yusef-Zadeh, F.\ 2008, \apjs, 175, 147

\bibitem[Reid et al.(2009)]{reid:09}
Reid, M.~J., Menten, K.~M., Zheng, X.~W., Brunthaler, A., \& Xu, Y.\ 2009, \apj, 705, 1548

\bibitem[Robitaille et al.(2006)]{robitaille:06}
Robitaille, T.~P., Whitney, B.~A., Indebetouw, R., Wood, K., 
\& Denzmore, P.\ 2006, \apjs, 167, 256

\bibitem[Robitaille et al.(2007)]{robitaille:07}
Robitaille, T.~P., Whitney, B.~A., Indebetouw, R., \& Wood, K.\ 2007, \apjs, 169, 328

\bibitem[Roche \& Aitken(1985)]{roche:85}
Roche, P.~F., \& Aitken, D.~K.\ 1985, \mnras, 215, 425

\bibitem[Rothman et al.(1996)]{hitemp}
Rothman, L.~S., et al.\ 1996, Journal of Quantitative Spectroscopy and
Radiative Transfer, 60, 665

\bibitem[Rothman et al.(2005)]{hitran}
Rothman, L.~S., et al.\ 2005, Journal of Quantitative Spectroscopy and
Radiative Transfer, 96, 139

\bibitem[Schuller et al.(2006)]{schuller:06}
Schuller, F., Omont, A., Glass, I.~S., Schultheis, M., Egan, M.~P.,
\& Price, S.~D.\ 2006, \aap, 453, 535

\bibitem[Schuller et al.(2003)]{schuller:03}
Schuller, F., et al.\ 2003, \aap, 403, 955 

\bibitem[Schultheis et al.(2003)]{schultheis:03}
Schultheis, M., Lan{\c c}on, A., Omont, A.,
Schuller, F., \& Ojha, D.~K.\ 2003, \aap, 405, 531

\bibitem[Schultheis et al.(2009)]{schultheis:09}
Schultheis, M., Sellgren, K., Ram{\'{\i}}rez, S., Stolovy, S.,
Ganesh, S., Glass, I.~S., \& Girardi, L.\ 2009, \aap, 495, 157 

\bibitem[Seale et al.(2011)]{seale:11} 
Seale, J.~P., Looney, L.~W., Chen, C.-H.~R., 
Chu, Y.-H., \& Gruendl, R.~A.\ 2011, \apj, 727, 36 

\bibitem[Serabyn \& Morris(1996)]{serabyn:96}
Serabyn, E., \& Morris, M.\ 1996, \nat, 382, 602 

\bibitem[Skrutskie et al.(2006)]{skrutskie:06}
Skrutskie, M.~F., et al.\ 2006, \aj, 131, 1163

\bibitem[Smith et al.(1978)]{smith:78}
Smith, L.~F., Biermann, P., \& Mezger, P.~G.\ 1978, \aap, 66, 65

\bibitem[Stolovy et al.(2006)]{stolovy:06}
Stolovy, S., et al.\ 2006, Journal of Physics Conference Series, 54, 176

\bibitem[Wannier(1980)]{wannier:80}
Wannier, P.~G.\ 1980, \araa, 18, 399

\bibitem[Warren et al.(2007)]{warren:07}
Warren, S.~J., et al.\ 2007, \mnras, 375, 213 

\bibitem[Werner et al.(2004)]{werner:04}
Werner, M.~W., et al.\ 2004, \apjs, 154, 1

\bibitem[White et al.(2005)]{white:05}
White, R.~L., Becker, R.~H., \& Helfand, D.~J.\ 2005, \aj, 130, 586 

\bibitem[Whittet et al.(2007)]{whittet:07} Whittet, D.~C.~B.,
Shenoy, S.~S., Bergin, E.~A., Chiar, J.~E., Gerakines, P.~A., Gibb, E.~L.,
Melnick, G.~J., \& Neufeld, D.~A.\ 2007, \apj, 655, 332

\bibitem[Whittet et al.(2009)]{whittet:09}
Whittet, D.~C.~B., Cook, A.~M., Chiar, J.~E., Pendleton, Y.~J., Shenoy, S.~S., 
\& Gerakines, P.~A.\ 2009, \apj, 695, 94

\bibitem[Whitney et al.(2003)]{whitney:03}
Whitney, B.~A., Wood, K., Bjorkman, J.~E., \& Cohen, M.\ 2003, \apj, 598, 1079

\bibitem[Whitney et al.(2004)]{whitney:04}
Whitney, B.~A., Indebetouw, R., Bjorkman, J.~E.,
\& Wood, K.\ 2004, \apj, 617, 1177

\bibitem[Yusef-Zadeh et al.(2004)]{yusefzadeh:04}
Yusef-Zadeh, F., Hewitt, J.~W., \& Cotton, W.\ 2004, \apjs, 155, 421 

\bibitem[Yusef-Zadeh et al.(2009)]{yusefzadeh:09}
Yusef-Zadeh, F., et al.\ 2009, \apj, 702, 178 

\bibitem[Zasowski et al.(2009)]{zasowski:09} 
Zasowski, G., Kemper, F., Watson, D.~M., Furlan, E., 
Bohac, C.~J., Hull, C., \& Green, J.~D.\ 2009, \apj, 694, 459 

\bibitem[Zoonematkermani et al.(1990)]{zoonematkermani:90} 
Zoonematkermani, S., Helfand, D.~J., Becker, R.~H., White, R.~L., 
\& Perley, R.~A.\ 1990, \apjs, 74, 181 

\end{thebibliography}
\end{document}